\pdfoutput=1
\documentclass[11pt]{article}

\usepackage[affil-it]{authblk} 
\usepackage[authoryear]{natbib}
\usepackage{gensymb}
\usepackage{amsmath}
\usepackage{amssymb}
\usepackage{graphicx}
\usepackage{float}
\usepackage{booktabs}
\usepackage[a4paper, top=1in,bottom=1in,left=1in,right=1in,marginparwidth=0cm]{geometry}
\usepackage[title]{appendix}
\usepackage{pdflscape}
\usepackage{longtable}
\usepackage{setspace}
\usepackage{adjustbox}
\usepackage{array}
\usepackage{caption}
\usepackage{chngcntr}
\usepackage{mathtools}
\usepackage{lineno}
\usepackage{makecell}
\usepackage{hyperref}
\hypersetup{colorlinks=true,allcolors=blue}

\newcolumntype{R}[2]{%
    >{\adjustbox{angle=#1,lap=\width-(#2)}\bgroup}%
    l%
    <{\egroup}%
}
\newcommand*\rot{\multicolumn{1}{R{45}{1em}}}

\newcommand{\citepos}[1]{\citeauthor{#1}'s \citeyear{#1}}
\renewcommand{\eqref}[1]{Eq.\,\ref{#1}}

\title{Explaining excitable population dynamics in bark beetles: From life history to large, episodic outbreaks}
\author[1,*]{Evan C. Johnson}
\author[2]{Antonia Musso}
\author[3]{José F. Negrón} 
\author[4,5]{Mark A. Lewis}
\affil[1]{Mathematical and Statistical Sciences; University of Alberta; Edmonton, Alberta, Canada}
\affil[2]{Biological Sciences; University of Alberta; Edmonton, Ontario, Canada}
\affil[3]{USDA Forest Service; Rocky Mountain Research Station; Fort Collins, CO, USA}
\affil[4]{Department of Mathematics and Statistics; University of Victoria; Victoria, British Columbia, Canada}
\affil[5]{Department of Biology; University of Victoria; Victoria, British Columbia, Canada}
\affil[*]{Corresponding author: Evan Johnson, ecjohns1@ualberta.ca}

\date{} 

\begin{document}

\maketitle 

\section*{Open Research statement}

All code and data will be made available upon publication at \url{TBD}


\newpage

\tableofcontents 

\newpage 

\section*{Abstract}  

Bark beetles are significant forest pests, with some species capable of causing widespread tree mortality. Among these, the mountain pine beetle (MPB) stands out for its exceptionally destructive outbreak in the 2000s. We use MPB as a case study to explore the concept of \textit{excitable dynamics}, where ephemeral perturbations produce large excursions from equilibrium. Our empirically-calibrated model of reveals five features of MPB biology: stand density-dependent dispersal, an Allee effect, time-scale separation between beetle and tree life cycles, tree size-dependent fecundity, and a large-tree preference. The first three features explain MPB's characteristic boom-bust dynamics, while the latter two explain outbreak magnitude. Other bark beetles lack one or more of these traits, partially explaining their generally lower impact. However, predicting bark beetle impact requires consideration of both life history and landscape factors: total damage increases linearly with host-tree biomass, but this relationship holds only for irruptive (i.e., excitable) beetle species. We distill our findings into a minimal mechanistic model that captures the essence of irruptive bark beetle dynamics. This model firmly establishes MPB as one of the first empirical examples of excitable dynamics in ecology.

Keywords: bark beetle, mountain pine beetle, outbreak, excitable dynamics, lodgepole pine, forest pest, population dynamics, density dependence

\newpage

\section{Introduction} \label{Introduction}


Bark beetle (Coleoptera: Curculionidae, Scolytinae) outbreaks occur worldwide \citep{lantschner2023spatiotemporal}, with recent epidemics in western North America causing widespread tree mortality \citep{meddens2012spatiotemporal, taylor2006forest}. These outbreaks exhibit a distinctive pattern: unpredictable and explosive population growth followed by rapid declines (Fig.\,\ref{fig:sample_time_series}). The episodic ``boom-bust'' pattern contrasts sharply with the population dynamics of most species, which typically show either stochastic fluctuations around a weakly-attracting equilibrium or cycles driven by resource-consumer interactions \citep{louca2015detecting, hassell1976patterns, ellner1995chaos}. In this paper, we demonstrate that certain bark beetle species are unique in that their population dynamics can be classified as ``excitable''. 




Excitable population dynamics are characterized by several features that set them apart from other types of dynamic behaviors \citep{murray1989mathematical, morozov2009excitable, hastings2021effects}. Perturbations trigger large excursions from equilibrium, leading to dramatic population increases from minor or transient environmental changes. These excursions involve both positive and negative feedback mechanisms, creating a self-reinforcing cycle of growth followed by an eventual crash. Excitable systems often exhibit ``slow-fast dynamics'' \citep{hastings2018transient}, where one state variable changes rapidly (e.g., beetle population size), while another changes slowly (e.g., tree population size). This results in a \textit{refractory period}, the minimum time between excursions. In summary, the notion of excitability (while lacking a formal mathematical definition; \citealp[pg. 118]{strogatz2018nonlinear}) involves perturbations causing large but ultimately transient deviations from a stable equilibrium. The canonical example of excitable dynamics is the action potential of a neuron \citep{hodgkin1952quantitative}: A perturabation (neurotransmitters from neighboring neurons) and positive feedback causes a neuron to fire, and slow-fast dynamics (related to conformal changes in ion channels) result in a refractory period. In the case of neurons, excitability is adaptive, ensuring unidirectional signal propagation and regulating firing frequency \citep{johnston1994foundations}.


Bark beetle outbreaks display these hallmarks of excitability. A perturbation such as drought or wind damage can increase tree susceptibility to beetle attack \citep{singh2024understanding}. This initiates a positive feedback loop: larger beetle populations overcome bigger, better-defended trees, further increasing their numbers. Irruptive bark beetles must mass attack host trees to overcome toxic resin defenses \citep{thalenhorst1958grundzuge, raffa1983role, boone2011efficacy}, using aggregation pheromones and tree volatiles to focus their efforts \citep{raffa2001mixed, seybold2006pine}. Many bark beetle populations are thought to exhibit a \textit{strong Allee effect}, wherein per capita growth rates become negative at sufficiently low population densities \citep{taylor2012allee}. Negative feedback when beetles run out of suitable or detectable host trees. Post-outbreak, trees must grow until their phloem or cambium is thick enough to support beetle development, which generates a slow-fast dynamic and a minimum interval between outbreaks. The perturbation that triggers the outbreak, however, is effectively random --- there is no maximum interval. Therefore, bark beetle outbreaks occur irregularly (episodic) rather than on a fixed schedule (cyclical).

\textbf{We will demonstrate that bark beetles are unique in ecology as the only known taxonomic group known to exhibit excitable population dynamics.} While the spruce budworm (\textit{Choristoneura fumiferana}) is sometimes cited as an example, this is probably incorrect. An influential mathematical paper \citep{ludwig1978qualitative} proposed that excitability in the spruce budworm system could result from interactions between trees, spruce budworms, and their avian predators. Specifically, birds with a type III functional response \citep{holling1959components} led to low predation rates at low budworm density due to prey-switching (preventing budworm extirpation between outbreaks), and saturating rates at high density due to handling time constraints (generating positive density dependence during incipient outbreaks). The slow part of the slow-fast dynamics is attributed to the fact forest quality rebounds slowly, and that avian predation is assumed to be proportional to forest quality. However, current evidence indicates that spruce budworms undergo noisy cycles due to their interactions with parasitoids \citep{royama1984population, berryman1996causes, royama1997population, regniere2007ecological, pureswaran2016paradigms} rather than excitable dynamics. Although budworms may face mate-finding challenges at low densities \citep{regniere2013mate}, there is little evidence for a demographic Allee effect over larger spatial scales. 


Examples of excitable dynamics in ecology have been explored theoretically but lack empirical support. Models of algal blooms suggest the possibility of excitable dynamics triggered by a temporary influx of nutrients, with positive feedback arising from either top-down or bottom-up control. The top-down mechanism is a saturating functional response where zooplankton become satiated \citep{truscott1994equilibria, truscott1994ocean}; negative feedback comes from predation, and the slow-fast dynamics result from disparate generation times of zooplankton and phytoplankton. The bottom-up mechanisms involves the formation of algal mats, which improves nutrient cycling and accelerate algal growth \citep{huppert2002model, huppert2004bottom}; the negative feedback comes from nutrient depletion, and the slow-fast dynamics result from slow replenishment of nutrients from runoff, decomposition, etc. Excitable algal blooms are likely to occur somewhere, given the diverse character of algal blooms worldwide \citep{glibert2018global}, anecdotes of threshold-effects \citep{pearsall1932phytoplankton, lund1950studies}, and correlations between the bloom occurrence and environmental fluctuations (e.g., \citealp{freund2006bloom}). To the best of our knowledge, however, there is no well-justified and empirically calibrated model that provides clear evidence for excitable algal blooms. Small mammal populations can experience severe outbreaks, but these are typically classified as pulse-driven rather than irruptive \citep{andreassen2021population}. Pulse-driven outbreaks occur due to significant increases in resource availability, such as abundant vegetation following unusually high precipitation, rather than small or transient perturbations characteristic of excitable systems \citep{holt2008theoretical}.

Some bark beetle species undergo pulse-driven outbreaks triggered by logging, windfall, or drought. Definitionally, these outbreaks subside shortly after the disturbance ends. For example, prolonged drought in the western United States from 1995--2004 \citep{breshears2005regional} depleted trees' defensive resin reserves, which led to an outbreak of \textit{pinyon Ips} beetles. The outbreak peaked from 2002 to 2004, and infestation densities fell by nearly two orders of magnitude after the drought ended \citep{raffa2008cross}. In contrast, excitable bark beetle outbreaks persist long after the initial disturbance subsides. An outbreak of the North American Spruce Beetle (\textit{Dendroctonus rufipennis} Kirby) lasted 13 years following a single windthrow event in 1939 \citep{massey1954biology, anderson2010pollen, spruce_beetle_2011} and an outbreak of the mountain pine beetle (MPB; \textit{Dendroctonus ponderosae} Hopkins) continuing for 10--15 years after the initiating drought ended in 2000 \citep{creeden2014climate}. Bark beetles that exhibit excitable dynamics are termed irruptive or aggressive, while those with pulse-driven outbreaks are classified as semi-aggressive or opportunistically aggressive \citep{lantschner2023spatiotemporal, weed2015population}.


\textbf{In this paper, we ask: why do some bark beetles species display excitable population dynamics, and why are some outbreaks so large?} To answer these questions, we examine the mountain pine beetle as a case study, using insights from this species and comparative analysis to better understand population patterns across bark beetle species.  Our methodology involves four key steps:

\begin{enumerate}
\item Developing and validating a realistic model of mountain pine beetle dynamics, incorporating known biological features.

\item Performing a causal analysis through counterfactual simulations, manipulating various biological features to determine their impact on population dynamics. 
\item Generalizing the results to other irruptive bark beetle species based on shared life history characteristics.

\item Building a minimal mechanistic model which contains only the necessary processes for excitability 
\end{enumerate}

Additionally, we explore the relationship between excitability and the magnitude of bark beetle outbreaks (i.e., the peak number of beetles). In theory, the two concepts are orthogonal --- excitability only implies that outbreaks are large relative to the initiating perturbation. In practice, however, the two concepts are connected. Later, we show that the total area affected by a beetle species is proportional to the combined biomass of their host trees, but that this relationship only applies to excitable (also known as irruptive) species; the relationship does not hold for beetles with pulse-driven outbreaks.


The mountain pine beetle serves as our model species, chosen for its well-documented life history and its significant impact. The most recent outbreak of MPB (ca. 2000--2015) stands as the largest recorded insect outbreak in history  (\citealp{taylor2006forest}), killing up to 8.5 million hectares of tree crowns across western North America \citep{meddens2012spatiotemporal}, and killing $>50\%$ of merchantable pine volume in British Columbia \citep{BCFLNRO2016}. This event had far-reaching consequences, including transforming British Columbia's forests from a carbon sink to a source \citep{kurz2008mountain}, inflicting billions of dollars in economic damage \citep{corbett2016economic}, and altering hydrological functioning \citep{schnorbus2011synthesis}.

\begin{figure}[H]
\centering
\includegraphics[scale = 1]{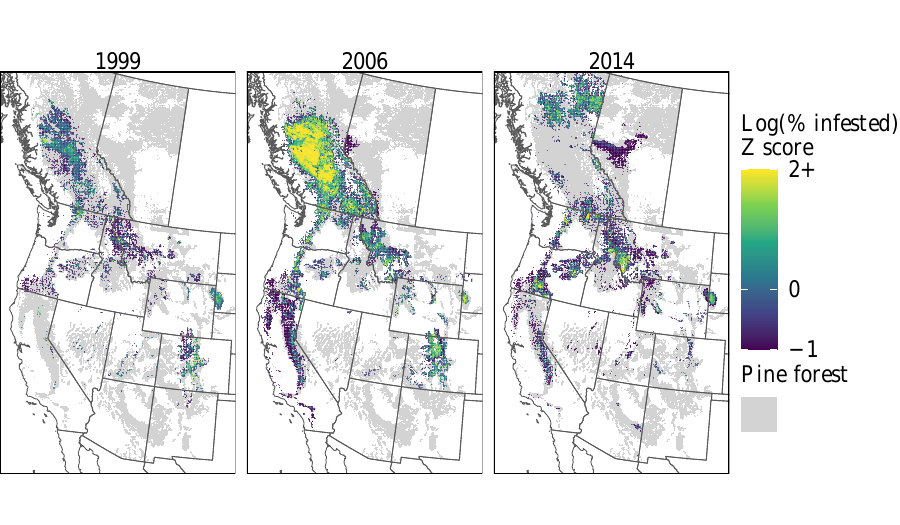}
\caption{Progression of the latest large-scale mountain pine beetle outbreak. The data comes from fixed-wing Aerial Overview Surveys (United States and British Columbia) and Heli-GPS surveys (Alberta). Pine forest data comes from \citet{beaudoin2014mapping} and \citep{ellenwood2015national}.}
\label{fig:liber_ero_map}
\end{figure}

Of course, the question of ``Why is mountain pine beetle so destructive?'' has been studied extensively. Climate change is the principal reason for the severity of the outbreak in the 2000s relative to past outbreaks \citep{carroll2006impacts, alfaro2009historical, creeden2014climate}. Widespread drought conditions caused synchronized increases in tree susceptibility. At the same time, warmer winter temperatures reduced overwintering mortality in beetle larvae. Fire suppression has also played a role, by enabling the buildup of large-diameter pines \citep{taylor2006forest}. Indeed, a comparative analysis of bark beetles found that the best predictor of outbreaks is the availability of susceptible host trees (\citealp{koricheva2012life}). Mountain pine beetle was far more destructive in British Columbia than the western United States \citep{meddens2012spatiotemporal}. Although this spatial pattern is not well-researched, we suspect that it involves the high density and spatial continuity of lodgepole pine in the Chilcotin Plateau –-- the epicenter of the last two outbreaks in British Columbia.




Our approach differs from previous research on bark beetle dynamics in several key aspects. Rather than focusing on the (proximal) environmental causes of outbreaks, we focus on the (ultimate) beetle life-history causes of outbreaks. Wee present quantitative and empirically-calibrated models that offer more precision than the verbal descriptions found in other reviews of bark beetle dynamics (e.g. \citealp{coulson1985forest, raffa2008cross, weed2015population,  biedermann2019bark, lantschner2023spatiotemporal, singh2024understanding}) and allow us to identify a minimal set of processes that generate characteristic boom-bust dynamics. Other studies have attempted to link life-history predictors to outbreaking behavior with statistical comparisons (e.g., \citealp{nothnagle1987forest, koricheva2012life, kozlov2010diverse}), but this approach faces significant limitations. It's often unclear how certain predictors are mechanistically connected to outbreaks. For example, larger bark beetles are more prone to outbreaks, contradicting $r-K$ selection theory's prediction that smaller beetles should have higher growth rates and thus greater outbreak potential \citep{koricheva2012life}. More critically, there is very little information available about the biology of non-outbreaking species \citep{nothnagle1987forest, johns2016population}. To circumvent these difficulties, we build a mechanistic model of the bark beetle species for which the most information is available --- the mountain pine beetle --- and then extrapolate to other species based on shared life history . 

The study of excitable dynamics in bark beetles has significant practical implications. The positive feedback mechanisms inherent in excitable systems mean that outbreaks can extend far beyond the spatial extent of an initial disturbance, making them widespread and difficult to manage. This suggests the existence of a ``Goldilocks zone'' of population densities where management efforts will be most effective –-- too low, and the strong Allee effect naturally limits population growth; too high, and the outbreak becomes unmanageable (\citealp{hodge2017strategic}, Fig. 6). After a major outbreak occurs, there is a known window of time --- a ticking clock --- before the next outbreak could occur. This period offers forest managers a timeline for implementing preventive measures, such as creating diverse stands in terms of age structure and species composition \citep{dymond2014diversifying, burton2010striving}.


\subsection{The biology of mountain pine beetle and lodgepole pine} \label{The mountain pine beetle life-cycle}

Adult beetles emerge in July or august and After emergence these beetles fly in search of new host trees, typically targeting large pines during the epidemic population phase. Mountain pine beetle has infested Jack pine and Rocky Mountain bristlecone pine, but the general suitability of these species is a matter of ongoing research \citep{bentz2019mountain, bentz2022great, bleiker2023suitability}. Individual beetles may fly around for several days, but few beetles survive more than 5--10 days outside of the tree \citep{reid2008fluorescent, fuchs1985pre}. Emergence from natal trees is synchronous due to temperature-dependent development thresholds (\citealp{powell2009connecting, logan2001ghost}), but the synchronicity is not perfect, and so the population-level flight period may last up to 6 weeks \citep{bleiker2016flight}. 

Because individual beetles cannot be tracked, the dispersal phase of the life cycle is mildly mysterious. Available evidence points to three modes of dispersal: short-, medium-, and long-range dispersal. Most beetles disperse less than 30-50 meters under the canopy (\citealp{safranyik1992dispersal}; \citealp{robertson2007mountain}; for the Douglas-fir beetle \citealp{dodds2002sampling}). A minority of beetles disperse up to 5 km \citep{robertson2009spatial, powell2014phenology, howe2021climate, koch2021signature}, performing pioneer attacks and aggregating to form semi-regular patterns of ``spot infestations'' (clusters of infested trees) across the landscape \citep{strohm2013pattern}. The smallest fraction of beetles (0--2.5\%) fly above the canopy \citep{robertson2009spatial, safranyik1992dispersal}, where they may get carried up to 300 km by upper atmospheric winds \citep{Hiratsuka1982, Cerezke1989, jackson2008radar}. Some experts have speculated that the fraction may be higher in stands where most suitable host trees have already been killed \citep{carroll2003bionomics, de2011incoming, bleiker2019risk}. 




Upon locating a suitable host, MPB females initiate attacks. The trees fight back by exuding resin, which serves to entrap or \textit{pitch out} (i.e. expel) attacking MPB. Female pioneers release \textit{trans} and \textit{cis}-verbenol, aggregation pheromones that draw more beetles to the site \citep{pureswaran2000dynamics, seybold2018management}. As males arrive, they emit \textit{exo}-brevicomin, another aggregation pheromone that specifically attracts additional females \citep{rudinsky1974antiaggregative}. Many bark beetle species use aggregation pheromones and host volatiles to coordinate \textit{mass attacks}, where swarms of beetles overwhelm a single tree. Depending on the beetle population density, tree suitability, and the tree's defensive capability, successful mass attacks may last from a couple of hours to a week, continuing until the tree's resin supply is depleted.  


Following a successful mass attack, females bore into the tree and excavate egg galleries. Mating occurs within the tree, and males will either seek out additional females or ``settle down'' to help with gallery defense. Eggs hatch within one to two weeks, and larvae immediately start feeding on the tree's phloem. The beetles carry fungal spores in specialized mouthparts, thereby spreading and inoculating the phloem tissue with fungi. Beetle larvae and adults rely on fungal hyphae and spores for nutrition \citep{reid1958behaviour, bleiker2007dietary}. Tree death is hastened by MPB's fungal symbionts, blue stain fungi \textit{spp.}, which colonize the tree's transport vessels \citep{arango2016differences}. Larvae mine horizontal galleries that terminate in an oval pupal chamber which is lined with spores.


Mature larvae overwinter after accumulating cyroprotectants that allow them to survive temperatures as low as -30C \citep{bentz1999ecology}. Other life stages are more susceptible to cold mortality particularly to cold snaps that are early or late in their overwintering period \citep{wygant1940effects, safranyik1991unseasonably, logan1995assessing}. Other life stages are more susceptible to cold mortality particularly to cold snaps that are early or late in their overwintering period \citep{regniere2007modeling, rosenberger2017cold}. Come spring, surviving larvae resume feeding and pupate in the late spring or early summer. By the time of adult emergence, approximately one year after the initial attack, the needles of successfully-attacked trees have turned a rust-red color. This telltale diagnostic makes it possible to gather extensive data through aerial surveys. 


Lodgepole pine and ponderosa pine are the predominant hosts of MPB, simply because they are the most abundant pine species in MPB's geographic range. Lodgepole pine dominates in British Columbia, western Alberta, and the American Rocky Mountain states, whereas ponderosa pine is more abundant in the American southwest and the Sierra mountains \citep{beaudoin2014mapping, ellenwood2015national}. Both species are shade-intolerant and fire-resistant. They tend to be seral (i.e., persisting as a non-climax species in a fire cycle), taking advantage of high-light conditions following a stand-leveling fire \citep{fryer2018pinus, anderson2003pinus, lotan1985role}. Lodgepole pine has closed cones with seeds that remain viable for decades, effectively creating a seed bank \citep{teste2011seed, teste2011viability}. Due to its abundant seeds and its ability to thrive in high-light conditions, lodgepole pine can take advantage of the light gaps created in the years following a mountain pine beetle attack, resulting in the recruitment of seedlings into approximately even-aged cohorts \citep{axelson2010changes}. However, this regeneration pattern is not universal across MPB's range; in Alberta, for example, the density of standing dead trees following MPB outbreaks prevents sufficient light penetration, leading to the replacement of lodgepole pine by more shade-tolerant species \citep{axelson2009influence}.


During epidemics, MPB seeks out large healthy trees with abundant phloem resources. However, in their low-density endemic state, there are not enough beetles to mass attack a single healthy tree, and thus there is a collective behavioral switch where beetles seek out trees that are weakened or stressed, typically due to root disease, wind damage, or lightning \citep{tkacz1986association, carroll2006mountain}. Endemic populations tend to remain at low density, primarily due to resource limitation, but also because of natural enemies and competition with other bark and woodboring beetle species \citep{amman1983mountain, safranyik2006biology}. Endemic populations of MPB can transition into an incipient-epidemic state when they successfully breed within trees that have thick phloem. This scenario can occur if a healthy tree experiences acute stress (such as wind damage or lightning) right before the MPB flight period, such that pulse-driven bark beetles have not yet extensively colonized the tree \citep{bleiker2014characterisation}. Such stressors can trigger local outbreaks (on the scale of square kilometers). However, widespread outbreaks are usually preceded by widespread drought conditions, which synchronously weaken the resin defenses of many trees \citep{ alfaro2009historical, creeden2014climate, kolb2016observed}. 

\subsection{Mountain pine beetle life history} \label{Mountain pine beetle life history traits}

A key element of this paper is a nonlinear dynamical model that captures key features of MPB biology.  In this section, we review life history with a view toward features that should be emphasized in the model, like the ways in which MPB  ``decides'' about what trees to attack, and how MPB responds to variation in the density of host trees and conspecific beetles. Our list is not exhaustive. Previous research has identified some biological processes that do not show up as a clear pattern in our data, or that we cannot feasibly incorporate into our models. These are discussed further in Appendix \ref{Additional model justification}. 



\subsubsection{Allee effect} \label{life_history:allee}

Mountain pine beetle is subject to a strong Allee effect, a phenomenon where per capita growth rates become negative at a sufficiently low population density. The Allee effect arises because a threshold number of beetles must attack a single tree to successfully kill it. Another common mechanism of an Allee effect, poor mate finding in small populations, has been ruled out by previous research \citep{amman1980incidence, bleiker2014characterisation}.

The Allee effect is mitigated by MPB's ability to aggregate. In general, aggregation counteracts the negative effects of positive density dependence, just as dispersal counteracts the negative effects of negative density dependence \citep{crowley1981dispersal, briggs2004stabilizing}. Aggregation pheromones and volatile monoterpene from attacked trees attract additional beetles. However, this chemical signal may be dampened by unsuitable host-tree chemistry \citep{hynum1980dendroctonus, boone2011efficacy}, unsuitable natal-tree chemistry (which negatively affects aggregation pheromone production; \citealp{chiu2018monoterpenyl}), or simply a limited ability to detect aggregation pheromones at-distance. Evidence from small-scale spatial maps, as well as mechanistic pheromone diffusion models, suggest that beetles primarily respond to aggregation pheromones that are 20-100 meters from their tree-of-origin ( \citealp[Fig. 4--5]{mitchell1991analysis}); \citealp{strohm2013pattern}). There is limited evidence that the detection limit is 500-1000 meters under the right wind conditions \citep{barclay1998trapping}. In summary, the Allee effect persists due to various physical and chemical factors that constrain the strength of aggregation.

It has long been recognized that the size disparity between beetles and trees means that many beetles must attack a single tree, and therefore, that outbreaks require a threshold or critical density of beetles \citep{thalenhorst1958grundzuge, safranyik1975interpretation, berryman1978synoptic}. However, explicit evidence for the Allee effect in MPB is not always forthcoming. Studies have detected an Allee effect when focusing on tree-level attack density (e.g., beetle entry holes per unit tree surface area). An Allee effect is implied by a positive relationship between attack density and the probability of tree death \citep{waring1983physiological, waring1985modifying, peterman1974some}, a positive relationship between attack density and the probability of reproduction \citep[Fig. 5]{raffa1983role}, or a positive relationship between the number of attacked trees and the probability of tree death \citep{boone2011efficacy}. On the other hand, studies \textit{do not} detect an Allee effect when they focus solely on the density of infested trees, treating this quantity as a proxy for beetle population size \citep{macquarrie2011density, tunnock1970chronic, trzcinski2009intrinsic}.

\subsubsection{Stand-Dependent dispersal mortality} \label{life_history:stand_dependent_dispersal}

Mountain pine beetle predominantly disperses from stands only after most large-diameter trees are killed. The vast majority of these emigrants die on their journey due abiotic elements or an inability to find a suitable host tree. Beetles that do survive may be less likely to attack suitable host trees, given the low energetic conditions resulting from a long flight \citep{jones2020mechanisms}. Since stand-dependent dispersal and the associated mortality are somewhat obscure features of MPB dynamics, we will take a moment to discuss supporting evidence. 

The most likely explanataion for stand-dependent dispersal is chemosensory constraints: MPB is simply unable to sense and aggregate  on small and medium-sized trees. Multiple lines of evidence support this explanation --- MPB primarily attacks small or non-host trees when they are adjacent to large host trees \citep{mitchell1991analysis, preisler1993colonization, amman1985mountain, huber2009successful}, and studies have shown large trees have higher monoterpene concentrations, which may be the main determinant of whether pioneer female beetles initiate an attack \citep{boone2011efficacy, seybold2006pine, howe2022landscape}.

Alternative explanations for stand-dependent dispersal have less support. One considers the fact that beetle infestations alter the forest microclimate \citep{bartos1989microclimate, amman1998silvicultural}. As stands are thinned, increased wind disrupts MPB aggregation pheromone plumes, and increased light may induce MPB to fly. However, the microclimate hypothesis is undermined by timing --- trees take 5-20 years to fall \citep{lewis2006rate, hansen2014forest} and 2--7 years to lose needles \citep{klutsch2009stand, audley2021dynamics}, while populations crash within 1--2 years (Fig.,\ref{fig:sample_time_series}). The inclusive fitness hypothesis suggests beetles refuse to consider medium-sized trees because large trees engender higher fitness, but this is unlikely given high dispersal mortality and positive fitness in medium trees (Fig.,\ref{fig:fitness_across_dbh}).

\begin{figure}[H]
\centering
\makebox[\textwidth]{\includegraphics[scale=0.8]{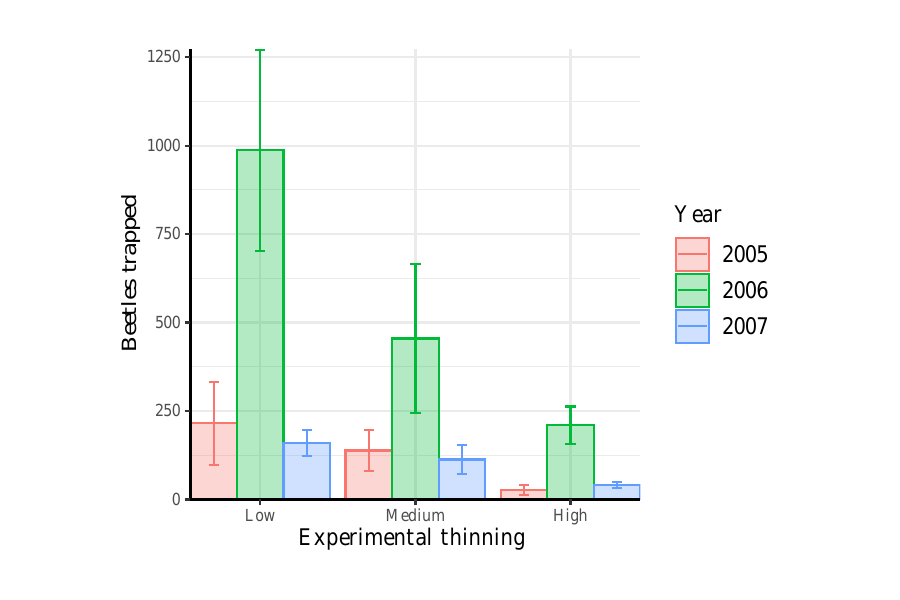}}
\caption{Fewer beetles are trapped in heavily thinned stands, suggesting that beetles disperse away from sparse forests. The y-axis displays the mean and standard errors of beetles in passive traps in Fraser Experimental Forest's (Colorado) managed stands over the full flight period. Thinning levels are categorized as 'Low', 'Medium', and 'High', with corresponding Growth Stocking Levels (GSL) of 120, 80, and 40, respectively. The small size of the experimental blocks (0.2 ha) suggests that captured beetles predominantly originated from adjacent unmanaged forest, in turn suggesting that the data reflect the impact of tree density on dispersal behavior, not on the within-block infestation density. Data comes from \citet{negron2018biological}.}
\label{fig:negron_trap}
\end{figure}

\begin{figure}[H]
\centering
\includegraphics[scale = 1]{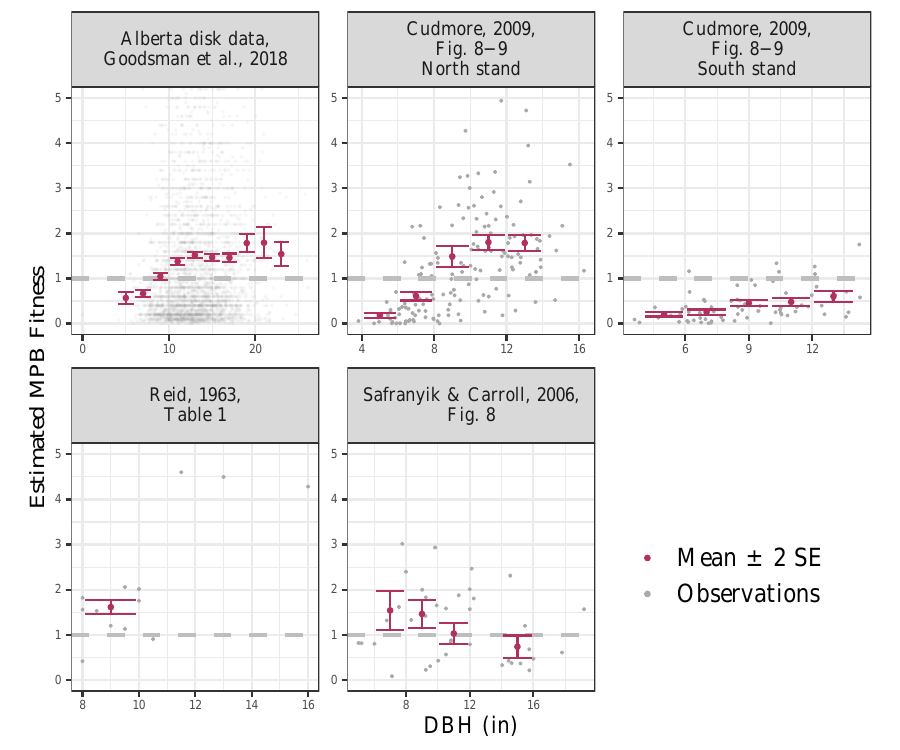}
\caption{Mountain pine beetle tends to have positive fitness when attacking medium-sized trees (approx. 8--14 inches DBH). Ecological fitness is estimated with $(0.3) \times$ \textit{females out of the tree} / \textit{females into the tree}, where the factor $0.3$ is the approximately the median dispersal-phase survival in the dataset from \citeauthor{klein1978attack} (See Fig.\,\ref{fig:mort2}). Depending on the study, gallery starts or entry holes serve as estimates for \textit{females into the tree} (note that only females initiate galleries). $(2/3) \times (3/2) \times$ Emergence holes, or  $(2/3)$ of the overwintered brood serve as estimates of \textit{females out of the tree}. Note that (2/3) is the typical proportion of females \citep{reid1962biology, cole1976mathematical}, and that (3/2) is the average number of beetles emerging from each hole \citep[Fig. 13]{safranyik1985relationship, peterman1974some}. Data comes from \citet{goodsman2018effect, cudmore2009geographic, reid1963biology, safranyik2006biology}. The means and standard errors were only computed for bins with at least 5 observations.}
\label{fig:fitness_across_dbh}
\end{figure}

Even though we cannot track individual beetle dispersal, there is robust evidence for stand-dependent dispersal. In addition to the previously discussed evidence, trapping experiments have shown that far more beetles are caught in unthinned stands than thinned stands, regardless of whether pheromone traps \citep{bartos1989microclimate, schmitz1989effect}, or passive traps are used (Fig. \ref{fig:negron_trap}). \citet{salle2007interactions} have shown that beetles with low body weight had proportionally more lipids. Since beetles are more reactive to aggregation pheromones once lipids are depleted, this explicates a physiological mechanism by which beetles can escape depleted stands. Several modeling studies explicitly infer substantial stand-dependent dispersal\citep{powell2014phenology, powell2018differential, howe2021climate}. Other, more complicated models implicitly assume stand-dependent dispersal, and then show that this assumption leads to realistic population dynamics (e.g., \citealp{powell2018differential, bone2014modeling, kautz2016dispersal}). 

\begin{figure}[H]
\centering
\makebox[\textwidth]{\includegraphics[scale=0.8]{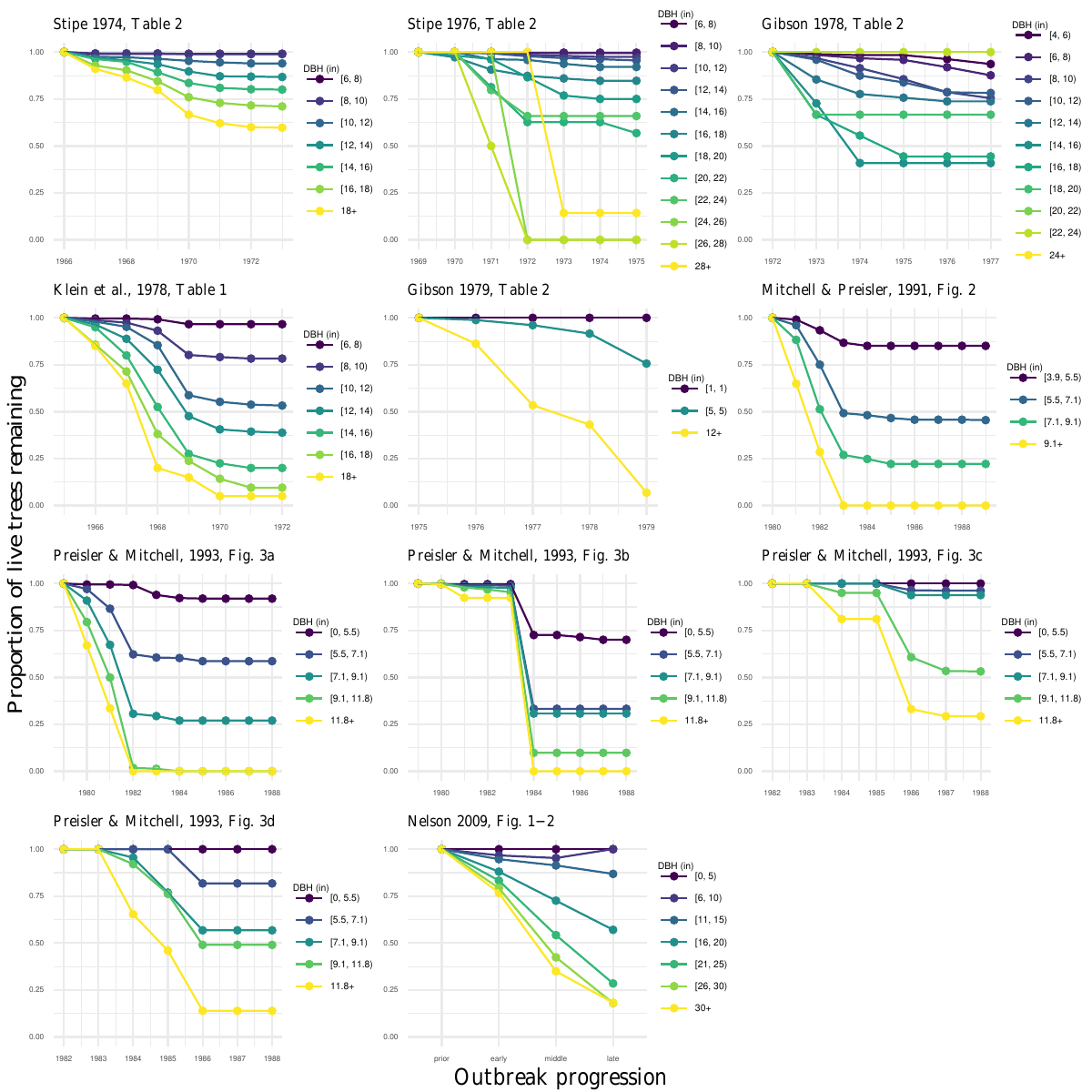}}
\caption{Outbreaks typically end before a majority of medium-sized trees (approx. 8--14 inches DBH) are killed. The figure depicts the proportion of remaining live trees as a function of time, stratified by diameter class (diameter at breast height, DBH). Data comes from \citet{stipe1974yearly, stipe1976trends, gibson1978damage, gibson1979an, klein1978attack, preisler1993colonization, nelson2009effect}.}
\label{fig:dbh_across_outbreaks}
\end{figure}

The most compelling evidence for stand-dependent dispersal is the fact that outbreaks invariably end before the vast majority of medium-sized trees are killed (Fig.\,\ref{fig:dbh_across_outbreaks}), even though beetles have positive fitness on medium-sized trees (Fig.\,\ref{fig:fitness_across_dbh}). Previous research \citep{goodsman2018effect} has explained this pattern —-- where an outbreak ends sooner than expected —-- as a result of \textit{overcompensatory density dependence}, a phenomenon where increased population size now leads to a decreased population size in the next generation. However, there is no evidence for overcompensatory density dependence in MPB. Attacking beetles pack themselves into trees at a density that roughly optimizes individual fitness \citep{raffa1983role}. When attack densities do exceed the theoretical optimum, there is no clear change in emergence densities (Fig.\,\ref{fig:disk_data_pop_map}; \citealp{berryman1974dynamics}, Fig. 2c; \citealp{berryman1976theoretical}; Fig. 2, \citealp{raffa1983role}, Fig. 6 shows a slight increase) and there is no clear change in population dynamics (Appendix \ref{subsec:dead}). 

There is also evidence that stand-dependent dispersal mostly results in mortality, although this evidence is indirect. First, it is our observation (based on Aerial Overview Survey data) that outbreaks do not last substantially longer in areas with high contiguous pine biomass. If dispersing beetles were not dying in such areas, emigrating beetles would be replaced by immigrating beetles, and thus outbreaks would be prolonged. Second, mortality is already substantial for beetles traveling short distances: baseline dispersal mortality is estimated to be around 40--80\% \citep{safranyik1989mountain, latty2007pioneer, pope1980allocation}; Fig.\,\ref{fig:mort2}). Third, dispersal puts beetles in a low energetic condition, leading to decreased aggregation pheromone production, and making them less likely to attack trees \citep{jones2019influence, jones2020mechanisms}. Fourth, lab experiments indicate that most beetles only survive several days outside of a tree \citep{fuchs1985pre, reid2008fluorescent}. Fifth, longer flights increase exposure to natural enemies, most notably avian predators \citep{rust1930relation, stallcup1963method}.

\subsubsection{Large-tree preference} \label{life_history:size_pref}


Mountain pine beetle exhibits a strong preference for large-diameter trees during their dispersal/host-selection phase, driven by both proximate sensory cues and ultimate evolutionary benefits. From an evolutionary perspective, large trees offer two key advantages: primarily, they have thicker phloem, which serves as the primary food source for MPB larvae, and secondarily, their thicker bark may provide better insulation during cold winters.

The proximate mechanisms driving a large-tree preference are evident across different stages of host selection. In the initial landing phase, beetles display a mix of random movement and slight preference for large, vertically-oriented black silhouettes \citep{shepherd1966factors, billings1976influence}, indicating that visual cues play a role. Chemo-sensory cues play a relatively minor role during the initial landing phase \citep{billings1976influence, hynum1980dendroctonus, moeck1991primary}. For pioneer attacks, females use physical and gustatory cues associated with tree diameter, such as terpene concentrations \citep{raffa1982gustatory} and bark roughness \citep{reid1963biology}. During mass attacks, larger trees become even more attractive due to increased concentrations of aggregation pheromones and host volatiles, both of which correlate positively with tree size. These sensory preferences align with the evolutionary advantages of large trees, ensuring that beetles select hosts that offer the best resources for their offspring.

Mountain pine beetle's large-tree attack preference is extremely well-established. The probability of tree death increases logistically with tree diameter (\citealp{buonanduci2023fine, rodman2022rocky, morris2022does, negron2020within, negron2017large, bjorklund2009diameter} and 15 sources therein, see Table 1), and stand susceptibility indexes invariably use variables that correlate with tree diameter: tree age, pine volume, or basal area (\citealp{shore1992susceptibility, chojnacky2000mountain, randall2011revised, bentz1993critical}, and sources therein). Depending on site-specific factors, the probability of tree mortality is near-zero for 4--8-inch trees and near-100\% for 12--16-inch trees.

Despite their preference for large trees, MPB have developed strategies to maintain a near-optimal attack density and avoid strong intraspecific competition. Mountain pine beetle will ``fill up'' trees to an approximately invariant density, measured in beetles per square meter of tree trunk surface area, before moving on to the next tree. Interestingly, this attack density is close to being optimal in terms of the population growth rate \citep{raffa1983role}. The mechanism for attaining (approximate) optimality is a suite of aggregation and anti-aggregation pheromones, and the relationship between such pheromones and host tree defenses. Beetles are attracted to the combination of terpenoid aggregation pheromones and volatilized terpenes. Because both are derived from tree resin, trees become less attractive as their resin defenses are depleted \citep{raffa1983role, raffa1988host}.  Anti-aggregation pheromones, which are only detected by beetles at short distances (around a meter; \citealp{bentz1996localized}), redirect beetles to nearby trees, while rivalry stridulations (noises produced by males) help ensure that galleries are evenly spaced. \citet{raffa2001mixed} puts forward several hypotheses for the apparent lack of cheaters --- beetles who ignore the anti-aggregation pheromones and enter a ``full'' tree. 

Beetle attack density remains relatively constant across tree diameters and beetle population sizes. Killed trees receive an approximately invariant number of attacks per unit surface area (Fig.,\ref{fig:size_dead_just3}, \citealp{reid1963biology}, Table 1, \citealp{peterman1974some}, Figs. 4--7). There is less literature on the relationship between attack density and beetle population size, but any such relationship is likely weak, given that 1) \citet{negron2018biological} found no relationship; 2) there is a clear mechanism for limiting attack density (see the previous paragraph), and 3) that attack density varies in the range of 50--100 females per square meter (see Fig. \ref{fig:hist_attack} in the \textit{Results}), despite beetle population sizes varying by several orders of magnitude.

\subsubsection{Tree size-dependent fecundity} \label{life_history:size_fecund}

Mountain pine beetle has higher effective per capita fecundity in larger trees. Large trees have thick phloem, which is the primary food source of larvae and teneral adults. Thin phloem can slow larval development, which can cause mortality if a cold snap occurs before the cold-resistant third \& fourth-instar stages are attained; if larvae do manage to fully develop within a small tree, they may have small body sizes and low energetic condition upon emergence \citep{amman1983mountain}. Additionally, the thick bark of large trees protects against cold winter temperatures. 

From a more abstract perspective, there are two ways in which large trees lead to more beetles: 1) they offer higher emergence densities (per unit of susceptible tree surface area); 2) they have lots of susceptible surface area. We use the term \textit{effective fecundity} specifically to the former, diameter-dependent emergence density; we use ``effective'' because tree diameter doesn't affect brood density per se, but rather brood density after overwintering but before the dispersal phase. Emergence density has a piecewise-linear relationship with tree diameter (\citealp{reid1963biology}, Table 1; \citealp{safranyik1968development}, Fig. 25; \citealp{cole1969mountain}, Fig. 5; \citealp{cudmore2009geographic}, Fig. 6). Small trees produce next-to-zero beetles, whereas large trees produce thousands of beetles per square meter. Susceptible tree surface area also increases linearly with tree diameter, implying that total beetle brood size increases quadratically with tree diameter.

\subsubsection{Time-scale separation} \label{life_history:time_scale}

Trees live longer than beetles. Although this is evident, it sets the MPB-lodgepole system apart from most predator-prey and host-parasitoid systems. Typically, predators live longer than prey, and parasitoids have lifespans that are similar to their hosts' lifespans.

\section{Methods} \label{Methods}

\subsection{Overview} \label{Overview1}

\begin{figure}[H]
\centering
\makebox[\textwidth]{\includegraphics[scale=0.9]{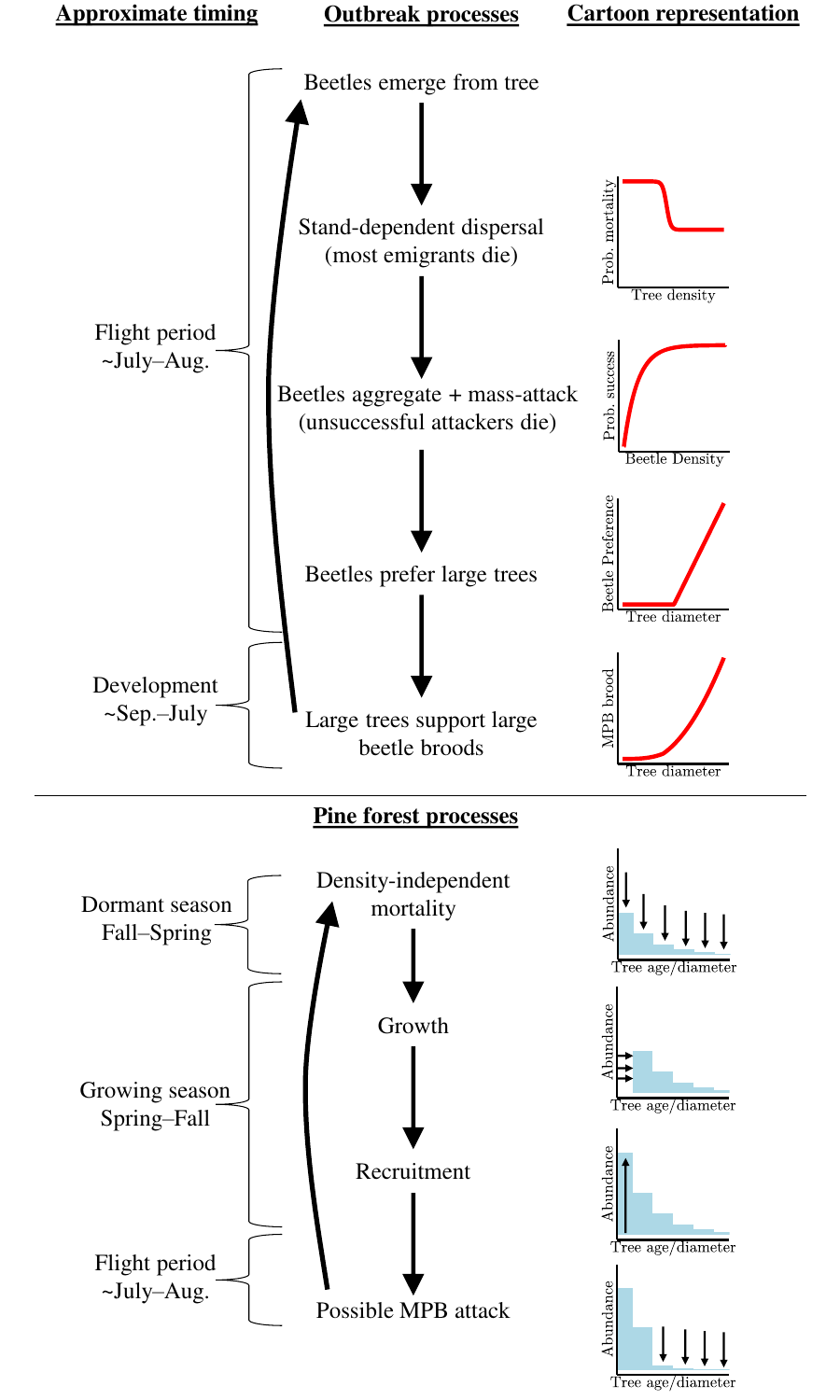}}
\caption{Depiction of the beetle/forest simulation model, and of the sub-models therein.}
\label{fig:model_concept_fig}
\end{figure}

To understand how MPB life history affects MPB population dynamics, we will build a mechanistic model of MPB dynamics (summarized in Fig.\,\ref{fig:model_concept_fig}), then simulate the model under counterfactual scenarios where various biological features have been turned on or off. If a population-dynamical phenomenon (e.g., episodic outbreaks) disappears when a certain feature is turned off (e.g., stand-dependent dispersal mortality), we can say that the feature is a necessary condition for the phenomenon. If two features being turned off makes the difference, then the two features are jointly necessary. As always, our inferences are conditional on the approximate veracity of the model.


``Our model'' is actually constituted by two types of related models: a simulation model and a set of statistical models. The statistical models are used for model-building and for parameter estimation. The parameter estimates feed into the simulation model, which is then used to explore system behavior under various scenarios. The description of the statistical model is quite long, due to a large number of sub-models, equations describing error distributions, and the extensive use of graphical evidence to justify model structure. Therefore, the statistical models are described in Appendix \ref{Statistical model description}, while the simulation model is described below (in Section \ref{Simulation model}).


Realism is the key feature of our model --- we intend to emulate the actual ecological system so that we can conduct virtual experiments. Many models of MPB have been published, but many are too simple for our purposes, having favored mathematical tractability over realism. Other models are too complex, containing functional forms that cannot be reasonably parameterized given available data. For further discussion of previous MPB models, see the \href{Discussion}{\textit{Discussion}} and Appendix \ref{Allee effect}. In contrast, our models are thoroughly justified via clear patterns in the data and/or a preponderance of evidence from previous research (see Appendices \ref{Statistical model description} \& \ref{Additional model justification}). All models were fit to data with \textit{Stan} \citep{stan2020rstan}, a Bayesian model-fitting software that implements Hamiltonian Monte Carlo (details in Appendix \ref{Fitting the models}).



\subsection{Data} \label{Data}

The model was parameterized using data from \citet{boone2011efficacy} and \citet{klein1978attack}. All data was extracted from figures and tables using the \textit{GraphGrabber} software and \textit{Microsoft Excel's} digitizing features.  Specific variables are described in Table \ref{tab:model_notation}. The \citeauthor{boone2011efficacy} dataset contains the number of mass-attacked trees and the number of \textit{successfully} mass-attacked trees, across a range of attack years, stands, and attack densities. This data justifies a functional form for the Allee effect. The second dataset, courtesy of \citeauthor{klein1978attack} (Fig.\,\ref{fig:mort1}), is one of the only datasets that contain beetle attacks, beetle emergence, and tree deaths, all stratified by year and tree diameter class. Our statistical model of MPB outbreak dynamics is primarily calibrated with this dataset, with the notable exception being that a Bayesian prior for the Allee effect parameter is derived from the \citeauthor{boone2011efficacy} data.

\subsection{Simulation model} \label{Simulation model}

\textbf{Mountain pine beetle outbreak dynamics (fast):}

\begin{enumerate}
    \item Beetle survival during the dispersal phase depends on forest conditions. The survival probability is a logistically increasing function of the total surface area of susceptible trees:
    \begin{equation} \label{eq:disp_surv}
        s_y = s_0 \left(1+\exp \left( -\lambda_1 \left( (\text{relative surface area})_y - \lambda_0 \right) \right)\right)^{-1}
    \end{equation}
    where $s_y$ is the survival probability in the year $y$, and $s_0$ is the maximum survival probability. The parameters $\lambda_0$ $\lambda_1$ respectively control the intercept and slope of the logistic function. The \textit{relative surface area} is the ratio of the current total surface area of trees with a diameter at breast height (DBH) of 8 inches or greater to the maximum possible surface area for such trees in an equilibrium forest without mountain pine beetle outbreaks.

    \item Beetle populations experience stochastic perturbations. To obtain the density of attacking beetles, we multiply the number of emerging females per acre, $B_y$, and the dispersal-phase survival probability, $s_y$. Potentially, with probability $P_{I}$, we also add $I$ beetles per acre:
\begin{equation} \label{eq:imm}
(\text{{attacks per acre}})_y = s_y \; B_y + I \times \text{Bernoulli}\left(P_{I}\right).
\end{equation}
The Bernoulli random variable simulates the stochastic endemic $\rightarrow$ incipient-epidemic transition. Because most outbreaks occur during regional epidemics, beetle immigration is the proximal catalyst for many endemic $\rightarrow$ incipient-epidemic transitions on the local scale. But because drought is responsible for region-wide outbreaks, drought is the ultimate catalyst for most local-scale outbreaks. Drought conditions are somewhat unpredictable, but the best predictor of drought in western North America is the Pacific Decadal Oscillation \citep{mccabe2004pacific, hidalgo2004climate}, which has a quasi-period of 15-30 years \citep{mantua2002pacific}. The last two drought periods preceded the last two continent-scale outbreaks and were 20 years apart, \citep[Fig. 5d]{creeden2014climate}. Therefore, we set $P_I = 1/20 = 0.05$. We also set $I =300$ based on a back-of-the-envelope calculation (Appendix \ref{Stochastic perturbation size}).


    \item The probability of a successful mass attack is an exponential saturating function of attack density (i.e., beetles per acre). The density of successful attacks is 
\begin{equation} \label{eq:allee}
\left(\text{succ. attacks per acre}\right)_y = (\text{{attacks per acre}})_y \overbrace{\left(1 - \exp\left(-2.3  \frac{{(\text{{attacks per acre}})_y}}{a}\right)\right)}^{\text{Probability of success}},
\end{equation}
    where $a$ is the Allee threshold parameter (defined as the attack density where the success probability equals 0.9). Unsuccessful beetles die without attacking additional trees.

    \item Beetles prefer larger trees. Beetle preference is a piecewise linear function of tree diameter where ``preference'' is operationalized as the proportion of beetle attacks within a particular diameter class, divided by the proportion of total tree surface area belonging to that class:
    \begin{equation} \label{eq:pref}
        (\text{preference})_j =
        \begin{cases}
            \beta_0 + \beta_1 \cdot \text{DBH}_j & \text{if } \text{DBH}_j > -\beta_0 / \beta_1 \\
            0 & \text{otherwise}.
        \end{cases}
    \end{equation}
    Here, $\beta_0$ and $\beta_1$ are intercept and slope parameters; and $\text{DBH}_j$ is the diameter at breast height of the $j^{\text{th}}$ discrete diameter class. Thus, $-\beta_0 / \beta_1$ is the susceptible DBH threshold for beetle attacks.

    \item Beetles distribute among tree diameter classes based on preference. The number of beetles per acre that successfully attack the $j^{th}$ DBH class is:
    \begin{equation} \label{eq:succ_attacks}
        (\text{succ. attacks per acre})_{j,y} = \frac{(\text{succ. attacks per acre})_{y} \cdot (\text{preference})_{j} \cdot (\text{prop. area})_{j,y}}{\sum_{k=1}^{J} (\text{preference})_{k} \cdot (\text{prop. area})_{k,y}}
    \end{equation}
    where $(\text{succ. attacks per acre})_{y}$ is the total density of successfully attacking beetles, and $(\text{prop. area})_{j,y}$ is the proportion of susceptible tree surface area belonging to the $j^{th}$ DBH class in year $y$.

    \item A tree is killed once it is ``filled-up'' to a density of $1/\gamma$ female beetles per $\text{m}^2$. The number of killed trees per acre (KTPA) in each diameter class is:
    \begin{equation} \label{eq:KTPA}
        (\text{KTPA})_{j,y} = \text{min}\left( \gamma \frac{(\text{succ. attacks per acre})_{j,y}}{(\text{surface area per tree})_{j,y}}, (\text{TPA})_{j,y} \right)
    \end{equation}
    where $\gamma$ a slope parameter, and $(\text{TPA})_{j,y}$ is trees per acre. If ``successfully'' attacking beetles kill all of the trees within a size class, the excess beetles are assumed to die (hence the ``min'' function). 

    \item The number of beetles produced by a single tree increases quadratically with tree diameter. Beetle emergence density (emerging female beetles per $\text{m}^2$) is a piecewise linear function of tree DBH:
    \begin{equation} \label{emerge_density}
        \left(\text{emerge density}\right)_{j,y} = 
        \begin{cases}
            \zeta_0 + \zeta_1 \cdot \text{DBH}_j & \text{if } \text{DBH}_j > -\zeta_0 / \zeta_1 \\
            0 & \text{otherwise}
        \end{cases}
    \end{equation}
    
    Here, $\zeta_0$ and $\zeta_1$ are intercept and slope parameters. Thus, $-\zeta_0 / \zeta_1$ is the  DBH threshold for beetle reproduction. The expected number of female beetles per acre in next year, $y+1$, is obtained with simple accounting:
    \begin{equation}
        \mathbb{E}\left[B_{y+1} \right] = \sum_{j=1}^{J}
         \left(\text{emerge density}\right)_{j,y} \cdot (\text{surface area per tree})_{j} \cdot (\text{KTPA})_{j,y}.   
    \end{equation}


    \item The realized number of beetles is stochastic. To capture temporal variation in overwintering mortality, we assume that the realized number of beetles is drawn from a zero-truncated normal distribution: 
    \begin{equation}
        B_{y+1} \sim \text{Normal}_{\scriptscriptstyle{Trunc.}}\left(\mathbb{E}\left[B_{y+1} \right], \sigma_{\text{env}} \cdot \mathbb{E}\left[B_{y+1} \right]\right). 
    \end{equation}
    The scale of environmental stochasticity is fixed at $\sigma_{\text{env}} = 0.2$, which is roughly consistent with the $\pm 40\%$ deviations of survival seen in the literature (\citealp{safranyik1971some}; \citealp{reid1963biology}, Tables 4 \& 7; \citealp{amman1983mountain}, Fig. 34; \citealp{regniere2007modeling}, Table 1).

\end{enumerate}

\textbf{Lodgepole pine dynamics (slow):}

\begin{enumerate}
    \item Trees grow at a constant rate until maturity. Tree growth is 4 mm (0.157 inches) DBH increase per year until trees reach 20 inches DBH, after which growth stops. 

    Although lodgepole pine trees are capable of indeterminate growth, intraspecific competition in dense stands constrains the maximum diameter. The vast majority of lodgepole pine trees in even-aged stands do not exceed 20--24 inches DBH \citep{rodman2022rocky, negron2020within, huang1999ecoregion}, and many U.S. Forest Service reports use a ``$> 18$ inches'' category to describe lodgepole diameters (e.g., \citealp{cole1969mountain, stipe1974yearly, mcgregor1973status}). The DBH growth rate is consistent with empirical estimates \citep{mata2003growth, heath1990growth, mclane2011climate}. However, like all pines, lodgepole pine grows faster in youth and slower in age, so 4mm here is a convenient approximation, representing the average increment over a tree's lifespan.

    \item Trees experience constant background mortality. A fixed proportion $(m)$ of trees die each year, regardless of size:
    \begin{equation} \label{eq:lodge1}
        (\text{TPA})_{j+1,y+1} = (1-m) \times (\text{TPA}^{\prime})_{j,y} - (\text{KTPA})_{j,y}
    \end{equation}
    where $(\text{TPA})_{j,y}$ is trees per acre in diameter class $j$ and year $y$, and $(\text{TPA}^{\prime})_{j,y}$ is trees per acre remaining after beetle-induced mortality.

    \item Forest maintains constant density through recruitment. Recruitment into the smallest diameter class (0.157 inches DBH) ensures a $K$ standing trees per acre, which includes live trees and MPB-killed trees that have not yet dropped their needles:
    \begin{equation} \label{eq:lodge2}
        (\text{TPA})_{1,y+1} = K - \sum_{j=1}^J \left(1-m \right) \times (\text{TPA}^{\prime})_{j,y} + \sum_{s=y-2}^{y} \sum_{j=1}^J (\text{KTPA})_{j,s}.
    \end{equation}

    \item Dead trees temporarily occupy canopy space. Trees killed by MPB retain needles for 3 years, occupying forest canopy and supressing recruitment as a live tree would.

\end{enumerate}

\subsection{Causal analysis} \label{Counterfactual simulations}

When the simulation model is instantiated with realistic parameter values, the five focal biological features are turned on. To turn the features off, we alter specific parameter values. 

To turn off the Allee effect, we set the Allee threshold, $a$, to infinity. As a consequence, the probability of a successful attack is always one, or equivalently, the number of initial attacks equals the number of successful attacks. To turn off stand-dependent dispersal mortality, we set $\lambda_0$ to infinity, such that the $s_y = s_0$ in \eqref{eq:sddda}. To turn off the large-tree preference, the piecewise linear function for attack preference, \eqref{eq:pref}, becomes a step function,
\begin{equation}
(\text{preference})_{j} =
\begin{cases}
1 & \text{if } \text{{DBH}}_j > -\beta_0 / \beta_1 \\
0 & \text{otherwise}.
\end{cases}
\end{equation}
All trees above the attack threshold, $-\beta_0 / \beta_1$, are equally attractive.  An alternative approach would be to make all trees (susceptible or not) equally attractive, but this approach would vastly increase the number of trees available to the beetles. In doing so, we would be modulating two key ecological factors --- size-dependent host-feeding \textit{and} resource availability --- which which diverges further from the real-world scenario and is therefore less causally relevant \citep{Lewis1979}

To turn off tree size-dependent fecundity, we set $\zeta_1 = 0$ and set $\zeta_0$ to the average emergence density (across killed tree surface area in the \citeauthor{klein1978attack} dataset), which is calculated as
\begin{equation}
\sum_{y=1}^{Y^{\prime}} \sum_{j=1}^{J^{\prime}} \begin{cases}
  \frac{\left(\zeta_0 + \zeta_1 \text{DBH}_{j}\right)(\text{KTPA}_{j,y} \times (\text{surface area per tree})_{j})}{\sum_{z=1}^{Y^{\prime}} \sum_{k=1}^{J^{\prime}} (\text{KTPA}_{k,z} \times (\text{surface area per tree})_{k})}, & \text{if } \text{DBH}_{j} > -\frac{\zeta_0}{\zeta_1} \\
  0, & \text{otherwise}.
\end{cases}
\end{equation}
Note here that $J^{\prime}$ and $Y^{\prime}$ are the number of diameter classes and years in the \citeauthor{klein1978attack} dataset, and should not be confused with the more granular diameter classes and the indeterminate number of years in the simulation model.

Finally, we turn off time-scale separation by forcing lodgepole pine to grow quickly. To accomplish this, we iterate density-independent pine mortality, growth, and recruitment (\eqref{eq:lodge1}--\eqref{eq:lodge2}) 50 times between bouts of beetle attacks, such that year-old trees have 8-inch diameters.

We create simulation scenarios where either 0, 1, or 2 of features have been turned off. The result is 10 simulation scenarios: 9 counterfactual and 1 realistic (i.e., all of the features are turned on). For each scenario, we draw 1,000 sets of parameters from the joint posterior distribution of the statistical models. For each parameter set, we simulate 3,000 years of data, starting with zero beetles and equilibrium forest conditions.

\subsection{Describing population dynamics} \label{Population metrics}

We track 5 key metrics of MPB population dynamics: peak beetle density, length of the outbreak growth phases, outbreak period, large tree density, and proportion of time spent in the endemic state. To compare model predictions to the real world, we also calculated \textit{observation-based estimates} of these 5 features.

As a preliminary step to quantifying the 5 features, we must distinguish persistent fluctuations (i.e., outbreaks and/or population cycles) from noisy fluctuations. First, we create a smoothed time series by applying a centered moving average with a three-year window. Second, we find peaks and troughs in the smoothed time series; these serve as initial guesses for the true peaks and troughs. Third, we go back to the original time series and identify the true peaks and troughs, e.g., a true trough is the minimum value between successive ``guess peaks'' identified in the smoothed time series. To prevent sub-critical immigration from being classified as an outbreak, we disregard sets of peaks and troughs where the $B_\text{peak} - B_\text{trough} < 600$; note that $600$ is twice the stochastic immigration parameter $I$ (see \eqref{eq:imm}).


Two of the metrics require an operational definition of an endemic MPB population. There are multiple definitions in the literature: less than 40 female beetles per ha \citep{carroll2006mountain}, less than 5 attacked trees per ha \citep{boone2011efficacy}, and less than 0.025 infested large-diameter trees per ha \citep{amman1984mountain}. Of these, we select the smallest value ($40$ females/ha $\approx$ $16$ females/acre) to operationalize the endemic state. This choice is based on two considerations: a) smaller values align more closely with aerial survey data, where the majority of stands appear to have no infestation; and b) our simulation model does not explicitly represent endemic beetle populations, leading to a slight underestimate of beetle density.

\begin{enumerate}
    \item \textbf{Peak beetle density}. For each simulated time series, we take the third quartile of all peak values of $B_y$, the density of emerging female beetles. The third quartile represents a typical large outbreak, which is relevant because large local outbreaks disproportionaly contribute to spatial spread and range expansion (see the \textit{Discussion}, Section \ref{Discussion}).

    Because beetles or emergence holes are rarely observed across multiple years, the only \textit{observation-based estimate} of peak beetle density comes from the \citeauthor{klein1978attack} dataset. This dataset is also used to parameterize the simulation model, so the predictions and observations are not independent.

    \item \textbf{Outbreak growth phase length} For each outbreak in a simulated time series times, we calculate the difference between the year of peak beetle density and the year before beetle density first exceeds the endemic threshold of $16$ females/ha. Then, we take the mean of this quantity across all outbreaks.

    An observation-based estimate is calculated using British Columbia's Aerial Overview Survey data from 1990-2021. First, we estimate the number of infested trees using the methodology in Fig.\,\ref{fig:sample_time_series} caption, aggregate data into $1\text{km}^2$ pixels, and subset to locations with moderate-to-severe outbreaks (cumulative infestations in the upper quartile). Next, we calculate the difference between the year with the maximum number of infested trees and the year before infested trees were first detected. Finally, we take the mean of this quantity across all locations.

    \item \textbf{Outbreak period}. This quantity is simply the mean number of years between peaks (whether cycles or outbreaks) in the simulated time series. The observation-based estimate was sourced from studies that have reconstructed region-wide outbreaks with tree ring data \citep{alfaro2003dendroecological, negron2020reconstructing, jarvis2015long, alfaro2009historical,axelson2009influence}. More precisely, the observation-based estimate is the median (across studies) of the mean outbreak recurrence time.
    
    
    \item \textbf{Large tree density}. The large tree density is simply the temporal average of trees per acre with $DBH \geq 8$ inches. Here, there are two observation-based estimates. The first comes from the \citeauthor{klein1978attack} dataset, and is therefore not independent of the simulation-based predictions. The second is the median (across several studies) of trees per acre with $DBH \geq 8$ inches, prior to a MPB outbreak \citep{mcgregor1973status, gibson1978damage, hamel1975status, stipe1974yearly, stipe1976trends, mitchell1991analysis, amman1969mountain, negron2020within, evenden1940destructive}. This is an overestimate of the temporal average of large-tree density, but it may be a sensible approximation, given that MPB populations are usually in the endemic state.
    
    \item \textbf{Proportion of time spent in the endemic state}. This is straightforwardly computed as the proportion of years with less than 16 emerging females per acre. The observation-based estimate comes from British Columbia's Aerial Overview Survey data (years: 1962--2021). Specifically, we aggregated data into $1 \text{km}^2$ pixels, subsetting to locations with moderate-to-severe outbreaks (cumulative infestations in the upper quartile), and computed the proportion of years where no infestations were detected. 
\end{enumerate}

The five metrics above can be used heuristically to classify simulated time series wtih respect to their population dynamical behavior: excitable dynamics (i.e., episodic boom-bust dynamics), population cycles, or equilibrium dynamics. If the proportion of time spent in the endemic states is $\geq 1/2$, then the time series has \textit{excitable} dynamics. If the proportion of time spent in the endemic states is $< 1/2$, and the outbreak period is $< 50$ years, then the time series has \textit{Cycles}. If the proportion of time spent in the endemic states is $< 1/2$, and the outbreak period is $\geq 50$ years, then the time series has \textit{Equilibrium} dynamics.

\section{Results}

\subsection{Model diagnostics and parameter estimates}

The statistical models performed well across multiple assessment criteria. Various diagnostics indicated that the models converged to the global posterior distribution, with most prior distributions having negligible impact on the results (Appendix \ref{model-fitting}). The models successfully recreated the original outbreak trajectory and out-of-sample data (Appendix \ref{LOOCV}).

The parameter estimates produced by the statistical models align with previous research (Table \ref{tab:pars_full}). The posterior distributions of attack density and emergence density (from a 10-inch diameter tree) are consistent with over over a dozen studies across British Columbia, Alberta, and the western United States (Figures \ref{fig:hist_attack} \& \ref{fig:emerge_hist}). The mean overstory density, 338 trees per acre or 835 trees per hectare, agrees with estimates from regenerating stands in southern British Columbia (\citealp{nigh2008density}, mean $=$ 893 stems/ha) and lodgepole-dominated stands in the U.S. Rocky Mountains (\citealp{pfeifer2011observations, meddens2011evaluating, meddens2012spatiotemporal}; mean $=$ 1065 stems/ha). The minimum susceptible tree diameter is estimated at 7.46 inches, which is consistent with the often-cited 8-inch (or 20 cm) threshold \citep{hopping1948relation, cole1969mountain, safranyik1988estimating}. The baseline dispersal mortality (i.e., 1-$s_0$) is 0.52, which is consistent with the 0.5--0.7 range provided by previous studies of bark beetles \citep{schmid1977spruce, safranyik1989mountain, latty2007pioneer, pope1980allocation}. Finally, the mean attack density of 71 females per square meter in the \citeauthor{klein1978attack} dataset is close to the optimal value of 60 females per square meter, proposed by \citet{raffa1983role}.

\begin{table}[H]
\centering
\begin{tabular}{llllll}
  \hline
Parameter & Short description & Mean & SD & $\text{CI}_{2.5\%}$ & $\text{CI}_{97.5\%}$ \\ 
  \hline
a & Allee threshold & 432 & 363 & 13.5 & 1338.88825 \\ 
  $s_0$ & max dispersal survival & 0.477 & 0.0701 & 0.363 & 0.639 \\ 
  $\lambda_{\text{resid}}$ & dispersal threshold & 0.652 & 0.0353 & 0.579 & 0.708 \\ 
  $\lambda_1$ & dispersal sensitivity & 49.7 & 96.6 & 17.2 & 262 \\ 
  $\beta_0$ & pref.$\sim$DBH intercept & -3.42 & 0.773 & -4.84 & -1.93 \\ 
  $\beta_1$ & pref.$\sim$DBH slope & 0.453 & 0.0688 & 0.319 & 0.580 \\ 
  $\gamma$ & dead tree$\sim$attack slope & 0.0151 & 0.000878 & 0.0134 & 0.0168 \\ 
  $\zeta_0$ & emerge$\sim$DBH intercept & -9.86 & 49.2 & -106 & 87.8 \\ 
  $\zeta_1$ & emerge$\sim$DBH slope & 25.4 & 5.21 & 15.3 & 35.5 \\ 
  K & forest carrying capacity & 904 & 313 & 413 & 1630.37825 \\ 
  m & annual tree mortality & 0.0518 & 0.00637 & 0.0376 & 0.0634 \\ 
  \midrule
  Derived quantities & Short description  & Mean & SD & $\text{CI}_{2.5\%}$ & $\text{CI}_{97.5\%}$ \\
  \midrule 
 $-\beta_1 / \beta_0$ & susc. DBH threshold & 7.46 & 0.666 & 5.98 & 8.47 \\ 
  $1/\gamma$  & attack density & 66.3 & 3.88 & 59.4 & 74.4 \\ 
  $\zeta_0 + \zeta_1 * 10$ & emerge density (10 in. DBH) & 244 & 25.1 & 191 & 296 \\ 
  $K-\sum_{t=0}^{17} (1-m)^{t}mK$ & Overstory density & 338 & 86.0 & 193 & 529 \\ 
\midrule
  Fixed parameter & Short description & value & \multicolumn{3}{l}{} \\
  \midrule 
  $P_I$ & Prob. immigration & 0.05 & & & \\
  $I$ & immigrating females & 300 & & & \\
  $\sigma_{\text{env}}$ & Environmental stochasticity & 0.2 & & & \\
  Symbol N/A & max DBH & 20 & & & \\
  Symbol N/A & annual DBH growth & 0.157 & & & \\
  \hline
\end{tabular}

\caption{Summary statistics of marginal posterior distributions for model parameters, including the Mean, standard deviation, and credible intervals ($\text{CI}_{2.5\%}$, $\text{CI}_{97.5\%}$). The table also includes derived quantities, including the diameter at breast height (DBH) of the smallest susceptible trees; the model-based estimate of successful attack density (gallery starts per square meter); the emergence density (females per square meter) for 10-inch DBH tree; and the density of overstory trees (stems per acre). Overstory trees are operationalized with a standard cutoff: DBH $\geq 7\text{cm} \approx 2.76\text{in}$. This explains the summation from 0 to 18 in the formula: $\frac{2.76 \text{in cutoff}}{0.157 \text{in per year}} \approx 18 \;\text{years}$. Finally, the table contains fixed parameters, whose values were obtained via literature search. For more elaborate descriptions and parameter units, see supplementary Table \ref{tab:model_notation}. For the summary statistics of sub-model noise parameters, see supplementary Table \ref{tab:pars_fullest}.} 
\label{tab:pars_full}
\end{table}

\subsection{Effects of life history on population dynamics}

\begin{figure}[H]
\centering
\includegraphics[scale = 1]{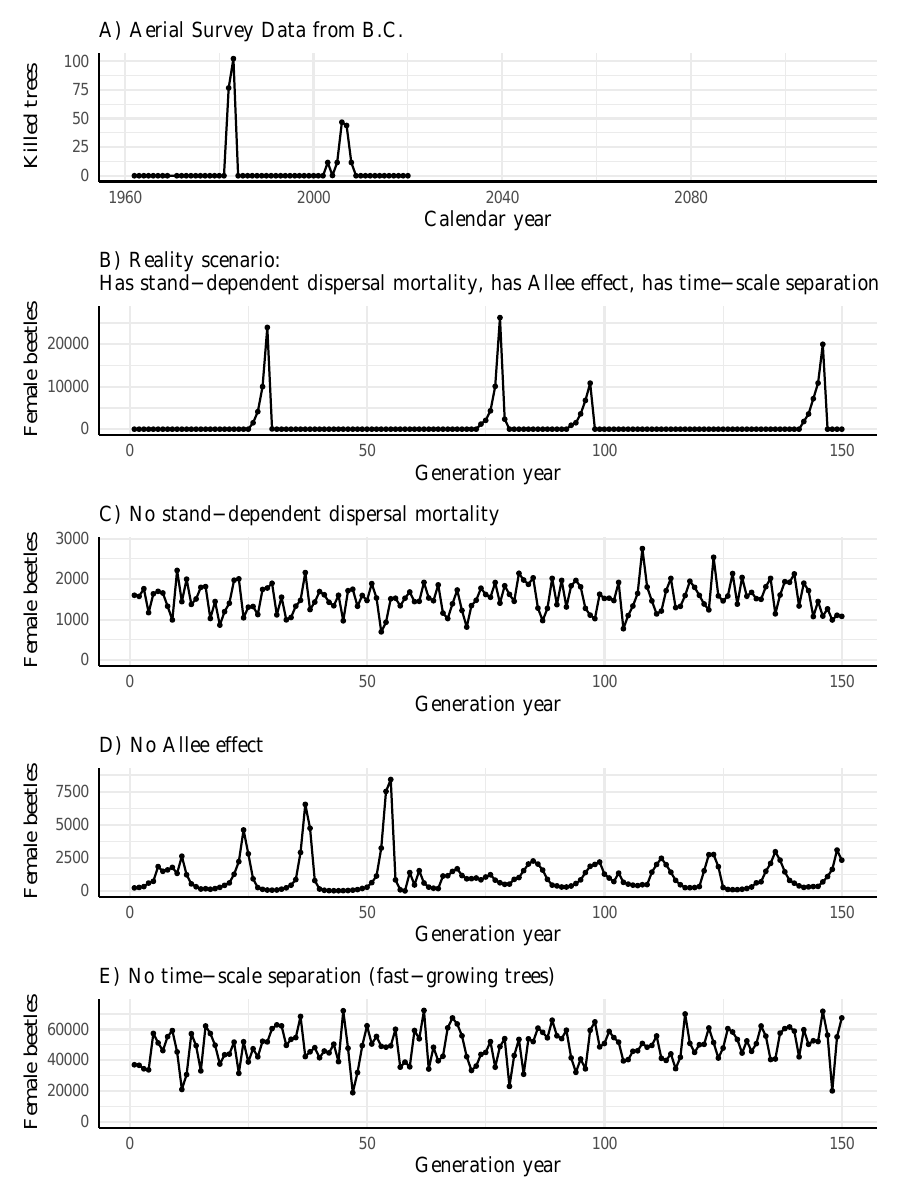}
\caption{Time series of mountain pine beetle infestations, real vs. simulated. The real data comes from Aerial Overview Survey data in British Columbia, aggregated across a representative 1 $\text{km}^2$ patch. Simulated time series are representative of various simulation scenarios.}
\label{fig:time_series}
\end{figure}

\begin{figure}[H]
\centering
\includegraphics[scale = 1]{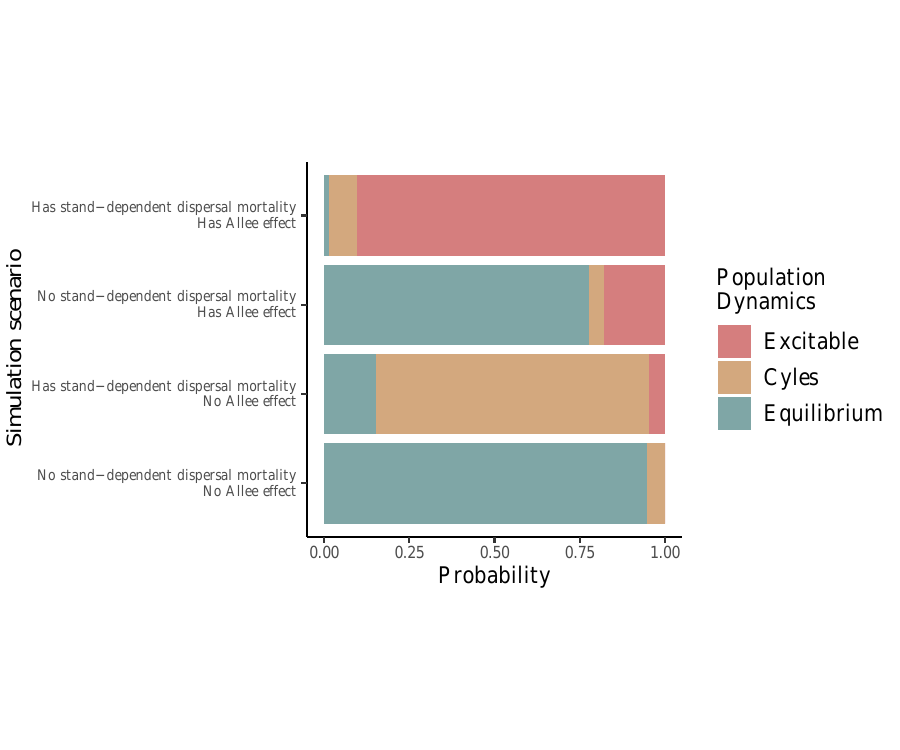}
\caption{Stand-dependent dispersal mortality and the Allee effect lead to excitable boom-bust dynamics. Barplot proportions represent the posterior probabilities of different dynamical behaviors, classified using the methodology in Section \ref{Population metrics}.}
\label{fig:counterfactual_dynamics_disp_allee}
\end{figure}

Three key features are essential for excitable population dynamics in MPB: stand-dependent dispersal mortality, an Allee effect, and time-scale separation (Fig.,\ref{fig:time_series}, Fig. \ref{fig:counterfactual_dynamics_disp_allee}). Without the stand-dependent dispersal mortality, MPB and lodgepole pine attain a noisy equilibrium where there is a consistently low density of beetles (Fig.\,\ref{fig:time_series}C), but not low enough to technically be endemic. Without the Allee effect, a classic population cycle appears (Fig.\,\ref{fig:time_series}D).  Without time-scale separation, MPB and lodgepole pine attain a noisy equilibrium with large numbers of beetles and susceptible trees (Fig.\,\ref{fig:time_series} E).

The process of stand-dependent dispersal mortality explains why beetles don’t overexploit trees. It is perhaps trivial to say that if beetles disperse away and die before they have killed all the trees, then the beetles won’t kill all the trees. However, the lack of over-exploitation, or equivalently, the seemingly premature “bust” phase of the “boom-bust cycle”, is a ubiquitous feature of outbreaks (Fig.\,\ref{fig:dbh_across_outbreaks}) and is thus worth connecting to MPB life-history. This also suggests an optimistic perspective: outbreaks are much less destructive than they theoretically could be. Stand-dependent dispersal probably occurs because chemo-sensory constraints prevent beetles from aggregating on small trees (Section \ref{life_history:stand_dependent_dispersal}), even though beetles have positive fitness when they do successfully attack small trees. Without stand-dependent dispersal mortality, the system would reach a stochastic equilibrium, creating a forest with few large pine trees.

The Allee effect explains why outbreaks are so extreme when they do occur. The Allee effect suppresses MPB populations, keeping them in their endemic state until enough large trees have accumulated that a small inoculum of beetles --- either from long-distance dispersal or a stochastic successful attack --– can grow explosively. Without the Allee effect, lodgepole pine and MPB enter a predator-prey cycle.

Classic predator-prey theory predicts that long-lived prey should make predator-prey cycles unlikely \citep{murdoch2003consumer}. One might naively expect that the time-scale separation between beetles and trees would allow MPB to grow rapidly enough to prevent an unchecked proliferation in pine abundance, thus preventing the \textit{prey-escape phase} of a population cycle \citep{de1990size} wherein the prey species can grow unimpeded by the predator. Even though pines live much longer than beetles, we still see population cycles in the counterfactual simulations (when the Allee effect is turned off). This can be explained by the phenomenon of cohort resonance \citep{bjornstad2004trends}. During an outbreak, many pines are killed at roughly the same time. Subsequently, pines are recruited at the same time, and then become susceptible to MPB at the same time. Thus, there are a brief few years where the pines that MPB can “see” are growing faster than exponentially. 

The two remaining features --- size-dependent tree preference and tree size-dependent fecundity --- do not affect the qualitative nature of MPB population dynamics, i.e., whether a simulation was classified as \textit{Excitable} or \textit{Cycles}. However, they do have quantitative effects on some of the fine-grained properties of outbreaks: the combination of both size-dependent tree preference and tree size-dependent fecundity led to a 40\% increase in the peak beetle density (from 15 to 21 thousand) and an 8\% decrease in the length of growth phase (from 5.8 years to 5.4 years). In other words, the combination of these life-history traits makes outbreaks moderately more explosive and slightly more rapid. 

\begin{landscape}
\begin{figure}[H]
\includegraphics[scale = 1]{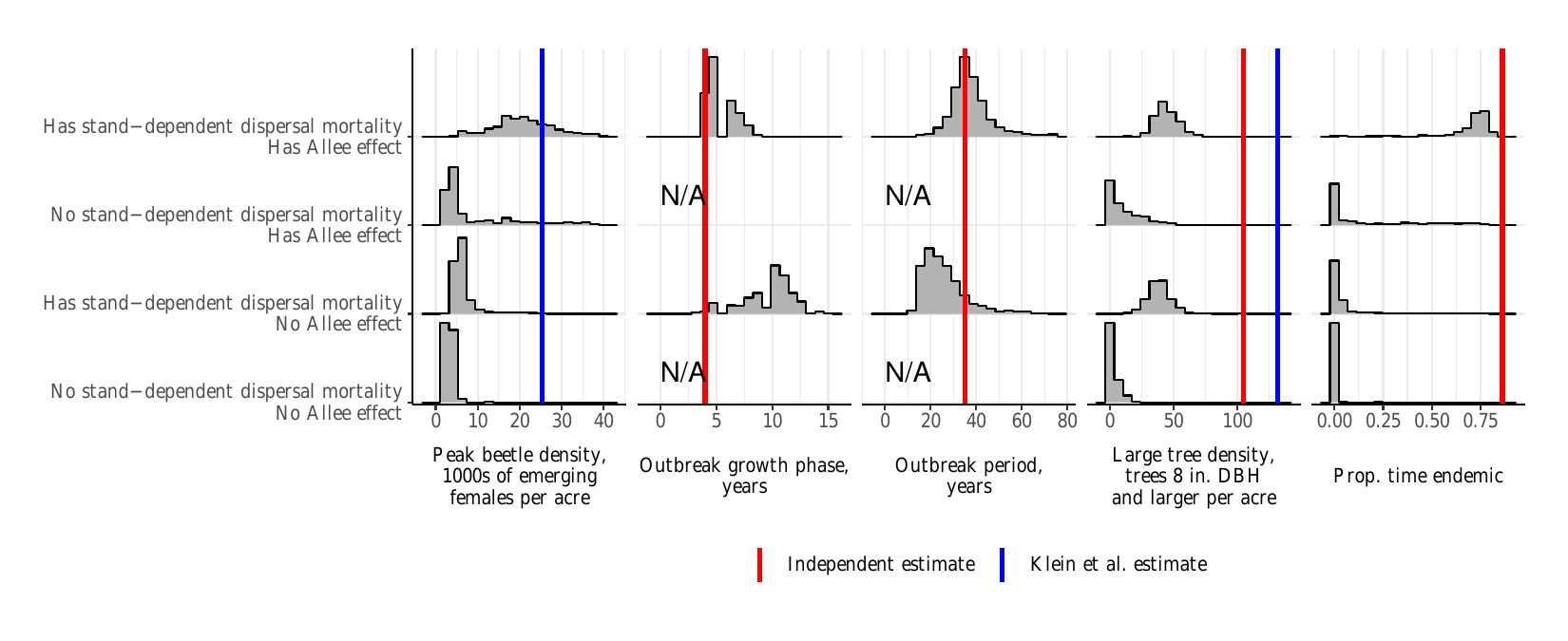}
\caption{A comparison of population dynamic metrics for various combinations of stand-dependent dispersal mortality and the Allee effect. The histograms show the frequency of predicted values across posterior samples. The vertical lines show observation-based estimates from the \citeauthor{klein1978attack} dataset or independent sources. Simulation scenarios without stand-dependent dispersal mortality resulted in few peaks and outbreaks (see Table \ref{tab:results}), hence the ``N/A'' label. The five metrics and the observation-based estimates are explained in Section \ref{Population metrics}.}
\label{fig:counterfactual_metrics_disp_allee}
\end{figure}
\end{landscape}

The inclusion of stand-dependent dispersal mortality leads to better predictions of outbreak magnitude and large-tree density (Fig.\,\ref{fig:counterfactual_metrics_disp_allee}). The Allee effect leads to better predictions of the outbreak recurrence time and growth phase length. Specifically, without the Allee effect, the cycles occur too frequently and the growth phase of each cycle is too long. Stand-dependent dispersal mortality primarily explains the fact that a portion of large trees remain despite MPB outbreaks.

\begin{landscape}
\renewcommand*\rot{\multicolumn{1}{R{50}{1em}}}
\begin{table}[H]
\centering
\begingroup\fontsize{10pt}{11pt}\selectfont
\begin{tabular}{llllll|llllllll|llll}
\rot{\underline{\textbf{Biological features}}} & \rot{Has stand-dependent dispersal mortality} & \rot{Has Allee effect} & \rot{Has large-tree preference} & \rot{Has tree size-dependent fecundity} & \rot{Has time-scale separation} & \rot{\textbf{\underline{Metrics} Mean $\pm$ (SD)}} & \rot{Peak beetle density, 1000s of emerging females per acre} & \rot{Outbreak growth phase, years} & \rot{Outbreak period, years} & \rot{Large tree density, trees $\geq$ 8 in DBH per acre} & \rot{Prop. time endemic} & \rot{Prop. sims without peaks} & \rot{Prop. sims without outbreaks} & \rot{\textbf{\underline{Probability of Dynamics}}} & \rot{Excitable} & \rot{Cycles} & \rot{Equilibrium} \\ 
  \hline
 & \checkmark & \checkmark & \checkmark & \checkmark & \checkmark &  & 21 (9.0) & 5.4 (1.2) & 44 (34) & 44 (9.7) & 0.68 (0.16) & 0.00 (0.00) & 0.00 &  & 0.90 & 0.08 & 0.01 \\ 
   &  & \checkmark & \checkmark & \checkmark & \checkmark &  & 10 (10) & 7.3 (1.1) & 174 (295) & 11 (13) & 0.18 (0.25) & 0.00 (0.06) & 0.44 &  & 0.18 & 0.05 & 0.78 \\ 
   & \checkmark &  & \checkmark & \checkmark & \checkmark &  & 6.4 (3.1) & 9.7 (2.2) & 26 (10) & 38 (9.2) & 0.07 (0.15) & 0.00 (0.00) & 0.06 &  & 0.05 & 0.80 & 0.15 \\ 
   & \checkmark & \checkmark &  & \checkmark & \checkmark &  & 16 (7.1) & 5.9 (1.4) & 57 (136) & 43 (10) & 0.66 (0.16) & 0.00 (0.00) & 0.00 &  & 0.89 & 0.07 & 0.04 \\ 
   & \checkmark & \checkmark & \checkmark &  & \checkmark &  & 16 (7.0) & 5.9 (1.4) & 60 (157) & 45 (10) & 0.65 (0.16) & 0.00 (0.00) & 0.00 &  & 0.88 & 0.10 & 0.03 \\ 
   & \checkmark & \checkmark & \checkmark & \checkmark &  &  & 72 (18) & 8.0 (2.4) & 4.8 (1.2) & 46 (9.1) & 0.00 (0.04) & 0.00 (0.00) & 0.99 &  & 0.00 & 1.0 & 0.00 \\ 
   &  &  & \checkmark & \checkmark & \checkmark &  & 3.3 (0.90) & 13 (1.4) & 293 (374) & 2.6 (4.0) & 0.00 (0.01) & 0.01 (0.08) & 1.0 &  & 0.00 & 0.05 & 0.95 \\ 
   &  & \checkmark &  & \checkmark & \checkmark &  & 11 (12) & 8.5 (1.2) & 221 (402) & 12 (15) & 0.19 (0.26) & 0.01 (0.08) & 0.45 &  & 0.19 & 0.07 & 0.74 \\ 
   &  & \checkmark & \checkmark &  & \checkmark &  & 4.6 (2.2) & 8.9 (1.9) & 1076 (876) & 4.8 (9.9) & 0.04 (0.15) & 0.09 (0.29) & 0.85 &  & 0.04 & 0.00 & 0.96 \\ 
   &  & \checkmark & \checkmark & \checkmark &  &  & 73 (16) & NA & 4.7 (0.12) & 46 (8.8) & 0.00 (0.00) & 0.00 (0.00) & 1.0 &  & 0.00 & 1.0 & 0.00 \\ 
   & \checkmark &  &  & \checkmark & \checkmark &  & 5.7 (2.7) & 10 (2.4) & 30 (17) & 37 (10) & 0.06 (0.14) & 0.00 (0.00) & 0.05 &  & 0.04 & 0.74 & 0.23 \\ 
   & \checkmark &  & \checkmark &  & \checkmark &  & 6.1 (2.9) & 9.9 (2.7) & 29 (36) & 39 (10) & 0.06 (0.15) & 0.00 (0.00) & 0.13 &  & 0.05 & 0.75 & 0.21 \\ 
   & \checkmark &  & \checkmark & \checkmark &  &  & 72 (16) & NA & 4.7 (0.12) & 47 (9.7) & 0.00 (0.00) & 0.00 (0.00) & 1.0 &  & 0.00 & 1.0 & 0.00 \\ 
   & \checkmark & \checkmark &  &  & \checkmark &  & 15 (7.0) & 5.8 (1.3) & 63 (200) & 42 (10.0) & 0.65 (0.16) & 0.00 (0.06) & 0.00 &  & 0.87 & 0.09 & 0.05 \\ 
   & \checkmark & \checkmark &  & \checkmark &  &  & 76 (20) & 7.4 (1.8) & 9.9 (125) & 42 (9.0) & 0.01 (0.07) & 0.00 (0.04) & 0.99 &  & 0.01 & 0.99 & 0.00 \\ 
   & \checkmark & \checkmark & \checkmark &  &  &  & 66 (15) & 13 (3.9) & 4.9 (0.30) & 46 (9.0) & 0.01 (0.05) & 0.00 (0.04) & 0.98 &  & 0.00 & 1.0 & 0.00 \\ 
   \hline
\end{tabular}
\endgroup
\caption{Results of the simulation study. A check mark indicates that a biological feature is present in the simulation scenario. The quantitative values take the format of ``Mean $\pm$ (SD)'', where both summary statistics are calculated across the posterior distribution. In addition to the 5 metrics of population dynamics (presented in Section \ref{Population metrics}), there are two extra metrics: 1) the proportion of simulations where neither peaks nor troughs were detected after applying the moving-average smoothing and various thresholds described in Section \ref{Population metrics}; 2) the proportion of simulations where no outbreaks were detected, where an outbreak is defined as a \textit{trough} $\rightarrow$ \textit{peak} $\rightarrow$ \textit{trough} sequence where the population attains the endemic state ($<16$ females per acre) before and after the peak. If a simulation scenario produced both NA and non-NA values for a particular metric, then the NA values were omitted when calculating that metric.} 
\label{tab:results}
\end{table}
\end{landscape}

\newpage

\section{Discussion} \label{Discussion}

Excitable population dynamics are related to five features of bark beetle ecology: stand-dependent dispersal mortality, an Allee effect, time-scale separation between beetle and tree life cycles, tree size-dependent fecundity, and a large-tree preference. Time-scale separation (i.e., trees grow slower than beetles) ensures that killed trees cannot be immediately replaced by new mature pines, thus preventing equilibrium dynamics. Stand-dependent dispersal (i.e., beetles disperse farther through sparse forests) leads to high levels of mortality, which is exacerbated by the Allee effect (i.e., not enough beetles to mass attack healthy trees), bringing beetle populations to an endemic, low-density state. The two inhibitory processes --- stand-dependent dispersal mortality and a strong Alle effect --- explain why populations spend the vast majority of time in their endemic state, and why subsequent outbreaks are so explosive: with suppressed beetle populations, large trees accumulate, providing proverbial fuel for the beetle fire.

The two remaining traits, tree size-dependent fecundity and a large-tree preference, explain the large magnitude of outbreaks, i.e., the peak number of beetles. Larger trees provide more resources for beetle reproduction and are preferentially attacked, leading to rapid population growth during outbreaks. While not essential for MPB's excitable population dynamics due to MPB's high baseline fecundity, these traits may be crucial for other bark beetle species that lie on the boundary between pulse-driven and irruptive behaviors. Bark beetles can be categorized as either ``outbreaking'' or ``non-outbreaking'', with the former category further divided into ``pulse-driven'' and ``irruptive'' (equivalent to excitable) categories. 

\subsection{Bark beetle comparative analysis}



Aggregation behavior is a key distinguishing factor between outbreaking and non-outbreaking bark beetle species. Non-outbreaking species are poor aggregators, with few individuals attacking simultaneously and pheromones used primarily for mate-finding. Examples include the greater European spruce beetle (\textit{Dendroctonus micans}) and the black turpentine beetle (\textit{Dendroctonus terebrans}); more are listed in \citet{reeve2012ancestral} Table 2. To survive in living trees, non-outbreaking beetles often have adaptations such as resistance to toxic tree resin \citep{weed2015population}. Interestingly, \textit{Dendroctonus micans} adults do not aggregate, but larvae do \citep{gregoire1988greater, deneubourg1990kinetics}, leading to greater larval survival in small pockets of tree phloem \citep{storer1997effect}. Although most bark beetles are non-outbreaking \citep{wood1982bark}, aggregation and mass-attacks appear to be the ancestral state for \textit{Dendroctonus} bark beetles \citep{reeve2012ancestral}.

The distinction between pulse-driven and irruptive species is less clear-cut. Instead, differences lie in the magnitude of various life history traits and environmental factors. For example, the Douglas-fir beetle (\textit{Dendroctonus pseudotsugae}), a pulse-driven species, prefers large trees that are downed \citep{hood2007predicting} or recently defoliated by the Douglas-fir tussock moth, \textit{Orgyia pseudotsugata} \citep{wright1984abundance, negron2014douglas}. During outbreaks, it may attack healthy trees, but these are typically adjacent injured trees \citep{cunningham2005attack, cole2022outbreaks}. This behavior, contrasting with the irruptive MPB's strong preference for healthy trees, contributes to the Douglas-fir beetle's classification as a pulse-driven species.

The western balsam bark beetle (WBBB; \textit{Dryocoetes confusus}) presents an intriguing case that further blurs the lines between these categories. The WBBB likely has excitable population dynamics: it is generally considered irruptive (e.g. \citealp{lantschner2023spatiotemporal}) and causes large destruction in response to drought perturbations \citep{maclauchlan2023infestation}. Further, WBBB shares many life history traits with the MPB, including a preference for large trees, tree-size-dependent fecundity, and the mass-attack-based Allee effect. Despite these similarities, their population dynamics look quite different. WBBB often attacks less than 5\% of a given stand in one year, with chronic infestations slowly killing many mature trees over time \citep{maclauchlan2003impact, mcmillin2003effects}. One possible explanation is that WBBB carries a fungus, \textit{Grosmannia dryocoetidis}, that is truly pathogenic to host trees \citep{molnar1965pathogenic, zipfel2006multi}. Therefore, sub-critical or initially unsuccessful attacks can kill trees simply by inoculating the fungus. Another possible explanation is that WBBB's primary host, subalpine fir, often grows in mixed-age stands \citep{antos2002structure, maclauchlan2016quantification} wherein younger trees are less susceptible. This contrasts with MPB's primary host, lodgepole pine, which typically experiences clearcut logging or stand-leveling fires, resulting in even-aged stands prone to explosive outbreaks. The highly clustered nature of WBBB's outbreaks \citep{howe2022landscape} suggest it may lack \textit{stand-dependent dispersal mortality}.


A beetle species' status as irruptive or pulse-driven can change over space and time in response to environmental conditions. One example is the red turpentine beetle (\textit{Dendroctonus valens}), which is considered non-outbreaking in its native range in North America but has become irruptive in China. This shift is attributed to novel fungal symbionts aiding host tree mortality, the higher susceptibility of the Chinese host \textit{Pinus tabuliformis}, more attractive host volatiles, prevalent monoculture plantations, and warmer weather \citep{yan2005red, sun2013red}. This case highlights the importance of considering traits (e.g., aggregation and large-tree preference due to attractive host volatiles), abiotic factors (e.g., climate), and idiosyncratic elements (e.g., fungal symbionts) in determining outbreak potential.

The European spruce beetle (\textit{Ips typographus}) provides another interesting comparison to MPB. \textit{I. typographus} experiences significantly higher intraspecific competition among larvae, resulting in consistently lower fitness \citep{karvemo2010comparison}. This difference may explain why the European spruce beetle is less destructive than MPB \citep{karvemo2010comparison}, despite both species being considered irruptive. The likely mechanism behind this disparity is \textit{I. typographus}' much lower emission of the anti-aggregation pheromone verbenone \citep{birgersson1984quantitative, ramakrishnan2022metabolomics}. Another way to conceptualize the disparity (in terms of our ``biological features''), is that the European spruce beetle lacks strong \textit{tree size-dependent fecundity}, since high competition undermines the benefits of thicker phloem.


Environmental conditions and interspecific interactions significantly influence outbreak potential. The spruce beetle (\textit{Dendroctonus rufipennis}) in Alaska's interior rarely outbreaks despite favorable conditions. Less precipitation in the interior leads to lower phloem moisture, thus favoring competitive non-eruptive insects like \textit{Ips perturbatus} \citep{werner2006dynamics}. Similarly, the roundheaded pine beetle (\textit{Dendroctonus adjunctus}) outbreaks in areas where trees are stressed by drought and other bark beetles, specifically western pine beetles (\textit{Dendroctonus brevicomis}) and \textit{Ips} species \citep{negron2000stand, negron2008roundheaded}.

\begin{figure}[H]
\centering
\includegraphics[scale = 1]{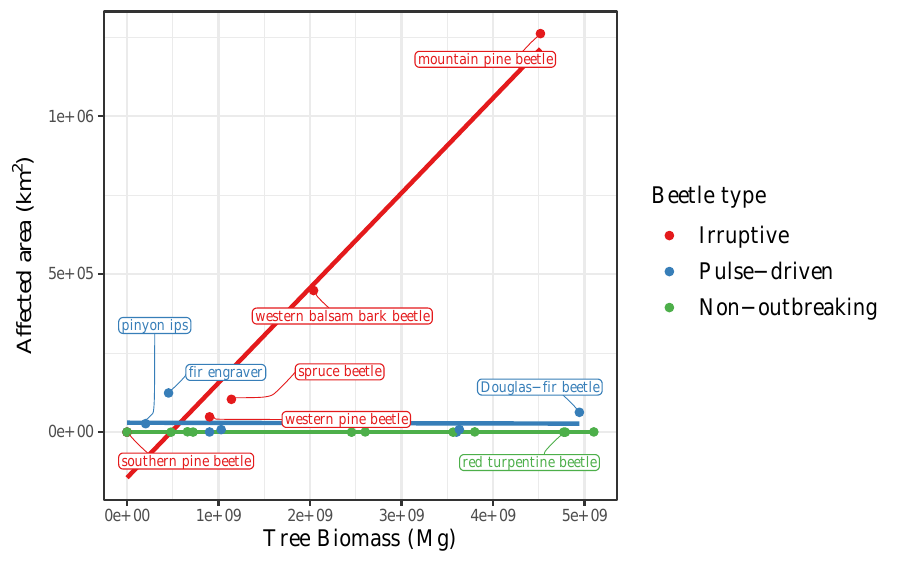}
\caption{Total bark beetle damage is determined by life-history traits and landscape characteristics. Cumulative damaged area is a linear function of the total biomass of host trees, across British Columbia and the western United States (i.e., regions 1--6 as defined by the US Forestry Service); however, this relationship only holds true for bark beetles classified as irruptive. Methodological details in Appendix \ref{landscape}. 
}
\label{fig:damage_by_species}
\end{figure}

We conducted a comparative analysis of bark beetle damage using aerial survey data from western North America and maps of host-tree biomass (Appendix \ref{landscape}). Our analysis reveals that classification as "irruptive" is a necessary condition for causing extensive damage, but the total amount of beetle damage scales linearly with the total biomass of host trees (Fig. \ref{fig:damage_by_species}). Previous research has found an approximately linear relationship between tree biomass and tree damage across sites, within species \citep{negron1999estimating, negron2017probability}; here we show that this relationship also applies \textit{between} species. Together, both life history traits and environmental factors determine if a bark beetle population is capable of overcoming an Allee effect (i.e., overcoming the defenses of healthy trees) and growing explosively; then landscape-level factors determine the overall impact of bark beetle outbreaks. This analysis adds a layer of nuance to Rudinsky's \citeyear{rudinsky1962ecology} claim that the ``Abundance of suitable breeding material has recently been definitely identified as the decisive factor in outbreaks of bark beetles.'' 


The western pine beetle (\textit{Dendroctonus brevicomis}) provides an illustrative example of how landscape characteristics can influence the impact of even highly irruptive species. Unlike MPB, which can attack any pine species within its native range, the western pine beetle's outbreaks are limited due to its high specificity to ponderosa pine \citep{homicz2022forest}. Recent trends in southern pine beetle (\textit{Dendroctonus frontalis}) damage illustrate how management practices can affect landscape characteristics and total beetle impact. Despite being considered irruptive, the total area affected by southern pine beetle has declined significantly over time \citep{fettig2022trends}. \citet{asaro2017have} attribute this decline to silvicultural practices, namely rapidly removing infested trees and limiting connectivity between high-risk host stands.


Other bark beetle species lack the unique life history of MPB, either in kind or in degree, resulting in pulse-driven outbreaks or less destructive irruptive outbreaks. We have examined examples of bark beetles lacking MPB's key traits: the great European spruce beetle lacks the ability to aggregate (implicating the Allee effect), the western balsam bark beetle possibly lacks stand-dependent dispersal mortality, and the European spruce beetle lacks size-dependent fecundity. The Douglas-fir beetle prefers large trees \citet{sturdevant2022douglas}, but prefers these to be downed or otherwise unhealthy. Consequently, these species have slow-moving outbreaks (e.g., the western balsam bark beetles) or are unable to achieve self-driving growth without sustained perturbations.

\subsection{Mountain pine beetle dynamics}

Our model accurately predicts the return time of MPB outbreaks, which is somewhere between 20--50 years according to tree-ring-based reconstructions. Our counterfactual simulations with no Allee effect, resulting in classic population cycles, predict an outbreak return time that is too short (approximately 20 years). Other models in the literature claim an outbreak return time that is too long \citep{powell2009connecting, duncan2015model, brush2023coupling}.

The period of MPB outbreaks is primarily determined by the time it takes lodgepole pine to grow into a susceptible age class. In an idealized scenario where MPB kills all trees, outbreaks would occur approximately every 55 years: 50 years for new trees to become vulnerable, plus a few years for the outbreak and subsequent growth of pine seedlings. However, there are two real-world complications. First, stand-dependent dispersal mortality results in a large number of medium-sized trees surviving the outbreak, so the observed periodicity of 35 years partially reflects the time it takes for medium-sized trees to become large trees, e.g., it takes approximately 25 years for an 8-inch tree to become a 12-inch tree. The 2000s outbreak in British Columbia killed more medium-sized trees than usual, which suggests a longer-than-usual wait time for the next outbreak. Second, widespread outbreaks require widespread perturbations (usually drought), which are functionally stochastic. Thus, tree growth and climate patterns imply a right-skewed distribution of waiting times between outbreaks, where the minimum time is determined by tree growth, and the shape of the right tail is determined by drought probability. 

Tree size influences both MPB fecundity and attack preference, explaining about 40\% of peak beetle numbers. The outbreak peak is relevant because it is thought to drive MPB range expansion \citep{bleiker2019risk}. Only a minuscule fraction of emerging beetles undergo long-distance dispersal (estimated at 2.5\%; \citealp{safranyik1992dispersal}), so outbreak expansion via long-distance dispersal requires severe local infestations, such that emigrating beetles are numerous enough to mass-attack trees at their destination. In other words, range expansion requires that peak beetle density exceeds the \textit{donor threshold} (terminology introduced by \citealp{johnson2006allee}).

The donor threshold concept may explain regional patterns of MPB spread. Outbreaks are spatially correlated at greater distances in British Columbia than in the western United States, where lower pine biomass leads to smaller peak beetle populations and fewer successful long-distance dispersers. The donor threshold may also explain why MPBs expanded into central Alberta from British Columbia, rather than gradually moving northward from Montana along the eastern Canadian Rockies. British Columbia is more climatically suitable than the eastern Rockies \citep{safranyik2010potential}, thus leading to large MPB populations and successful long-distance dispersal. 

Our research offers a revised perspective on the density-dependent dynamics of MPB, challenging established assumptions. In previous work, the number of new infestations is assumed to be proportional to both beetles and trees (e.g. \citealp{burnell1977dispersal, heavilin2007dynamics,kvrivan2016dynamical}), as in the classic Lotka-Volterra model. Yet, our analysis shows that for intermediate-to-high tree density (i.e., before stand-dependent dispersal mortality kicks in) the number of new infestations is only proportional to beetles, reflecting the fact that outbreaking beetles are limited by their ability to overcome tree defenses, not by their ability to find susceptible trees. While overcompensatory density dependence has been a common assumption in past studies (e.g. \citealp{goodsman2017positive, goodsman2018effect, lewis2010structured, duncan2015model}), we argue that MPB experiences near-perfect compensatory density dependence (see Section \ref{life_history:stand_dependent_dispersal}). Previous research has assumed that the probability of a successful mass attack depends on the quotient of beetle and tree density (e.g., \citealp{burnell1977dispersal, goodsman2016aggregation, brush2023coupling}). Instead, we propose that beetle density alone is a more significant determinant of the success probability, as detailed in Appendix \ref{Allee effect}.

Putting all of these results together, we can simplify our simulation model (Section \ref{Simulation model}) to formulate a minimal mechanistic model of MPB dynamics. Because the large-tree preference and tree size-dependent fecundity are non-essential (for excitability in MPB, specifically), we drop explicit age structure and only track the density of all susceptible trees (i.e., trees with DBH $>$ 8 inches). All susceptible trees are equally preferred for mass attacks, and all killed trees give rise to the same number of beetles. For simplicity, stochasticity is eliminated. The minimal mechanistic model can be written as
\begin{equation} \label{eq:minimal_A}
 \overbrace{A_{y+1}}^{\mathclap{\text{Attacking females per acre}}} =A_y \times \underbrace{(1-e^{\frac{-2.3 A_y}{a}})}_{\mathclap{\text{Allee effect}}} \times \overbrace{\left(\zeta^* \; \gamma^*\right)}^{\mathclap{\raisebox{2ex}{\scriptsize\text{Emerging females per succ. attack}}}} \times \underbrace{s_0 \left(1 + \exp\left(-\lambda_1 \left(\frac{S_y^\prime}{S_{max}} - \lambda_0\right) \right) \right)^{-1}}_{\mathclap{\text{Survival after stand-dependent dispersal}}} + \overbrace{\epsilon}^{\mathclap{\raisebox{2ex}{\scriptsize\text{Endemic beetles}}}}
\end{equation}
\begin{equation} \label{eq:minimal_S}
\underbrace{S_{y+1}}_{\mathclap{\text{Susceptible trees per acre}}} = \overbrace{g(K^*-S_y)}^{\mathclap{\text{Recruitment of susceptible trees}}} - \underbrace{A_y \times (1-e^{\frac{-2.3 A_y}{a}}) \times \overbrace{\gamma^*}^{\mathclap{\text{Trees killed per succ. beetle attack}}} }_{\mathclap{\text{Trees killed by MPB}}}.
\end{equation}
Here, $A_y$ is the female attack density in the $y^{\text{th}}$ year, $S_y$ is the density of susceptible trees, $S_y^\prime$ is the density of susceptible trees remaining after beetle attack (i.e., $S_y^\prime = S_y - A_y \times (1-e^{\frac{-2.3 A_y}{a}})$), the parameter $\zeta^*$ is the average emerging females per killed tree, $\gamma^*$ is the average number of killed trees per successful attack, $K^*$ is redefined as the carrying capacity of the susceptible trees, $\epsilon$ is the now-explicit number of endemic female beetles, $g$ modulates the rate of recruitment into the class of susceptible trees, and all other parameters have their original meanings (see Table \ref{tab:model_notation}). When simulating the model, a floor at $S = 0$ may be necessary to avoid negative trees. 

\begin{landscape}
\begin{figure}[H]
\centering
\includegraphics[scale = 0.5]{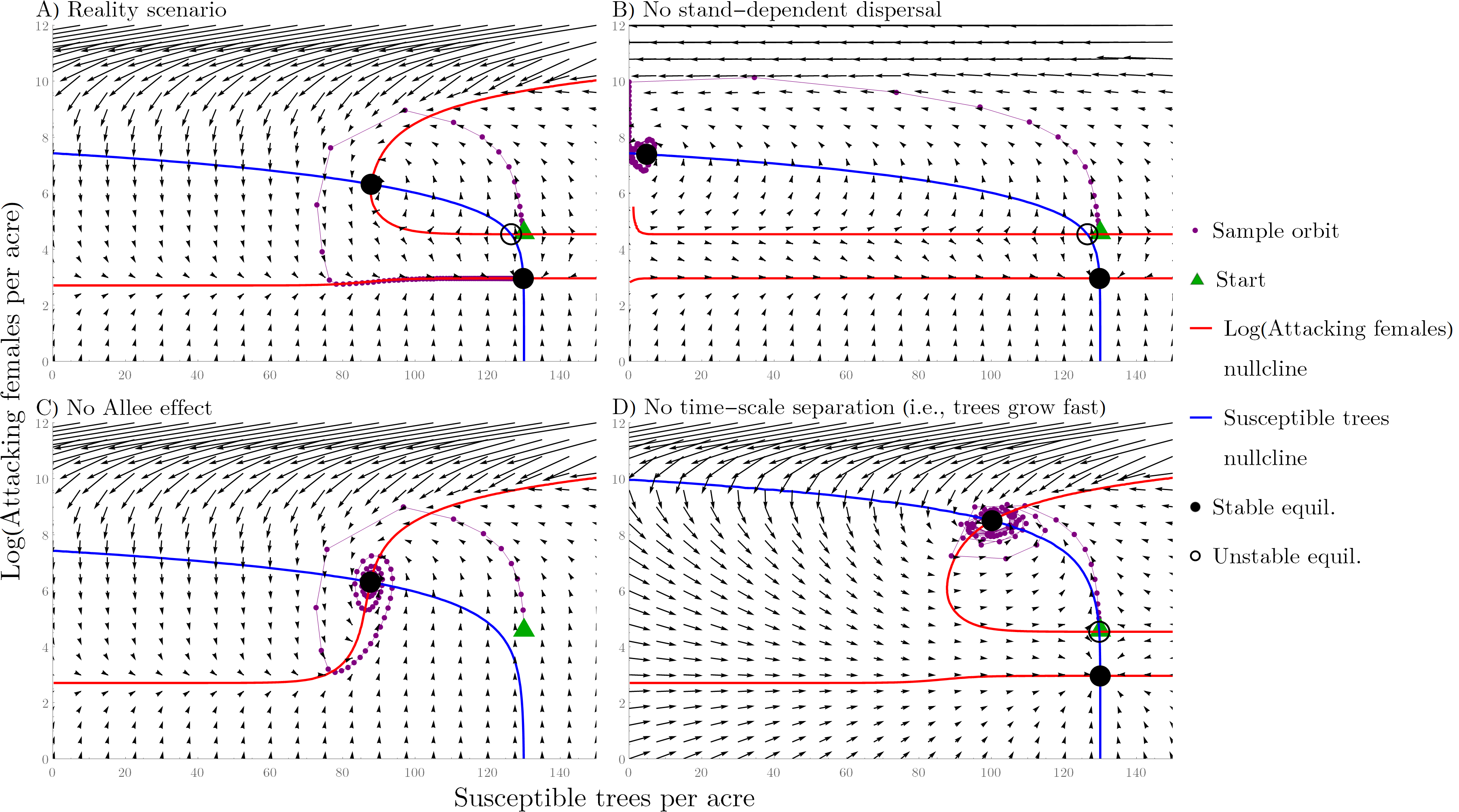}
\caption{Phase portraits of the minimal mechanistic model (\eqref{eq:minimal_A} \& \eqref{eq:minimal_S}), across various simulation scenarios \textbf{Panel A}: Stand-dependent dispersal, an Allee effect, and time-scale separation leads to excitable dynamics. Nullclines show curves where $\log(A_{y+1})=\log(A_y)$ and $S_{y+1}=S_y$, and arrow directions are $\log(A_{y+1})-\log(A_y)$ and $S_{y+1}-S_y$. Most model parameters were set to their respective posterior means (see Table \ref{tab:results}). The exceptions are $\gamma^* = 1/360$ and $\zeta^* = 1300$, which are approximately the respective means (across trees) of attacks and emergences in the \citeauthor{klein1978attack} dataset. The carrying capacity's value, $K^* = 130$, is approximately the density of trees with $\text{DBH} \geq 8\text{in}$ in the \citeauthor{klein1978attack} dataset. The recruitment parameter's value, $g = 0.036$, was obtained by solving the differential equation $dS(t)/dt = g(K-S(t))$ with initial condition $S(0)=0$, then solving for the $g$ where 90\% of carrying capacity is attained in $t=50$ years; lodgepole pine attains an 8-inch DBH and becomes susceptible after approximately 50 years.}
\label{fig:phase_panels}
\end{figure}
\end{landscape}

The minimal model has several purposes. 1) It is descriptive, compactly representing the density-dependent dynamics of MPB and lodgepole pine. 2) It demonstrates that there are two stable equilibria: an endemic equilibrium with few beetles, and an epidemic equilibrium with many beetles. This \textit{bistability} is typical of systems with a strong Allee effect \citep{taylor2012allee}, so it is no surprise that the endemic equilibrium disappears when the Allee effect is turned off (Fig.\,\ref{fig:phase_panels}C) 3) It shows the MPB/lodgepole system is an instance of an \textit{excitable dynamical system}, where perturbations can lead to long excursions in phase space, before returning to a resting state (Fig.\,\ref{fig:phase_panels}A). Typically, excitable systems have a single stable equilibrium \citep[pg. 118]{strogatz2018nonlinear}, whereas our system has two. However, the endemic equilibrium is globally-attracting as a practical matter, since the epidemic equilibrium has a minuscule basin of attraction (Fig.\,\ref{fig:phase_boa}). Previous papers proposed two stable equilibria for irruptive bark beetles (e.g., \citealp{berryman1979dynamics, mawby1989endemic, brush2024eruptive}). Our model corroborates this while adding crucial nuance: while it is possible to temporarily attain beetle densities at (or above) epidemic equilibrium, it is not possible to stay at the epidemic equilibrium.

\newpage


Through the development and analysis of our age-structured simulation model, we have shed light on the life history and biological processes underlying outbreaks, and challenged established ideas in the MPB modelling literature. A comparative analysis of bark beetles reveals that MPB's life history traits contribute to its uniquely destructive nature. However, to accurately predict total damage, it's necessary to consider host-tree biomass (Fig. \ref{fig:damage_by_species}) and idiosyncratic factors, such as fungal symbionts and interspecific competition. Our findings are succinctly summarized in a minimal mechanistic model (\eqref{eq:minimal_A} \& \eqref{eq:minimal_S}), which establishes the MPB/lodgepole system as the first empirical examples of an excitable dynamical system in ecology. This work underscores the importance of detailed case studies and looking beyond traditional predator-prey models to understand complex population dynamics. 


\section{Acknowledgements} 
We would like to thank Micah Brush for insightful discussions; and Xiaoqi Xie for obtaining the Biosim data (Figs. \ref{fig:year_weather} \& \ref{fig:summer_weather}). This research was funded through a grant to the TRIA-FoR Project to ML from Genome Canada (Project No. 18202).

\begin{appendices}
\counterwithin{figure}{section}
\counterwithin{table}{section}

\section{Statistical model description} \label{Statistical model description}

\subsection{Data}

\citeauthor{boone2011efficacy}'s study was conducted from 2000--2005 in the mature lodgepole pine stands of southern British Columbia. Six stands were selected for the study, with an average area of 15.3 hectares per stand, and an average of 1225 canopy trees per hectare. Each year, until the beetle population waned and attained its endemic condition, every tree in each stand was inspected to determine the number of attacked trees and the success or failure of these attacks. 

\begin{figure}[H]
\centering
\includegraphics[scale = 1]{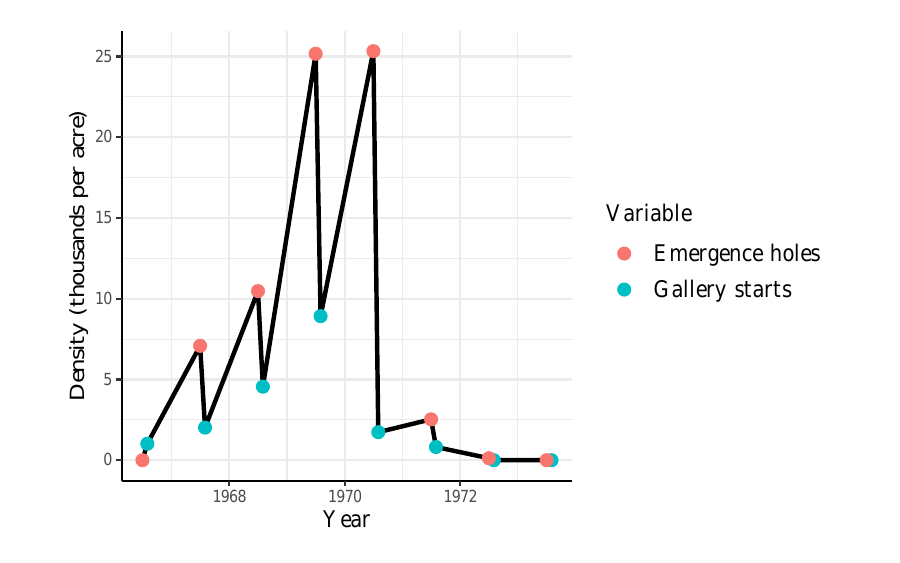}
\caption{Time series of emergence holes and gallery starts in the \citeauthor{klein1978attack} dataset. Emergence and subsequent attacks are assumed to take place in July and August respectively. There are always far fewer gallery starts than emergence holes, suggesting high levels of dispersal mortality.}
\label{fig:mort1}
\end{figure}

\citeauthor{klein1978attack}'s study was conducted from 1965--1972 in the southwest corner of Yellowstone National Park in Idaho. Data was collected and aggregated across 85 permanent plots that spanned a square mile. Trees were grouped into six 2-inch diameter classes, ranging from 8 to $\geq$ 16 inches. For each of the 106 sampled trees, paired 6x6-inch sample squares were marked on opposite sides of the bole at regular height intervals. Within these squares, emergence holes and successful attacks (operationalized as egg-gallery starts) were measured (Fig. \ref{fig:mort1}). Some of this data was originally presented by \citet{parker1973trend}, but a more thorough reporting is given by \citet{klein1978attack}.

\subsection{Model overview} \label{Overview2}

There are three statistical models. 1) The tree growth model (Appendix \ref{Lodgeple pine forest dynamics}); 2) The simple Allee effect model (Appendix \ref{subsec:allee}), which is fit to the \citeauthor{boone2011efficacy} dataset; and 3) the complex outbreak model (Appendices \ref{subsec:pref}--\ref{Integrator model description}), which is fit to the \citeauthor{klein1978attack} dataset. Table \ref{tab:model_notation} describes all relevant notation, including subscripts, data, intermediate quantities, and bonafide parameters.

The tree growth model (Appendix \ref{Lodgeple pine forest dynamics}) estimates the demographic parameters of lodgepole pine: density-independent mortality and forest carrying capacity ($m$ and $K$). This model was fit with the 1965 pre-outbreak tree diameter distribution in the \citeauthor{klein1978attack} dataset (sample size = 7). The simple Allee effect model (Appendix \ref{subsec:allee}) describes the Allee effect and was fit with \citeauthor{boone2011efficacy} data (sample size = 28). The posterior distribution of the Allee threshold parameter was used to create an informative prior for the Allee effect in our third statistical model. 

The third statistical model (Appendices \ref{subsec:pref}--\ref{Integrator model description}), which is by far the most complex, uses the \citeauthor{klein1978attack} dataset (sample size = 91) to model the outbreak dynamics of MPB. This outbreak model contains several \textit{sub-models}, simple models that are used to resolve one aspect of MPB's life cycle. For example, tree diameter and beetle attack preference, are used to estimate the parameters $\beta_0$ and $\beta_1$. A variety of such sub-models provide direct estimates of the parameters $\beta_0$, $\beta_1$, $\gamma$, $\zeta_0$, $\zeta_1$, and additional noise parameters. The remaining parameters ($a$, $s_0$, $\lambda_0$, and $\lambda_1$), are estimated by stringing together the deterministic aspects of the sub-models to predict the number of gallery starts in the following year; we call this the \textit{integrator model}. All of the sub-models, along with the integrator model are combined and fit simultaneously. The result is a hierarchical model structure, where information is partially pooled across sub-models, because all parameters must work cohesively to predict next year's gallery starts. 

In the third statistical model, the Allee threshold parameter is given a half-normal prior, with a standard deviation equal to half of the point estimate of $a$ (the mean of the marginal posterior) from the second model (which estimates the Allee effect in the \citeauthor{boone2011efficacy} data). This prior expresses the belief that the Allee threshold in the \citeauthor{boone2011efficacy} data is an upper bound on the threshold in the \citeauthor{klein1978attack} dataset, since the \citeauthor{klein1978attack} data are averages over a much larger spatial scale (259 hectares vs. 15 hectares) and MPB are known to be highly spatially aggregated. All other parameters are given weakly-informative priors, which are later shown to have negligible influence on the results.


In the main text, we presented the simulation model with each time-step beginning and ending with the density of emerging beetles. This approach was chosen for two reasons: Firstly, flying beetles are a noticeable aspect of actual outbreaks. Secondly, emergence ties directly into our discussion of the interaction between peak beetle density and long-distance dispersal. In contrast, the outbreak statistical model ends each step with beetles that have successfully mass-attacked a tree. This structure yields an extra year of data at the beginning of the outbreak's. Later years, dominated by stand-dependent dispersal, offer less insight into non-dispersal parameters.

Throughout this appendix, we will refer to the surface area over which beetles may attack as the \textit{susceptible surface area}; note that beetles do not attack the upper part of the bole due to thin phloem. To convert between tree diameter and susceptible surface area, we use a simple linear model,
\begin{equation} \label{DBH_to_SA}
 SA = -3.792 + 0.821 \times DBH,
\end{equation}
where $DBH$ is the diameter at breast height in inches, and $SA$ is the susceptible surface area in square meters. \eqref{DBH_to_SA} was first published by \citet{burnell1977dispersal}, who claims an excellent fit ($R^2 = 0.94$) and attributes the parameterization to William Klein.

\subsection{Model-fitting} \label{model-fitting}

The parameters of the simulation model are estimated with three \textit{statistical models}: the lodgepole-dynamics model, the Allee-effect model, and the outbreak-dynamics model. The combined statistical models are conceptually identical to the simulation model.


The first statistical model (Appendix \ref{Lodgeple pine forest dynamics}) estimates the demographic parameters of lodgepole pine: density-independent mortality and forest carrying capacity ($m$ and $K$). This model was fit with the 1965 pre-outbreak tree diameter distribution in the \citeauthor{klein1978attack} dataset (sample size = 7). Our second statistical model (Appendix \ref{subsec:allee}) describes the Allee effect and was fit with \citeauthor{boone2011efficacy} data (sample size = 28). The posterior distribution of the Allee threshold parameter was used to create an informative prior for the Allee effect in our third statistical model. 

The third statistical model (Appendices \ref{subsec:pref}--\ref{Integrator model description}), which is by far the most complex, uses the \citeauthor{klein1978attack} dataset (sample size = 91) to model the outbreak dynamics of MPB. This outbreak model contains several \textit{sub-models}, simple models that are used to resolve one aspect of MPB's life cycle. For example, tree diameter and beetle attack preference, are used to estimate the parameters $\beta_0$ and $\beta_1$. A variety of such sub-models provide direct estimates of the parameters $\beta_0$, $\beta_1$, $\gamma$, $\zeta_0$, $\zeta_1$, and additional noise parameters. The remaining parameters ($a$, $s_0$, $\lambda_0$, and $\lambda_1$), are estimated by stringing together the deterministic aspects of the sub-models to predict the number of gallery starts in the following year; we call this the \textit{integrator model}. All of the sub-models, along with the integrator model are combined and fit simultaneously. The result is a hierarchical model structure, where information is partially pooled across sub-models, because all parameters must work cohesively to predict next year's gallery starts. 

In the third statistical model, the Allee threshold parameter is given a half-normal prior, with a standard deviation equal to half of the point estimate of $a$ (the mean of the marginal posterior) from the second model (which estimates the Allee effect in the \citeauthor{boone2011efficacy} data). This prior expresses the belief that the Allee threshold in the \citeauthor{boone2011efficacy} data is an upper bound on the threshold in the \citeauthor{klein1978attack} dataset, since the \citeauthor{klein1978attack} data are averages over a much larger spatial scale (259 hectares vs. 15 hectares) and MPB are known to be highly spatially aggregated. All other parameters are given weakly-informative priors, which are later shown to have negligible influence on the results. 

\subsection{Lodgeple pine forest dynamics} \label{Lodgeple pine forest dynamics}

We begin by deriving tree ages in a stable population, undisturbed by mountain pine beetle. Let \( K \) be the carrying capacity (units: trees/acre). Let \( m \) be the annual probability of density-independent mortality, constant across all ages. When a tree dies, a seedling is immediately recruited in its place, such that there are always $K$ trees per acre. The asymptotic density of trees of age \( t \) is given by:

\begin{equation}
N_t = K  m (1 - m)^{t}
\end{equation}

Here, \( K \times m \) is the density of recruits into the first age class, and \( (1 - m)^{t} \) is the probability that a tree survives up to age \( t \).

The \citeauthor{klein1978attack} dataset does not contain tree ages, but it does contain densities of trees within diameter classes. To convert diameter to age, we divide DBH (in inches) by 0.15748, the average annual diameter increment of lodgepole pine.

The response variable is the pre-outbreak (1965) density of trees in the \( j \)-th age bin, which is defined by the age range interval $\left(\text{(lower age)}_j, \text{(upper age)}_j \right]$. The expected density $ \mu_{j,\text{tree}} $ for the $j^{\text{th}}$ age bin can be defined as:

\begin{equation}
\mu_{j,\text{tree}} = \sum_{t = \text{(lower age)}_j + 1}^{\text{(upper age)}_j}  K  m (1 - m)^t 
\end{equation}

Finally, the likelihood for the observed tree density, denoted, $ (\text{TPA})_j $, is given by the normal PDF:

\begin{equation}
\log\left(\text{TPA}\right)_j \sim \text{Normal}(\log(\mu_{j,\text{tree}}), \sigma_{\text{tree}}).
\end{equation}

\begin{figure}[H]
\centering
\includegraphics[scale = 1]{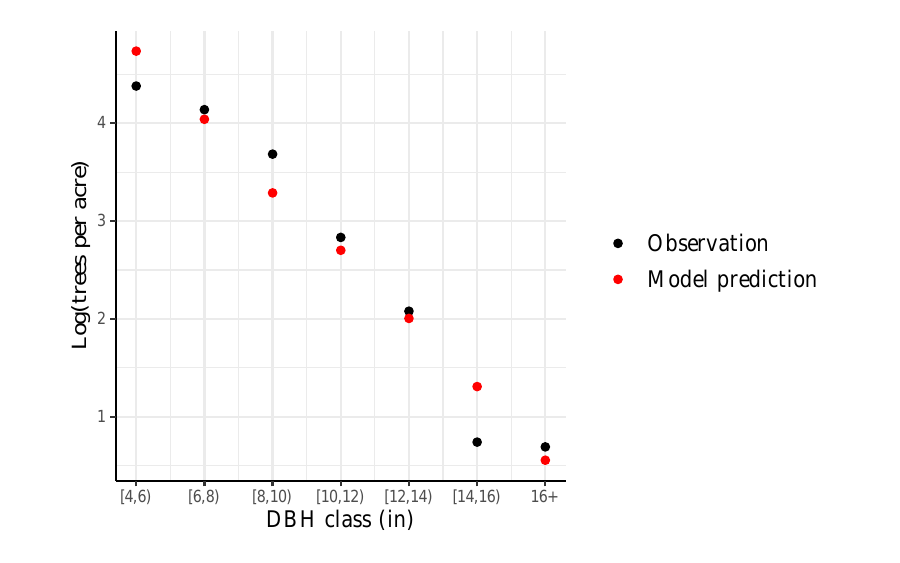}
\caption{The pre-outbreak diameter distribution of lodgepole pine is effectively modeled by assuming density-independent mortality and instant recruitment. The theoretical distribution of trees per acre in the $j^{th}$ diameter class is $(\text{TPA})_{j,\infty} = (1-m)^{(j-1)}mK$, so the logarithm of trees per acre within should be approximately linear with respect to diameter at breast height (DBH). Data from \citet{klein1978attack}. The predictions were generated with posterior means of model parameters.}
\label{fig:tree0}
\end{figure}

\begin{figure}[H]
\centering
\includegraphics[scale = 1]{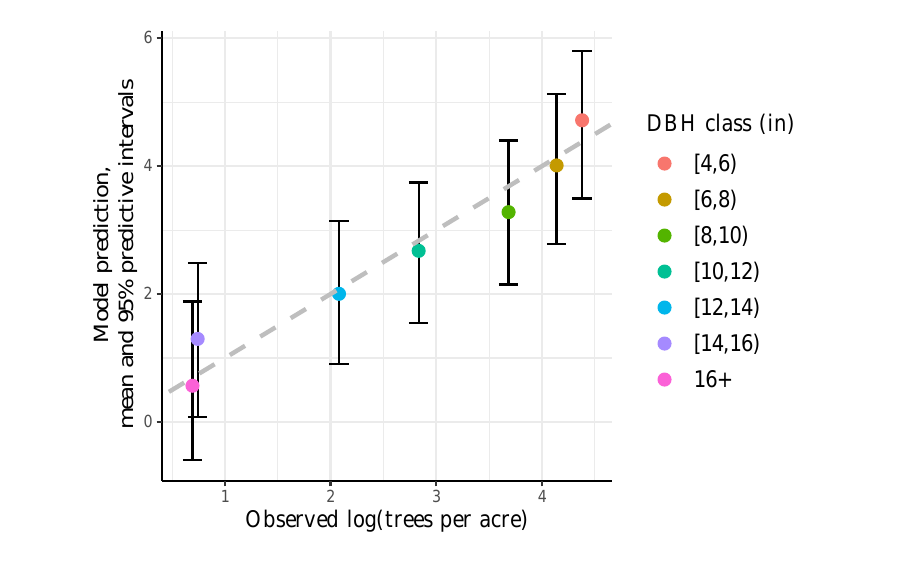}
\caption{Observations vs. predictions for the lodgepole pine forest model. The error bars give the 95\% predictive interval, representing both parameter and sampling uncertainty. The dashed grey line represents perfect agreement between observed and predicted values. Data from \citet{klein1978attack}.}
\label{fig:tree1}
\end{figure}

Residual errors in the densities of small and large-diameter trees are made commensurate through the logarithmic transformation. The noise parameter is $\sigma_{\text{tree}}$. The model provides a good fit to the data (Fig.\,\ref{fig:tree0}), though there is substantial uncertainty (Fig. \ref{fig:tree1}) due to the high quotient of parameters and data (i.e., $3/7$).


\subsection{Tree size-dependent preference sub-model} \label{subsec:pref}

We define \textit{mass attack preference} for trees in the the $j^{th}$ diameter class, in the $y^{th}$ year, as
\begin{equation}
(\text{{preference}})_{j,y} = \frac{(\text{{proportion of successful attacks received}})_{j,y} }{(\text{{proportion of total tree surface area}})_{j,y}}.
\end{equation}

\begin{figure}[H]
\centering
\includegraphics[scale = 1]{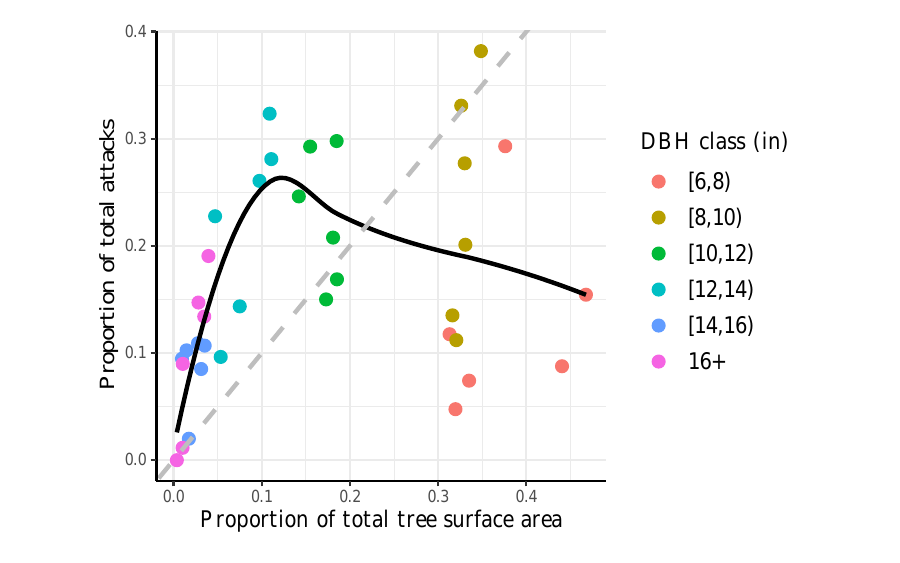}
\caption{Beetles disproportionately kill large trees. The x-axis is the proportion of susceptible surface area belonging to live trees of a particular diameter class, within a year. The y-axis is the proportion of total attacks received by the same diameter class, within a year. The black line is a LOESS model prediction. The grey dashed line is the one-to-one line, representing the \textit{no-mass-attack-preference hypothesis}. Data from  \citet{klein1978attack}.}
\label{fig:size_pref0}
\end{figure}

\begin{figure}[H]
\centering
\includegraphics[scale = 1]{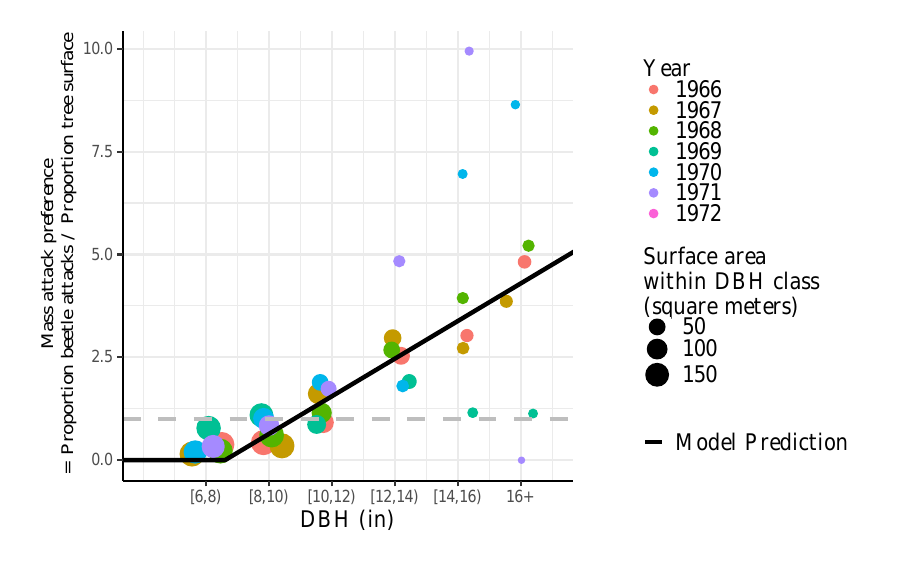}
\caption{Beetle preference is an increasing function of tree diameter. The data (from \citealp{klein1978attack}) is well-modeled by a piecewise linear function (see \eqref{eq:pref0}). There is clear heteroskedasticity in the data, which is likely due to smaller sample sizes within classes of large-diameter trees.  The exact number of samples is not available, so we use the within-DBH-class surface area as a proxy for sample size, visualized as scatter-plot point size. The prediction was generated using posterior means of model parameters.}
\label{fig:size_pref1}
\end{figure}

High values of preference indicate that beetles like to attack a DBH class disproportionately, in relation to the surface area that is ``owned'' by that DBH class. The preference variable can be calculated directly from the \citeauthor{klein1978attack} data. The data clearly show that that preference increases linearly with tree diameter (Fig. \ref{fig:size_pref0}, \ref{fig:size_pref1}).

Preference is normally distributed with mean $\mu_{j,\text{{pref}}}$ and standard deviation $\sigma_{j,y,\text{{pref}}}$:
\begin{equation}
(\text{{preference}})_{j,y} \sim \text{Normal}(\mu_{j,\text{{pref}}}, \sigma_{j,y,\text{{pref}}}).
\end{equation}
The mean is a piecewise linear function,
\begin{equation} \label{eq:pref0}
\mu_{j,\text{{pref}}} =
\begin{cases}
\beta_0 + \beta_1 \cdot \text{{DBH}}_j & \text{if } \text{{DBH}}_j > -\beta_0 / \beta_1 \\
0 & \text{otherwise}.
\end{cases}
\end{equation}
Above, we assume that $DBH_j$ is the diameter at breast height for the middle of a diameter class, e.g., if we are considering all trees where $DBH \in [10,12)$, then $DBH_i$ in the above equation is 11 inches. In the \citeauthor{klein1978attack} dataset, there are six 2-inch DBH classes. In the simulation model, each tree age is associated with a unique DBH, separated by intervals of 0.157 inches.

The standard deviation of preference is
\begin{equation} \label{eq:sigma_pref}
\sigma_{j,y,\text{{pref}}} =  \sqrt{ \frac{{\sigma_{\text{{pref0}}}^2}}{(\text{total surface area})_{j,y}} + \sigma_{\text{{pref1}}}^2},
\end{equation}
where $\sigma_{\text{{pref0}}}$ accounts for a baseline level of variability, and $\sigma_{\text{{pref1}}}$ accounts for increasing variability with smaller sample sizes (clearly visible in Fig. \ref{fig:size_pref1}), with the typical inverse square root scaling that would be expected under the central limit theorem. We do not know the exact sample sizes, so we treat the total tree surface area  (within the $j^{th}$ diameter class) as a proxy for sample size.



\subsection{Converting beetle attacks to tree deaths} \label{subsec:dead}

\begin{figure}[H]
\centering
\includegraphics[scale = 1]{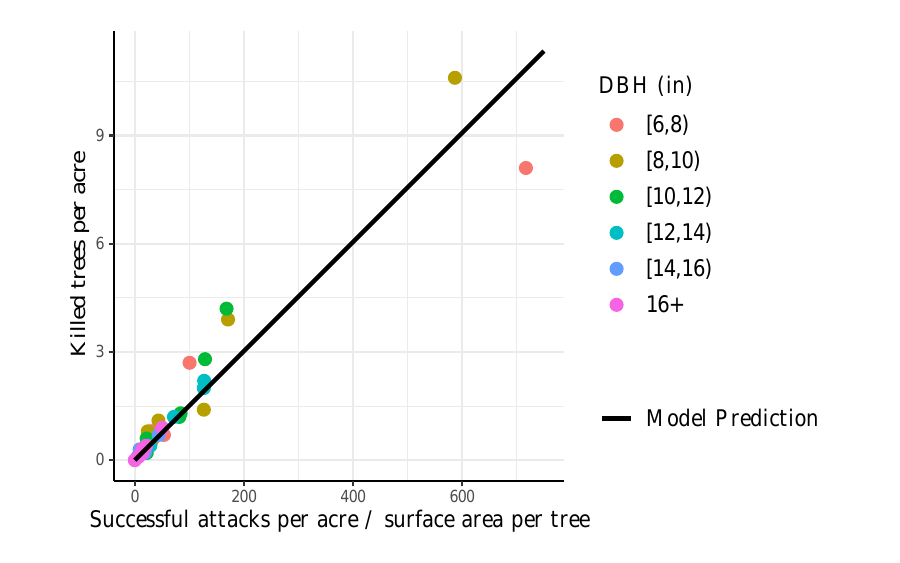}
\caption{The number of killed trees is a linear function of successful beetle attacks (corrected for DBH-specific surface area), suggesting that beetles ``fill up'' a tree to a universal density, and then move on to the next tree. The prediction was generated using posterior means of model parameters. Data from \citet{klein1978attack}.}
\label{fig:dead1}
\end{figure}

We model killed trees as a linear regression through the origin, with a dynamic noise term to account for some obvious heteroskedasticity (Fig.\,\ref{fig:dead1}). The killed trees per acre, denoted $(KTPA)_{j,y}$ is normally distributed:
\begin{equation}
(\text{{dead}})_{j,y} \sim \text{Normal}(\mu_{j,y,\text{{dead}}}, \sigma_{j,y,\text{{dead}}}).
\end{equation}
The mean and standard deviations are given by the equations, 

\begin{equation}
\mu_{j,y,\text{{dead}}} = \gamma  \frac{(\text{succ. attacks per acre})_{j,y}}{(\text{surface area per tree})_j}, \quad \text{and}
\end{equation}

\begin{equation}
\sigma_{j,y,\text{{dead}}} = \sigma_{\text{{dead0}}}  \frac{(\text{succ. attacks per acre})_{j,y}}{(\text{surface area per tree})_j},
\end{equation}

We fit variants of the above model where the standard deviation increased as a square of the independent variable, where the standard deviation was constant, and where the standard deviation contained both constant and dynamics parts (as in the \eqref{eq:sigma_pref}). The model presented above was the best according to Leave-One-Out Cross Validation with expected log predictive density.

\subsection{Tree size-dependent fecundity} \label{subsec:exit}

\begin{figure}[H]
\centering
\includegraphics[scale = 1]{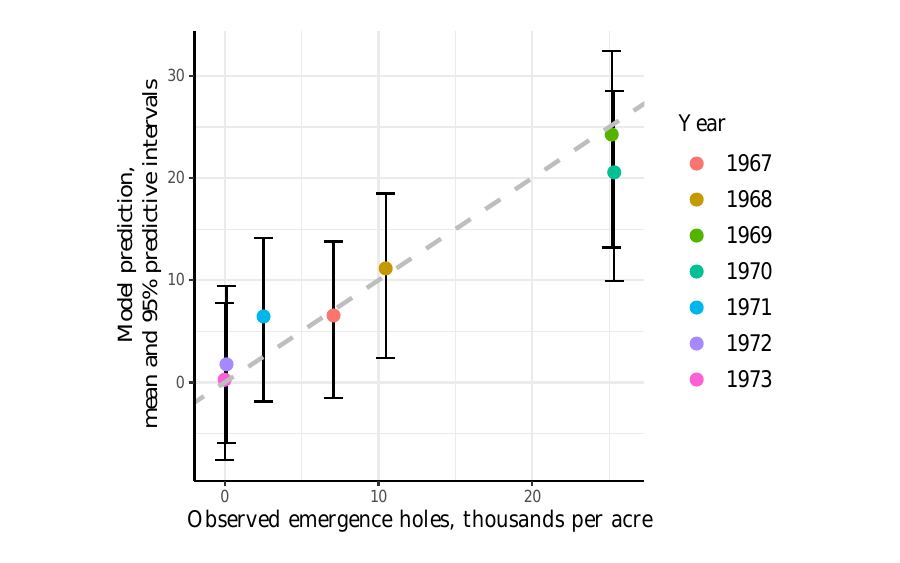}
\caption{Observations vs. predictions for the tree size-dependent fecundity sub-model. The error bars give the 95\% predictive intervals, representing both parameter and sampling uncertainty. The dashed grey line diagonal line represents perfect agreement between observed and predicted values. Data from \citet{klein1978attack}.}
\label{fig:exit1}
\end{figure}

The \citeauthor{klein1978attack} dataset does not contain DBH-specific emergence data. The total number of emergence holes per acre is normally distributed:
\begin{equation}
(\text{{exit density}})_{y+1} \sim \text{Normal}(\mu_{y,\text{{exit}}}, \sigma_{\text{{exit}}}),
\end{equation}
where the mean is a sum over DBH classes:
\begin{equation} \label{eq:emerge2}
\mu_{y+1, \text{exit}} = \sum_{j=1}^{J} \begin{cases}
  \overbrace{\left(\zeta_0 + \zeta_1 \cdot \text{DBH}_{j}\right)}^{\text{Emergence density}} \cdot \overbrace{(\text{KTPA})_{j,y} \cdot (\text{surface area per tree})_{j}}^{\text{surface area of killed trees}}, & \text{if } \text{DBH}_{j} > -\frac{\zeta_0}{\zeta_1} \\
  0, & \text{otherwise}.
\end{cases}
\end{equation}
Emergence occurs after the MPB brood overwinters, so the calendar year index for the response variable and the mean prediction is $y+1$. The standard deviation is simply a constant.  Refining the error structure would require unjustifiable assumptions, given the absence of DBH-specific emergence data. The model predictions are in agreement with the data (Fig.\,\ref{fig:exit1}).

We cannot justify the functional form of \eqref{eq:emerge2} with graphical evidence, since the \citeauthor{klein1978attack} dataset only contains total emergence holes (not stratified by diameter class). However, the linear nature of the relationship has been well-established (\citealp{reid1963biology}, Table 1; \citealp{safranyik1968development}, Fig. 25; \citealp{cole1969mountain}, Fig. 5; \citealp{cudmore2009geographic}, Fig. 6). The linear relationship also implicitly accounts for the fact that the male:female sex ratio of emerging beetles increases with tree diameter \citep[Fig. 5]{cole1976mathematical}. Note that emergence only depends on killed trees, not attacking beetles, because a slightly above-optimal attack density is balanced by intraspecific density dependence, manifesting as a decrease in per capita oviposition or early-instar survival. In other words, there is perfectly compensatory density dependence at the tree scale (Fig.\,\ref{fig:disk_data_pop_map}; \citealp{berryman1974dynamics}, Fig. 2c; \citealp{berryman1976theoretical}, Fig. 2).

\subsection{Allee effect} \label{subsec:allee}

The \citeauthor{boone2011efficacy} dataset contains two pertinent variables: density of trees attacked and proportion of attacked trees that are killed. As a first step, we convert the number of attacked trees to the number of attacking beetles. This is done primarily for practical reasons, as we aim to model the \citeauthor{klein1978attack} data, which lacks information on the number of attacked trees (only successfully attacked trees are reported) 


To estimate the number of attacking beetles, we multiply the attacked trees by 360, the average number of attacks per successfully attacked tree in the \citeauthor{klein1978attack}. Since successfully attacked trees receive more attacks, this is probably an overestimate of beetles per attacked tree, which would ultimately lead to a less informative prior for the Allee threshold parameter, $a$, when modeling the \citeauthor{klein1978attack} data.

The probability of a female successfully killing a tree, denoted by $(\text{{prob. of success}})_y$, is modeled as a normal random variate with mean $\mu_{y,\text{{allee}}}$ and a standard deviation parameter $\sigma_{\text{{allee}}}$, as shown by the following equation,
\begin{equation} \label{eq:prob_success_stat}
(\text{{prob. of success}})_y \sim \text{Normal}(\mu_{y,\text{{allee}}}, \sigma_{\text{{allee}}}).
\end{equation}

The mean probability of success is an exponential saturating function of $(\text{{attacks per acre}})_y$, parameterized with the Allee threshold, denoted $a$:

\begin{equation} \label{eq:mean_allee}
\mu_{y,\text{{allee}}} = 1 - \exp\left(-2.3 \cdot \frac{{(\text{{attacks per acre}})_y}}{a}\right).
\end{equation}

\begin{figure}[H]
\centering
\includegraphics[scale = 1]{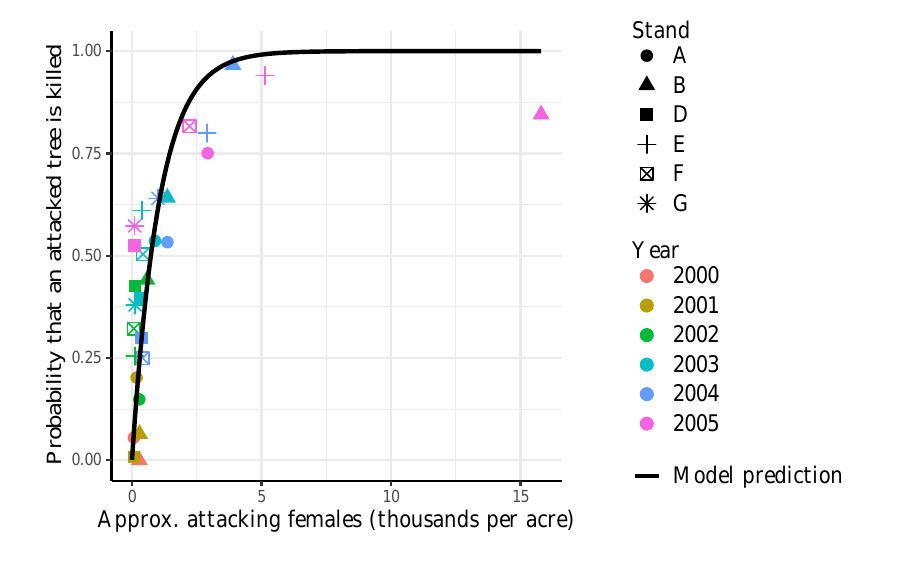}
\caption{The probability of an attacked tree being killed is an increasing function of beetle density. The data is effectively modeled by an exponential saturating function (see \eqref{eq:allee}, ``Probability of success'' term). The prediction was generated using posterior means of model parameters. Data from the \citeauthor{boone2011efficacy} dataset.}
\label{fig:beetle_density_vs_tree_mortality2}
\end{figure}

The exponential saturating function above is justified by the \citeauthor{boone2011efficacy} dataset (Fig.\,\ref{fig:beetle_density_vs_tree_mortality2}). The Allee effect sub-model is first fit to the \citeauthor{boone2011efficacy} data; the subsequent estimate of $a$ is used to set an informative prior for the full statistical outbreak model (constituted by the Allee effect sub-model, other sub-models, and the integrator model), which is fit to the \citeauthor{klein1978attack} data. Specifically, we choose a half-normal prior with a standard deviation equal to half of the posterior mean of $a$.

The normal distribution in \eqref{eq:prob_success_stat} is unrealistic because it implies that probabilities outside of the interval $[0,1]$ are possible. Additionally, the normal distribution assumes constant variance, whereas the \citeauthor{boone2011efficacy} data clearly shows heteroskedasticity. That being said, a more realistic error structure would unnecessarily complicate the model. The normality of errors and constant variance are the least important assumptions when trying to estimate the conditional expectation \citep[Sec. 3.6]{gelman2007data}. This is exactly what we are trying to do: only the mean probability of success (i.e., \eqref{eq:mean_allee}) is used to model the \citeauthor{klein1978attack} data.

Another unrealistic aspect of the Allee effect sub-model is that the probability of success approaches an asymptote of 100\%, whereas visual inspection of the data suggests an approximate maximum of 80-90\% Fig. \ref{fig:beetle_density_vs_tree_mortality2}. However, in the integrator model (Section \ref{Integrator model description}), an additional parameter for the maximum probability of successful attacks would be non-separable with the maximum dispersal-phase survival, $s_0$. Thus, our estimate of $s_0$ not only represents maximum survival during the dispersal phase, but also the fact that some mass attacks are unsuccessful despite high beetle density. Additionally, $s_0$ captures the phenemenon where low-vigor females successfully attack trees but do not oviposit (5--30\% of females could fall into this category; \citealp{amman1975abandoned, amman1980incidence}).


\subsection{Integrator model} \label{Integrator model description}

The integrator model combines the deterministic skeletons of all sub-models (Appendices \ref{subsec:pref}--\ref{subsec:allee}) to predict the next year's gallery starts, solely from this year's gallery starts. In doing so, we can estimate several parameters that cannot be estimated from singular sub-models: $a$, $s_0$, $\lambda_0$, $\lambda_1$, and a residual noise parameter, $\sigma_{\text{resid}}$.



First, we predict the distribution of successful beetle attacks among diameter classes. The predicted mass attack preference is
\begin{equation}
\hat{\text{preference}_{j,y}} =
\begin{cases}
\beta_0 + \beta_1 \cdot \text{{DBH}}_j & \text{if } \text{{DBH}}_j > -\beta_0 / \beta_1 \\
0 & \text{otherwise}.
\end{cases}
\end{equation}
Note that the $\text{preference}_{j,y}$ has a ``hat'', which will be used throughout this section to distinguish model-based predictions from data.

Next, we use the total number of successful beetle attacks --- denoted $(\text{succ. attacks per acre})_{y}$ and operationalized as gallery starts per acre --- to predict the number of successful beetle attacks within the $j^{th}$ DBH class, denoted $(\text{succ. attacks per acre})_{j,y}$: 
\begin{equation}
\hat{(\text{succ. attacks per acre})_{j,y}} = \frac{(\text{succ. attacks per acre})_{y} \cdot \hat{(\text{preference})_{j,y}} \cdot (\text{prop. area})_{j,y}}{\sum_{k=1}^{J} \hat{(\text{preference})_{k,y}} \cdot (\text{prop. area})_{k,y}},
\end{equation}
The quantity $(\text{prop. area})_{j,y}$ is the proportion of total tree surface area belonging to a particular diameter class.

\begin{figure}[H]
\centering
\includegraphics[scale = 1]{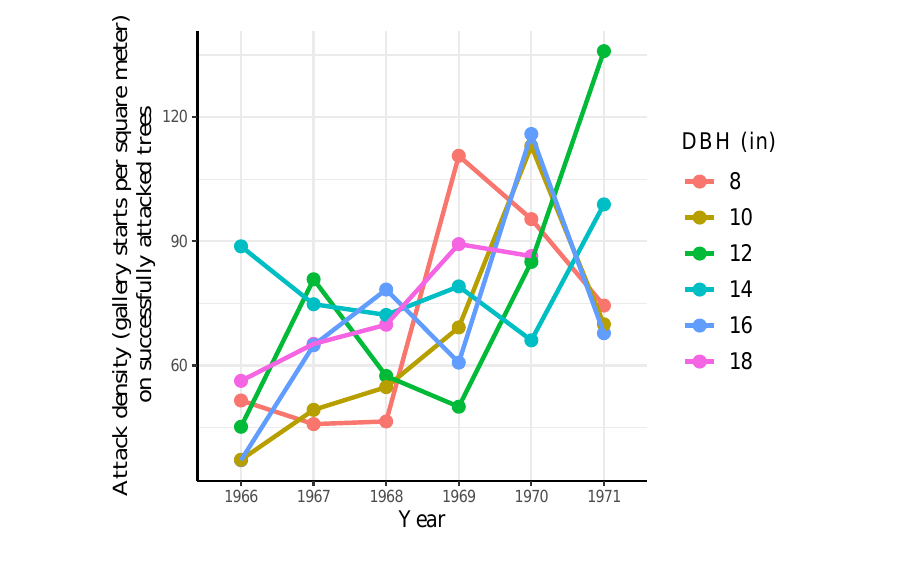}
\caption{Successful attack density (gallery starts per square meter) increases slightly over time, but does not depend on tree diameter at breast height (DBH). Data from \citet{klein1978attack}. }
\label{fig:size_dead_just3}
\end{figure}

The equation for successful attacks, \eqref{eq:pref}, shows how beetles are distributed among tree diameter classes, but not how beetles are distributed among individual trees. As it turns out, killed trees receive an approximately invariant number of attacks per unit surface area (Fig.\,\ref{fig:size_dead_just3}; see further discussion in Section \ref{life_history:size_pref}).

If beetles are unsuccessful, the tree survives and the beetles are assumed to die. This is a simplifying assumption because unsuccessful beetles could attack other trees. The assumption is perhaps justifiable because some beetles will be killed by the first tree, the remainder face increased mortality risk during secondary dispersal, and late-arriving beetles have low fitness \citep{raffa2001mixed}. 

Successful beetle attacks are converted into killed trees with
\begin{equation}
\hat{\text{dead}}_{j,y} = \text{min}\left( \gamma \; \frac{\hat{(\text{successful attacks per acre})}_{j,y}}{(\text{susceptible surface area per tree})_{j,y}}, \text{TPA}_{j,y} \right),
\end{equation}
and killed trees are converted into emerging females with

\begin{equation} \label{eq:emerge3}
\hat{(\text{{exit density}})_{y+1}} = \sum_{j=1}^{J} \begin{cases}
  \overbrace{\left(\zeta_0 + \zeta_1 \cdot \text{DBH}_{j}\right)}^{\text{Emergence density}}(\overbrace{(\text{KTPA})_{j,y} (\text{surface area per tree})_{j}}^{\text{surface area of killed trees}}), & \text{if } \text{DBH}_{j} > -\frac{\zeta_0}{\zeta_1} \\
  0, & \text{otherwise}.
\end{cases}
\end{equation}

To predict the number of attacking beetles, we multiply the exit hole density by the predicted probability of survival during the dispersal phase. This quantity is a logistically increasing function (from 0 to the maximum $s_0$) of ``relative surface area'' which is defined as the susceptible surface area of all trees with $DBH \geq 8\,\text{inches}$ in year $y$, divided by the total susceptible surface area when MPB is absent and the forest is at equilibrium. We use the \textit{relative} surface area mostly for interpretability (e.g., 0.5 means the forest is half-depleted) but also for computational reasons (the Bayesian model-fitting program \textit{Stan} prefers all parameters to be on the unit-scale). Unlike the simulations, there are no stochastic influxes of beetles:

\begin{equation} \label{eq:sddda}
\hat{(\text{{attacks per acre}})_{y+1}} =  \frac{s_0 \; \hat{(\text{{exit density}})_{y+1}} }{1+\exp \left( -\lambda_1 \left( (\text{relative surface area})_{y+1} - \lambda_0 \right) \right)}.
\end{equation}

We cannot directly justify the functional form of \eqref{eq:sddda} --- say, by plotting total emergence against subsequent total attacks --- because \citeauthor{klein1978attack} only reported attacks on successfully killed trees. However, we note that the logistic function is a standard way to model probabilities, and that the extreme limit of the function (i.e., dispersal mortality is near 100\% when there are few trees) is supported by both the \citeauthor{klein1978attack} dataset (Fig.\,\ref{fig:mort2}) and the literature (see Section \ref{life_history:stand_dependent_dispersal}). In particular, \citet{powell2014phenology} estimated that mean dispersal distance increased from several meters to several kilometers as tree density decreased.

To complete the virtual life-cycle and predict the number of successful attacks next year, denoted $\mu_{y,\text{full}}$, the Allee effect is applied to the predicted attack density:

\begin{equation}
\mu_{y+1,\text{full}} = \hat{(\text{{attacks per acre}})_{y+1}} \times \left(1 - \exp\left(-2.3 \cdot \frac{{\hat{(\text{{attacks per acre}})_{y+1}}}}{a}\right)\right).
\end{equation}

Finally, the likelihood of the integrator model is given by a normal distribution with a residual noise parameter.

\begin{equation}
(\text{{succ. attacks per acre}})_{y+1} \sim \text{Normal}(\mu_{y+1,\text{{full}}}, \sigma_{\text{{resid}}}).
\end{equation}

\newpage
\begin{landscape} 
\begingroup\fontsize{8pt}{9pt}\selectfont
\renewcommand*{\arraystretch}{1.4}
    \begin{longtable}{p{0.2\textwidth} p{0.7\textwidth} p{0.2\textwidth} p{0.2\textwidth}}
    \caption{Statistical model notation}\\
    \label{tab:model_notation}\\
        \toprule
        \textbf{Symbol} & \textbf{Description} & \textbf{Units} & \textbf{submodel(s)} \\

        \midrule
        \textbf{Subscript} & & & \\
        \midrule
        
        \(t\) & Tree age in years & N/A & Lodgepole, \ref{Lodgeple pine forest dynamics}\\
        j & The DBH-class index, uniquely identifies a tree diameter class & N/A & All \\
        y & The year index, & N/A & All\\
        
        tree & Pertaining to the lodgepole pine dyanamics sub-model & N/A & Lodgepole, \ref{Lodgeple pine forest dynamics} \\
        pref & Pertaining to the tree size-dependent preference sub-model & N/A & Attack pref., \ref{subsec:pref} \\
        dead & Pertaining to the tree to killed tree sub-model & N/A & Killed trees, \ref{subsec:dead} \\
        exit & Pertaining to the tree size-dependent fecundity sub-model & N/A & Fecundity, \ref{subsec:exit} \\
        allee & Pertaining to the Allee effect sub-model & N/A & Allee effect, \ref{subsec:allee} \\
        
        \midrule
        \textbf{Data} & & & \\
        \midrule

        \(\text{(upper age)}_j\) & Upper age limit of trees within a particular diameter class & years & Lodgepole, \ref{Lodgeple pine forest dynamics} \\
        \(\text{(lower age)}_j\) & Lower age limit of trees within a particular diameter class  & years & Lodgepole, \ref{Lodgeple pine forest dynamics} \\        
        \( \text{TPA}_j \) & Density of trees within the ages corresponding to the $j^{th}$ diameter class & $\text{trees} \cdot$ $\text{acres}^{-1}$ & Lodgepole, \ref{Lodgeple pine forest dynamics} \\

        $(\text{{preference}})_{j,y}$ & Mass attack preference, defined as the attack density within a particular diameter class, divided by the attack density across all trees (within a year) & unitless & Attack pref., \ref{subsec:pref}\\
        \(\text{{DBH}}_j\) & Diameter at breast height & inches & Attack pref., \ref{subsec:pref} \\ 
        
        $(\text{{KTPA}})_{j,y}$ & Killed trees per acre & $\text{trees}\cdot$ $\text{acres}^{-1}$ & Killed trees, \ref{subsec:dead} \\
        $(\text{succ. attacks per acre})_{j,y}$ & Gallery starts per acre & $(\text{gallery starts})\cdot$ $\text{acres}^{-1}$ & Killed trees, \ref{subsec:dead} \\
        $(\text{surface area per tree})_j$ & Susceptible surface area per individual tree, defined by \eqref{DBH_to_SA} & $\text{meters}^2 \cdot$ $\text{trees}^{-1}$ & Killed trees, \ref{subsec:dead} \\        
        
        $(\text{{exit density}})_y$ & Emergence holes per acre & ${(\text{exit holes})\cdot}$ $\text{acres}^{-1}$ & Fecundity, \ref{subsec:exit} \\
        \(J\) & Total number of unique diameter classes,  & unitless & Fecundity, \ref{subsec:exit} \\
        \(Y\) & Total number of years in the data set & unitless & Fecundity, \ref{subsec:exit} \\
        
        $(\text{{attacks per acre}})_{j,y}$ & Attacking female beetles per acre & $\text{beetles}\cdot$ $\text{acres}^{-1}$ & Allee effect, \ref{subsec:allee} \\
        
        $(\text{prop. area})_{j,y}$ & Proportion of total susceptible tree surface area belonging to a particular diameter class & unitless & Integrator, \ref{Integrator model description} \\
        
        \hline
        \textbf{Derived quantities} & & & \\
        \midrule
        \(\mu_{j,\text{tree}}\) & Expected tree density in the $j^{th}$diameter class & $\text{trees} \cdot$ $\text{acres}^{-1}$ & Lodgepole, \ref{Lodgeple pine forest dynamics} \\
        
        \(\mu_{j,y,\text{{pref}}}\) & Predicted mean of the mass attack preference & unitless & Attack pref., \ref{subsec:pref} \\
        \(\sigma_{j,y,\text{{pref}}}\) & Standard deviation of the mass attack preference & unitless  & Attack pref., \ref{subsec:pref} \\

        \(\mu_{j,y,\text{{dead}}}\) & Predicted mean of the killed trees density & $\text{trees} \cdot$ $\text{acres}^{-1}$ & Killed trees, \ref{subsec:dead} \\
        \(\sigma_{j,y,\text{{dead}}}\) & Standard deviation of the killed trees density & $\text{trees} \cdot$ $\text{acres}^{-1}$ & Killed trees, \ref{subsec:dead} \\ 
        
        \(\mu_{y,\text{{exit}}}\) & Predicted mean of emergence hole density & $(\text{exit holes}) \text{acres}^{-1}$ & Fecundity, \ref{subsec:exit} \\  

        \(\mu_{j,y,\text{{allee}}}\) & Mean proportion of successful attacks & unitless & Allee effect, \ref{subsec:allee} \\
        
        \(s_y\) & Actual survival probability for beetles during the dispersal phase & unitless & Integrator, \ref{Integrator model description}  \\

        \(\mu_{y,\text{full}}\) & Expected density of successful beetle attacks (across all DBH classes) & $(\text{gallery starts})\cdot$ $\text{acres}^{-1}$ & Integrator, \ref{Integrator model description} \\

        \hline
        \textbf{Model parameters} & & & \\
        \midrule
        
        \(K\) & Tree carrying capacity & $\text{trees}\cdot$ $\text{acres}^{-1}$  & Lodgepole, \ref{Lodgeple pine forest dynamics} \\
        \(m\) & Annual probability of tree death (not caused by MPB) & unitless & Lodgepole, \ref{Lodgeple pine forest dynamics} \\
        \(\sigma_{\text{tree}}\) & Noise parameter & $\text{trees}\cdot$ $\text{acres}^{-1}$  & Lodgepole, \ref{Lodgeple pine forest dynamics} \\        

        \(\beta_0\) & Y-intercept of the piecewise linear preference-DBH function & unitless & Attack pref., \ref{subsec:pref} \\
        \(\beta_1\) & Slope of the piecewise linear preference-DBH function; the DBH threshold for attacks is $-\beta_0/\beta_1$ & unitless & Attack pref., \ref{subsec:pref} \\             \(\sigma_{\text{{pref0}}}\) & noise parameter; accounts for a baseline level of variability in mass attack preference & unitless & Attack pref., \ref{subsec:pref} \\
        \(\sigma_{\text{{pref1}}}\) & Noise parameter; accounts for increasing variability with smaller sample sizes in mass attack preference &  unitless & Attack pref., \ref{subsec:pref} \\

        \(\gamma\) & Proportionality constant for number of killed trees; the reciprocal of the average attack density &  $\text{meters}^2\cdot$ $(\text{gallery starts})^{-1}$ & Killed trees, \ref{subsec:dead} \\
        \(\sigma_{\text{{dead0}}}\) & noise parameter; the effect of the relative attack density on the standard deviation &  $\text{meters}^2\cdot$ $(\text{gallery starts})^{-1}$  & Killed trees, \ref{subsec:dead} \\

        \(\zeta_0\) & Y-intercept of the piecewise linear emergence-DBH function & $(\text{exit holes})\cdot$ $(\text{meters})^{-2}\cdot$ $(\text{acres})^{-1}$ & Fecundity, \ref{subsec:exit} \\   
        \(\zeta_1\) & Slope of the diameter-emergence regression; the DBH threshold for emergence is $-\zeta_0/\zeta_1$ & $(\text{exit holes})\cdot$ $(\text{meters})^{-2}\cdot$ $(\text{acres})^{-1}\cdot$ $(\text{inches})^{-1}$ & Fecundity, \ref{subsec:exit} \\
        \(\sigma_{\text{{exit}}}\) & Standard deviation parameter for emergence holes & $(\text{exit holes})\cdot$ $\text{acres}^{-1}$ & Fecundity, \ref{subsec:exit} \\

        \(a\) & Allee threshold (density of beetles at which probability of attack success is 0.9) & $\text{beetles}\cdot$ $\text{acres}^{-1}$ & Allee effect, \ref{subsec:allee} \\
        \(\sigma_{\text{{allee}}}\) & Standard deviation parameter for proportion of successful attacks & unitless & Allee effect, \ref{subsec:allee} \\
        
        \(s_0\) & Maximum survival probability for beetles during the dispersal phase & unitless & Integrator, \ref{Integrator model description} \\
        \(\lambda_1\) & Sensitivity of dispersal mortality on tree surface area; controls the speed of the transition from low to high dispersal mortality as trees are killed. & unitless & Integrator, \ref{Integrator model description} \\
        \(\lambda_0\) & Transition point parameter; controls the level of tree death at which dispersal mortality becomes significant & unitless & Integrator, \ref{Integrator model description} \\        
        \(\sigma_{\text{resid}}\) & Residual noise parameter, to account for error in the predicted density of successful beetle attacks next year & $(\text{gallery starts})\cdot$ $\text{acres}^{-1}$  & Integrator, \ref{Integrator model description} \\

        \bottomrule
  \end{longtable}
  \endgroup
\end{landscape} 
\newpage

\section{Additional model justification} \label{Additional model justification}

\subsection{Stochastic perturbation size} \label{Stochastic perturbation size}

To simulate the influx of female beetles corresponding to the incipient-epidemic transition, we set $I =300$ (see \eqref{eq:imm} in the main text). This number was obtained with a back-of-the-envelope calculation: the median emergence density of females across several studies and Forest Service reports  (\textit{Other Studies} in Fig.\,\ref{fig:emerge_hist}), multiplied by the susceptible surface area of typical killed tree (10 inches DBH as an input to \eqref{DBH_to_SA}), multiplied by the median estimate of the dispersal-phase survival probability (using gallery-starts/emergence-holes quotient from the \citeauthor{klein1978attack} dataset, Fig.\,\ref{fig:mort2}), multiplied by 0.81, which is the number of successfully attacked trees per acre that corresponds to an incipient-epidemic state (according to \citealp{carroll2006direct}). The chosen values of $P_I$ and $I$ are educated guesses, but we re-ran our analysis with several different values and obtained qualitatively identical results.

\subsection{Mass attack preference does not change} \label{Tree size-dependent preference}

\begin{figure}[H]
\centering
\includegraphics[scale = 1]{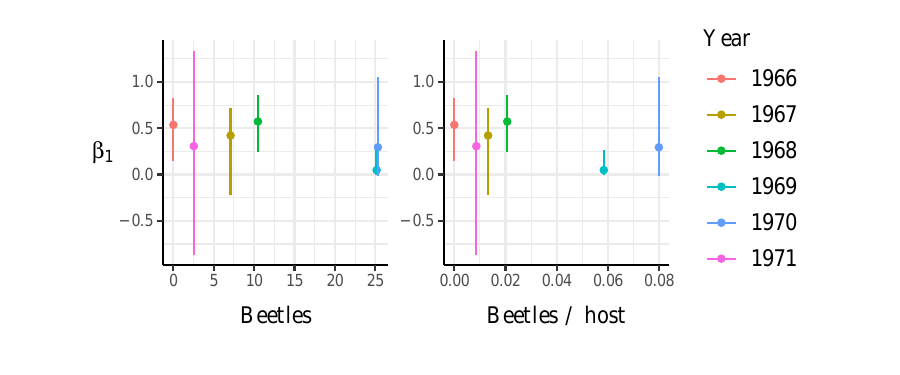}
\caption{The mean and 95\% credible intervals of $\beta_1$ --- the slope parameter in the preference$\sim$DBH model --- as a function of beetle pressure. To obtain multiple estimates of $\beta_1$, the tree size-dependent preference sub-model was fit independently to each year of data. Two metrics of beetle pressure are used: \textit{Beetles}, i.e., recently emerging female beetles (units: thousands per acre), and \textit{Beetles/host}, i.e., the quotient of emerging female density and within-DBH-class surface area (units: females per acre per square meter). There is no clear relationship between $\beta_1$ and beetle pressure. Data from \citet{klein1978attack}.}
\label{fig:size_pref_just1}
\end{figure}

\begin{figure}[H]
\centering
\includegraphics[scale = 1]{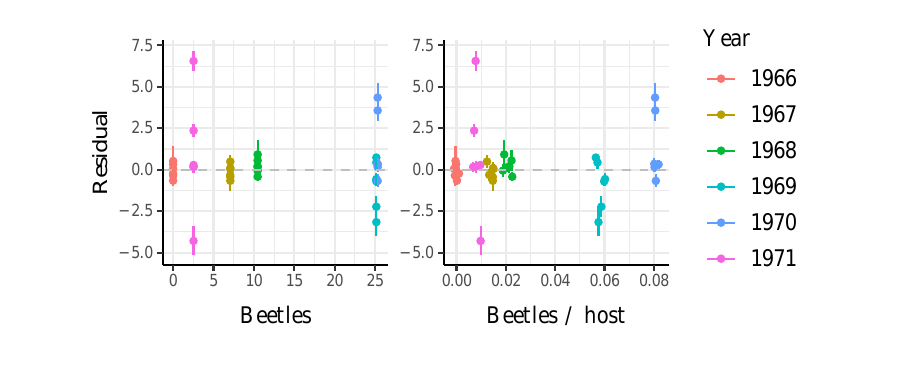}
\caption{The mean and 95\% credible intervals of the $(\text{preference})_{j,y}$ residuals --- from the tree size-dependent preference sub-model --- as a function of beetle pressure. Two metrics of beetle pressure are used: \textit{Beetles}, i.e., recently emerging female beetles (units: thousands per acre), and \textit{Beetles/host}, i.e., the quotient of emerging female density and within-DBH-class surface area (units: females per acre per square meter). There is no clear relationship between residual values and beetle pressure. Data from \citet{klein1978attack}.}
\label{fig:size_pref_just2}
\end{figure}

One potential shortcoming of our model is that preference only depends on tree diameter, whereas previous papers have shown that MPB's preference for large-diameter trees increases as outbreaks progress \citep{boone2011efficacy, howe2022numbers}. To investigate further, we separately fit our preference sub-model to each year of data and plotted $\beta_1$ (the slope of the preference $\sim$ DBH regression) against beetle pressure, as measured by the density of emerging females, and the number of emerging females per unit of susceptible tree surface area (Fig.\,\ref{fig:size_pref_just1}). There was no clear pattern, indicating that dynamic changes to the DBH preference are small or nonexistent. Similarly, there was no relationship between beetle pressure and the residuals of our original preference model (Fig.\,\ref{fig:size_pref_just2}).

There are a couple of ways to explain the absence of evidence for dynamic DBH-preferences. First, it could be that MPB's large-tree preference ``kicks in'' at an early, unobserved stage of the \citeauthor{klein1978attack} outbreak. In Fig. 6. of the publication by \citet{boone2011efficacy}, the ratio of DBH of trees entered and not entered increases from 1.2 to 1.6 as an outbreak progresses from endemic to eruptive conditions. However, based on \citeauthor{boone2011efficacy}'s classification scheme, the \citeauthor{klein1978attack} outbreak is approximately in the eruptive condition until the very end of the outbreak (calculations in supplementary files). Second, dynamic changes to DBH-preferences could be small. In Fig. 3D of \citet{howe2022numbers}, the mean DBH of attacked trees only increased a few centimeters as infestation severity increased from 1 to 1000 attacked trees per hectare.

\subsection{Increasing attack density does not affect population dynamics} \label{Converting beetle attacks to tree death}

\begin{figure}[H]
\centering
\includegraphics[scale = 1]{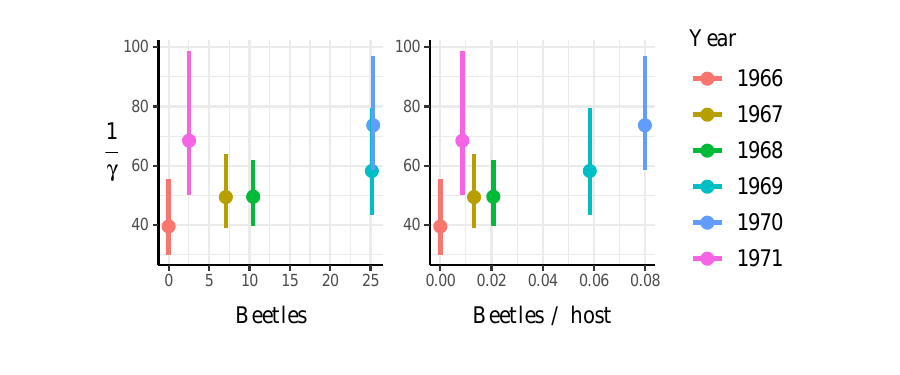}
\caption{The mean and 95\% credible intervals of $1/\gamma$ --- the idealized attack density in killed-tree sub-model --- as a function of beetle pressure. To obtain multiple estimates of $\gamma$, the killed-tree sub-model was fit independently to each year of data. Two metrics of beetle pressure are used: \textit{Beetles}, i.e., recently emerging female beetles (units: thousands per acre), and \textit{Beetles/host}, i.e., the quotient of emerging female density and within-DBH-class surface area (units: females per acre per square meter). Data from \citet{klein1978attack}.}
\label{fig:size_just_dead3}
\end{figure}

Figure \ref{fig:size_dead_just3} in the main text shows that the density of attacks on killed trees (e.g., gallery starts per square meter) increases over time. This pattern can potentially be explained by a positive relationship between attack density and beetle population size, an assumption made by several models in the literature. To investigate further, we fit the killed-tree sub-model (Appendix \ref{subsec:dead}) to each year of data and plotted $1/\gamma$ (the model-based estimate of attack density) against beetle pressure, as measured by the density of emerging females, and the number of emerging females per unit of susceptible tree surface area (Fig.\,\ref{fig:size_just_dead3}). The plots are suggestive of a positive relationship if the data from 1971 are excluded, but the nature of the relationship is ultimately inconclusive.

\begin{figure}[H]
\centering
\includegraphics[scale = 1]{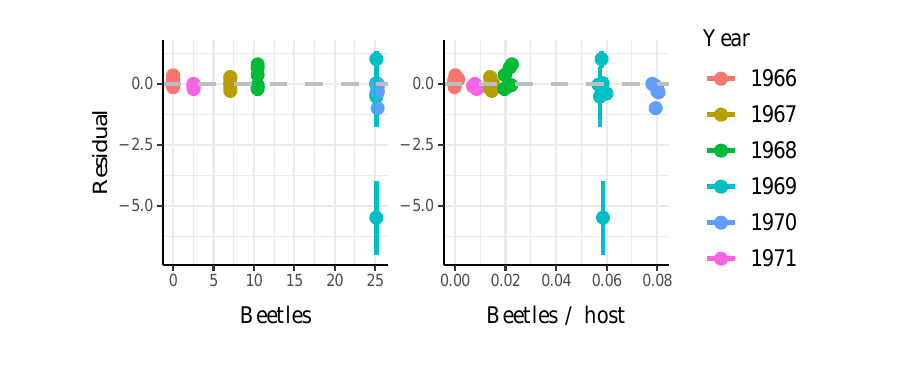}
\caption{Mean and 95\% credible intervals of $(\text{KTPA})_{j,y}$ residuals from the killed-tree sub-model, as a function of beetle pressure. Two metrics of beetle pressure are used: \textit{Beetles}, i.e., recently emerging female beetles (units: thousands per acre), and \textit{Beetles/host}, i.e., the quotient of emerging female density and within-DBH-class surface area (units: females per acre per square meter). Data from \citet{klein1978attack}.}
\label{fig:size_dead_just2}
\end{figure}

Even if beetle pressure does affect attack density, this has negligible influence on population dynamics. The killed-tree sub-model provides an excellent fit to the data, and the residual variation cannot be explained by beetle pressure (Fig.\,\ref{fig:size_dead_just2}). This is likely because variation in attack density is swamped by variation in $\frac{(\text{succ. attacks per acre})_{j,y}}{(\text{surface area per tree})_{j,y}}$ (see \eqref{eq:KTPA} in the main text).

\subsection{Leave one year out cross validation} \label{LOOCV}

\begin{figure}[H]
\centering
\includegraphics[scale = 1]{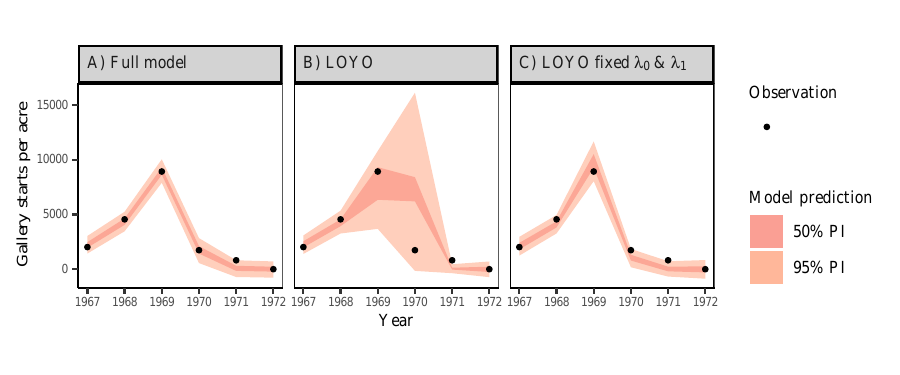}
\caption{Observations of successful attack density in the \citeauthor{klein1978attack} dataset (operationalized as gallery starts), with superimposed model-based predictive intervals (PIs). Panel A shows in-sample predictions for the full statistical model; panel B shows out-of-sample, leave-one-year-out (LOYO) predictions, where the model is not trained on the focal year's data; and panel C shows LOYO predictions when the stand-dependent dispersal mortality parameters ($\lambda_0$ and $\lambda_1$) are fixed at the medians of their respective marginal posterior distributions (see Table \ref{tab:pars_full}). Panels B \& C demonstrate that data from a single beetle generation (1969--1970) has an overwhelming influence on the estimation of $\lambda_0$ and $\lambda_1$, but that all other parameter estimates are robust.} 
\label{fig:integrator_pred_panel}
\end{figure}

The statistical model successfully predicts in-sample data (Fig.\,\ref{fig:integrator_pred_panel}, panel A) and out-of-sample data. The exception is that the out-of-sample predictive intervals for 1970 are wide (Fig.\,\ref{fig:integrator_pred_panel}, panel B), which upon further investigation, was found to be the result of uncertain estimates of the stand-dependent dispersal mortality parameters, $\lambda_0$ and $\lambda_1$ (Fig.\,\ref{fig:integrator_pred_panel}, panel C).

\begin{figure}[H]
\centering
\includegraphics[scale = 1]{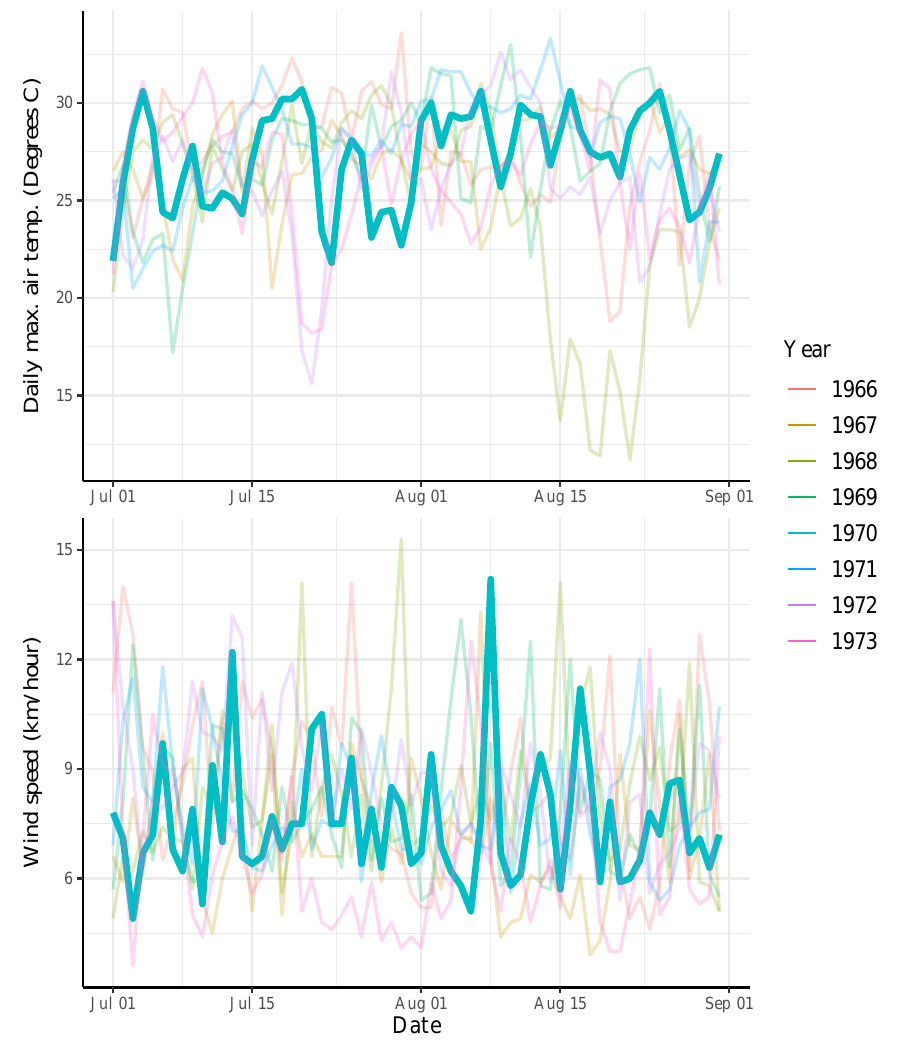}
\caption{In 1970, emergence density was high but subsequent attack density was low. Anomalous weather conditions during the flight period could be responsible for this pattern, but the relevant weather variables were unexceptional in 1970 (the bold, teal line). The maximum summer temperature was consistently greater than 18.3\textdegree C, the lower limit for beetle flight \citep{mccambridge1971temperature}. The estimated wind speed often exceeded 8 km/hour, the proposed upper threshold for beetle flight \citep{gray1972emergence}; however, the wind speeds in 1970 were not anomalous in comparison to the other outbreak years. Weather data was obtained from the \textit{Biosim} program \citep{regniere2014biosim}.}
\label{fig:summer_weather}
\end{figure}

\begin{figure}[H]
\centering
\includegraphics[scale = 1]{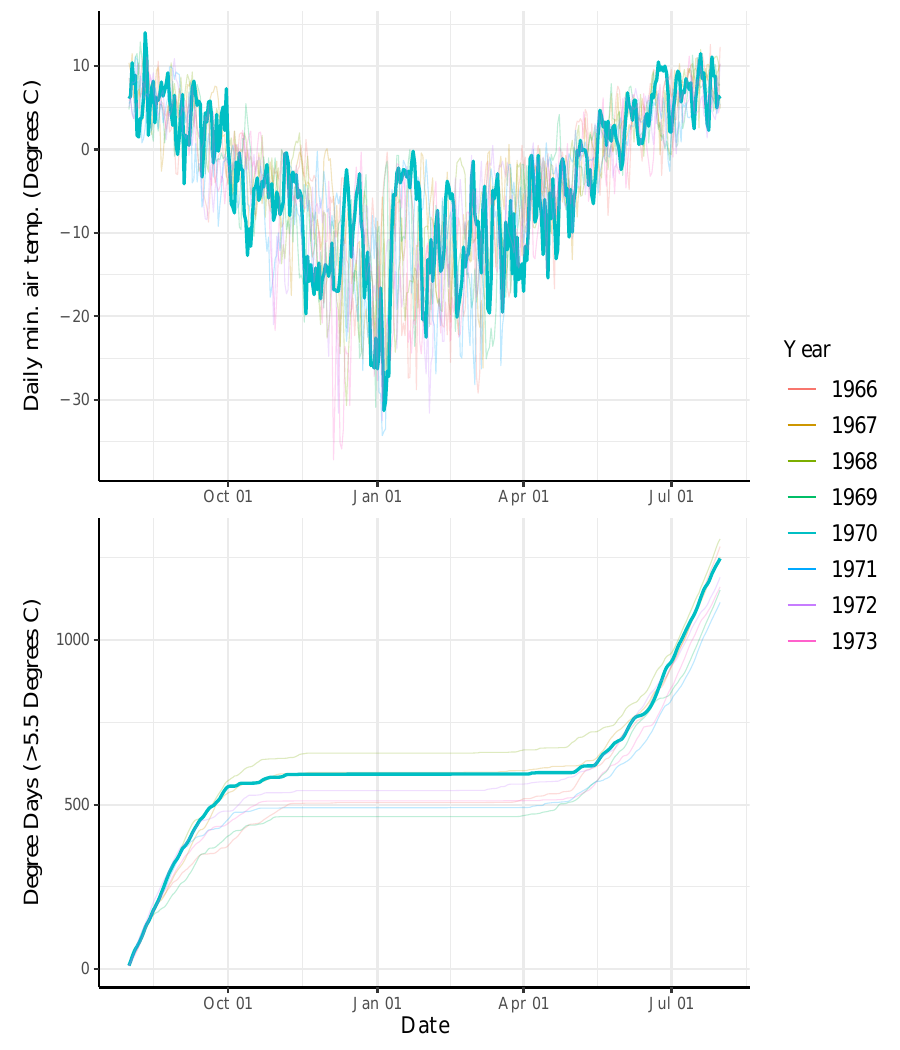}
\caption{In 1970, emergence density was high but subsequent attack density was low. The decline could theoretically be attributed to poor environmental conditions that weakened the beetles, despite not being lethal (because they still emerged). However, the developmental period leading up to 1970 had unexceptional weather conditions. The minimum winter temperature is related to larval mortality (here, sub-lethal damage), and degree days are related to emergence and univoltinism. The minimum winter temperature is approximately -30\textdegree C, which is substantially greater than the proposed threshold of -40\textdegree C \citep{safranyik1998mortality} for significant overwintering mortality. There are approximately 1250 degree days from Aug. 1 to Jul. 31, far beyond the proposed threshold of 833 for the population to be univoltine \citep{reid1962biology}. Weather data was obtained from the \textit{Biosim} program \citep{regniere2014biosim}.}
\label{fig:year_weather}
\end{figure}

The estimation of stand-dependent dispersal mortality parameters is highly dependent on the data from 1969--1970. This would be indicative of overfitting (i.e., fitting the noise) \textit{if} the decrease in attack density in 1970 was the result of noise, rather than stand-dependent dispersal mortality. The obvious source of statistical noise is the weather, but the weather leading up to the summer of 1970 was unexceptional (Figs. \ref{fig:year_weather} \& \ref{fig:summer_weather}). Taken together, evidence for the presence of stand-dependent dispersal mortality (Section \ref{life_history:stand_dependent_dispersal}) and the absence of anomalous weather patterns imply that the model did not overfit.

\subsection{The Allee effect depends only on beetle density} \label{Allee effect}

Our model states that the success probability (i.e., the probability that an attacked tree is successfully killed) only depends on beetle density. Several previous papers have modeled the success probability as an increasing function of the quotient of beetles and susceptible trees, a simplifying assumption made for mathematical tractability. In this appendix, we argue that these models 1) are unrealistic, and 2) themselves show that the quotient of beetles and trees is superfluous for modeling the Allee effect, as a practical matter.

A good example of a model with non-separable beetles and trees is provided by \citet{goodsman2017positive}. In this model, the densities of beetles and susceptible trees are denoted $B$ and $S$, respectively. In each generation, only an $\alpha$ proportion of susceptible trees are attacked. As the authors write, “In reality, because of the clustered nature of attacks in space, most trees in a stand are not attacked and attacks are focused on a small subset of trees \citep{logan1998model} some of which may receive many more attacks than others.” The probability that an attacked tree is killed is an increasing function of $m=B/S$.

Each attacked tree receives a number of beetles that is drawn from a negative binomial distribution. The distribution is parameterized with the mean number of beetles, $m = B/(\alpha S)$, and a dispersion parameter $k$ (note that we are redefining some parameters from the main text). Both parameters modulate aggregation --- small $k$ and small $\alpha$ both correspond to high aggregation, where a small number of trees receive a large number of beetles. Finally, a tree is killed if the number of beetles it receives is greater than the threshold parameter $\phi$. With these dynamics, the probability of success (i.e., the probability that an attacked tree is killed) is the probability mass across all attacks with more than $\phi$ beetles:
\begin{equation} \label{eq:goodsman}
 F(m; \phi, k) = \sum_{i=\phi+1}^\infty NB(i;m,k). 
\end{equation}
This is the complementary CDF of the negative binomial distribution.

While this model may be attractive from a mathematical point of view, it has several unrealistic assumptions: the proportion of attacked trees, $\alpha$, does not scale with the number of beetles; the per-surface-area attack density may be arbitrarily large, when in reality it is constrained by gallery defense and anti-aggregation pheromones; and brood size does not increase once the attack threshold $\phi$ is met (see \citealp{raffa1983role}, Fig. 6 for counter-evidence). More seriously, the model leads to unrealistic population dynamics. If aggregation is too strong (i.e. $k \ll 1$), then the beetles will always be able to find the few fractional trees that slip into the susceptible age class each year; this leads to a large stochastic equilibrium of beetles that is much larger than endemic population seen in nature. The solution, it would seem, is to accept that aggregation is not so strong. But weak aggregation makes it difficult for beetles to invade a mature forest; $m = B/S$ is small, and the beetles are spread evenly across trees, so the threshold $\phi$ is never exceeded. Counter-intuitively, the beetles get worse at invading as a forest becomes denser since $m=B/S$ becomes smaller. In our experience, with realistic values of $\phi$ and $\alpha$, it is difficult to find values of $k$ where beetles experience realistic boom-bust dynamics --- where beetle populations will be able to invade, but will later recede to an endemic state.

While acknowledging certain limitations in the \citet{goodsman2017positive} model, we will accept it as a premise to illustrate that the probability of mass attack success is predominantly influenced by beetle density. Our Allee effect sub-model with a single parameter (Appendix \ref{subsec:allee}) is well-approximated by the comparatively parameter-rich model of \citet{goodsman2016aggregation}, when the latter model's parameters are given realistic values.

\begin{figure}[H]
\centering
\includegraphics[scale = 1]{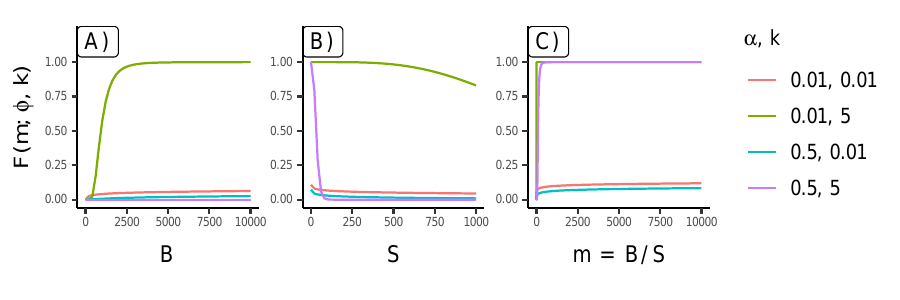}
\caption{The probability that an attack tree is killed, calculated via \citepos{goodsman2017positive} model (\eqref{eq:goodsman}), as a function of $B$ (female beetles per acre), $S$ (susceptible trees per acre), and $m=B/S$. The range for beetle density was selected based on the \citeauthor{klein1978attack} dataset, and the range for susceptible tree density was selected based on the literature: the density of lodgepole pine forests with susceptible trees rarely exceeds 1500 trees/ha, or equivalently, $\approx 600$ trees/acre \citep{nigh2008density, pfeifer2011observations, meddens2011evaluating}. For panels A and B, the other state variable is fixed at $S = 500$ and $B=3000$ respectively. To reduce the parameter space that we must numerically explore, we fix $\phi = 176$, which is the attack density threshold given by \citet{raffa1983role} --- approximately 40 females per square meter ---- times the susceptible surface area of a 10-inch tree --- approximately 4.4 square meters.}
\label{fig:goodsman1}
\end{figure}

\begin{figure}[H]
\centering
\includegraphics[scale = 1]{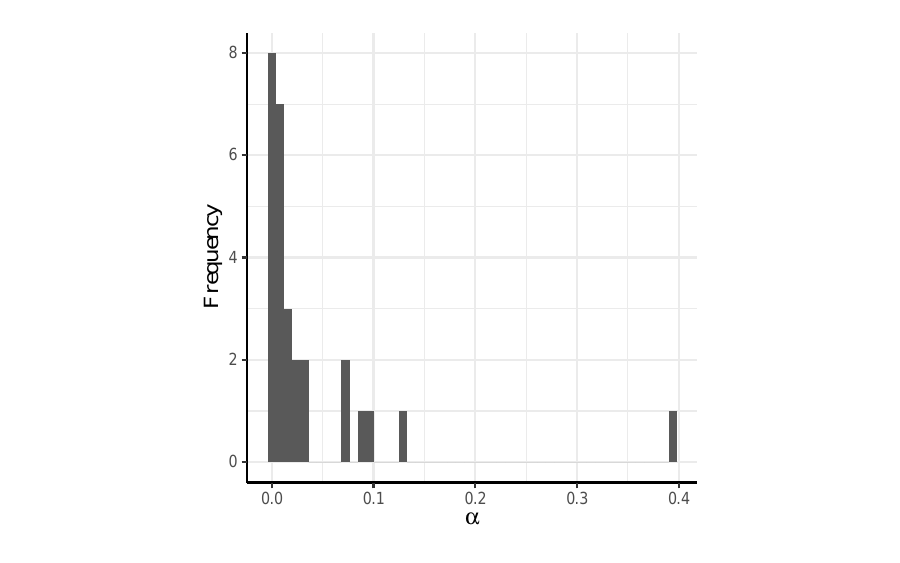}
\caption{Estimates of $\alpha$, the proportion of susceptible trees that are attacked. The data are sourced from \citet[Table 1]{boone2011efficacy}. The median is $0.009$.}
\label{fig:boone_alpha}
\end{figure}

We examine the relationship between the success probability, here given by \eqref{eq:goodsman}, and our three state variables: $B$, $S$, and $m=B/S$. More specifically, we examine how the nature of this relationship changes as the parameters $\alpha$ and $k$ change. Figure \ref{fig:goodsman1} shows that the probability of success only shows a weak dependence on $S$ when the dispersion parameter is large (i.e.  $k > 1$) and the attacked fraction of trees is small (i.e. $\alpha \ll 1$). This is precisely the parameter combination that best corresponds to reality (Table \ref{agg_stats}; Fig.\,\ref{fig:boone_alpha}): beetles spread themselves evenly (i.e. $k > 1$) across a small fraction of susceptible trees (i.e. $\alpha \ll 1$), which they can successfully kill. Success probability remains high unless $S$ is unrealistically large.


\begin{table}[H]
\centering
\begin{tabular}{lllllll}
  \hline
Source & $\overline{\mu}$ & $\mu_{\text{2.5\%}}$ & $\mu_{\text{97.5\%}}$ & $\overline{k}$ & $k_{\text{2.5\%}}$ & $k_{\text{97.5\%}}$ \\ 
  \hline
Peterman, 1974, Thesis, Fig. 4, Plot: Elk & 119.6 & 104.6 & 136.6 & 9.7 & 5.2 & 16.1 \\ 
  Peterman, 1974, Thesis, Fig. 4, Plot: Lean & 84.8 & 75.3 & 95.6 & 3.9 & 2.8 & 5.2 \\ 
  Peterman, 1974, Thesis, Fig. 4, Plot: Pars & 75.1 & 63.3 & 89.1 & 2.9 & 1.9 & 4.1 \\ 
  Peterman, 1974, Thesis, Fig. 4, Plot: Terr & 76.5 & 66.5 & 87.7 & 3.6 & 2.5 & 5.1 \\ 
  Safranyik, 1968, Thesis, Fig. 32, Plot: Elk creek  & 32.1 & 23.3 & 43.9 & 6.5 & 1.7 & 16.9 \\ 
  Safranyik, 1968, Thesis, Fig. 32, Plot: Horsethief creek & 23.8 & 20.9 & 26.9 & 23.8 & 7.3 & 68.8 \\ 
  Shepard, 1965, Can. Ento., Fig. 4 & 46.4 & 37.4 & 57.7 & 3.2 & 1.7 & 5.3 \\ 
  Waring and Pitman, 1985, Ecol. Fig. 1 & 92.6 & 78.6 & 110.2 & 2.1 & 1.5 & 2.9 \\ 
   \hline
\end{tabular}
\caption{Mean and dispersion parameters ($\mu$ and $k$ respectively) for the negative binomial distribution, fit to the distribution of attack density (units: females per square meter) across successfully and unsuccessfully attacked trees. Posterior means and 95\% credible intervals endpoints are displayed. The data are sourced from \citet{peterman1974some, safranyik1968development, shepard1965distribution, waring1985modifying}. The dispersion parameter $k$ may be scale-dependent, so it is important to note the sizes of the study areas. All studies were conducted at the stand level, but the sizes of the study areas are not directly comparable due to disparate sampling methodologies (i.e., transect vs hierarchical spatial sampling).} 
\label{agg_stats}
\end{table}






\section{Model fitting details and diagnostics} \label{Fitting the models}

We checked a standard suite of summary statistics to provide evidence that the Markov Chains had converged to the global posterior and were efficiently sampling. In particular, $R < 1.1$ for all parameters (i.e., all chains had mixed), the effective sample size per iteration was greater than 0.001 (i.e., efficient sampling), the energy Bayesian fraction of missing information (E-BFMI) was less than 0.2 (i.e., the model was not severely misspecified), and the proportion of divergent trajectories was far less than 1\% (i.e., the posterior geometry did not lead do biased estimation). This diagnostic analysis is provided in the R markdown file \textit{model\_diagnostics.Rmd} and its output \textit{model\_mcmc\_diagonstic\_printoff.txt}, in the supplementary files. See \citet[Ch. 6]{gelman2014bayesian} for more information on the aforementioned summary statistics.

To show that the marginal prior distributions had negligible influence on the posterior distribution, we computed the posterior contraction,  
\begin{equation}
\text{post. contraction} = 1 - \frac{ \mathbb{V}_{\mathrm{post}}} { \mathbb{V}_{\mathrm{prior}}}.
\end{equation}
This is a parameter-specific metric that shows the influence of the data on parameter estimation. For nearly every parameter in every model, the prior contraction was above 0.9, implying negligible prior influence. The exception is the Allee threshold parameter in the complex outbreak dynamics threshold. The posterior contraction analysis is provided in the R markdown file \textit{model\_diagnostics.Rmd} and its output \textit{model\_post\_contract\_printoff.txt}, in the supplementary files.

The low value of the posterior contraction for the parameter $a$ indicates that the \citeauthor{klein1978attack} data is not informative with respect to estimating the Allee effect. This is unsurprising, as discerning the Allee effect is challenging without data from trees that were attacked but not successfully killed (see Section \ref{life_history:allee}), data which is absent in \citepos{klein1978attack} publication. However, the Allee threshold estimate is not devoid of information, since the prior distribution was derived from the Allee effect sub-model fit to the \citeauthor{boone2011efficacy} dataset (see Table \ref{tab:pars_allee} \& Appendix \ref{subsec:allee}). 

\begin{table}[H]
\centering
\begin{tabular}{llllll}
  \hline
Parameter & Short description & Mean & SD & $\text{CI}_{2.5\%}$ & $\text{CI}_{97.5\%}$ \\ 
  \hline
$\beta_0$ & pref.$\sim$DBH intercept & -3.42 & 0.773 & -4.84 & -1.93 \\ 
  $\beta_1$ & pref.$\sim$DBH slope & 0.453 & 0.0688 & 0.319 & 0.580 \\ 
  $\sigma_{\text{pref0}}$ & pref. baseline noise & 0.628 & 0.0830 & 0.484 & 0.814 \\ 
  $\sigma_{\text{pref1}}$ & pref. sampling noise & 0.154 & 0.112 & 0.00663 & 0.417 \\ 
  $\gamma$ & dead tree$\sim$attack slope & 0.0151 & 0.000878 & 0.0134 & 0.0168 \\ 
  $\sigma_{\text{dead}}$ & dead tree noise & 0.00508 & 0.000657 & 0.00398 & 0.00653 \\ 
  $\zeta_0$ & emerge$\sim$DBH intercept & -9.86 & 49.2 & -106 & 87.8 \\ 
  $\zeta_1$ & emerge$\sim$DBH slope & 25.4 & 5.21 & 15.3 & 35.5 \\ 
  $\sigma_{\text{exit}}$ & emerge noise & 3.61 & 1.52 & 1.76 & 7.42 \\ 
  $s_0$ & max dispersal survival & 0.477 & 0.0701 & 0.363 & 0.639 \\ 
  $\sigma_0$ & residual noise & 363 & 88.7 & 225 & 568 \\ 
  $\lambda_{\text{resid}}$ & dispersal threshold & 0.652 & 0.0353 & 0.579 & 0.708 \\ 
  $\lambda_1$ & dispersal sensitivity & 49.7 & 96.6 & 17.2 & 262 \\ 
  a & Allee threshold & 432 & 363 & 13.5 & 1338.88825 \\ 
  K & forest carrying capacity & 904 & 313 & 413 & 1630.37825 \\ 
  m & annual tree mortality & 0.0518 & 0.00637 & 0.0376 & 0.0634 \\ 
  \midrule
  Derived quantity & Short description  & Mean & SD & $\text{CI}_{2.5\%}$ & $\text{CI}_{97.5\%}$ \\
  \midrule 
  $\sigma_{\text{tree}}$ & tree abund. noise & 0.482 & 0.211 & 0.240 & 1.02 \\ 
  $-\beta_1 / \beta_0$ & susc. DBH threshold & 7.46 & 0.666 & 5.98 & 8.47 \\ 
  $1/\gamma$  & attack density & 66.3 & 3.88 & 59.4 & 74.4 \\ 
  $\zeta_0 + \zeta_1 * 10$ & emerge density (10 in. DBH) & 244 & 25.1 & 191 & 296 \\ 
  $K-\sum_{t=0}^{17} (1-m)^{t}mK$ & Overstory density & 338 & 86.0 & 193 & 529 \\ 
  \bottomrule
\end{tabular}
\caption{Summary statistics for \textit{all} model parameters, including the Mean, standard deviation, and parameter credible intervals ($\text{CI}_{2.5\%}$, $\text{CI}_{97.5\%}$). Here, the parameters come from the full hierarchical model (Appendix \ref{Statistical model description}) fit to the \citeauthor{klein1978attack} dataset. The table also includes derived quantities, including the DBH of the smallest susceptible trees, the model-estimated attack density (gallery starts per square meter), and the emergence density (females per square meter) on a tree with a 10-inch DBH, which is approximately the average DBH of attacked trees in the \citeauthor{klein1978attack} dataset. Finally, the table contains fixed parameters, whose values were obtained via literature search. For parameter units, and the summary statistics of sub-model noise parameters, see supplementary Table \ref{tab:model_notation}.} 
\label{tab:pars_fullest}
\end{table}

\begin{table}[H]
\centering
\begin{tabular}{llllll}
  \hline
Parameter & Short description & Mean & SD & $\text{CI}_{2.5\%}$ & $\text{CI}_{97.5\%}$ \\ 
  \hline
a & Allee threshold & 2492 & 569 & 1631 & 3809 \\ 
  $\sigma_{\text{Allee}}$ & Pr. success noise & 0.207 & 0.03 & 0.158 & 0.275 \\ 
   \hline
\end{tabular}
\caption{Summary statistics for Allee effect model parameters, including the Mean, standard deviation, and parameter credible intervals ($\text{CI}_{2.5\%}$, $\text{CI}_{97.5\%}$). Here, the parameters come from the Allee effect sub-model (Appendix \ref{subsec:allee}) fit to the \citeauthor{boone2011efficacy} dataset.} 
\label{tab:pars_allee}
\end{table}

\section{Landscape analysis of bark beetle damage} \label{landscape} 

Here, we analyze the relationship between bark beetle damage and host tree biomass across western North America, resulting in Figure \ref{fig:damage_by_species} (in the main text). We focused on entirety of British Columbia and regions 1--6 of the United States (defined by the United States Forest Service), as these areas contain the majority of bark beetle data and initiated data collection earlier than other regions. The temporal scope of our analysis begins with the earliest start date for Aerial Detection Survey (ADS) data in each region, or 1962 when ADS data preceded this date, since surveys were not comprehensive before 1962. Specifically, the start years are: Region 1 (1962), Region 2 (1994), Region 3 (1997), Region 4 (1991), Region 5 (1996), Region 6 (1962), British Columbia (1962), and Alberta (1998). All regions have an end date of 2021.

Bark beetle damage is estimated using publicly available Aerial Detection Survey (ADS) data. This comprehensive dataset is collected annually through aerial surveillance of all coniferous forests within defined regions. Trained surveyors fly over these areas, delineating polygons that represent the extent of observed forest damage while simultaneously estimating damage severity (e.g., percentage of trees affected within each polygon). Due to variations in severity metrics across space and time, we simplified our analysis by measuring bark beetle damage as the total affected area, summing affected polygon areas across all years and regions. Host tree species information primarily comes from \citet{wood1982bark}, with additional references for the North American spruce beetle \citep{USDA_SpruceBeetle, BC_SpruceBeetle} and the southern pine beetle \citep{NC_SouthernPineBeetle, USDA_southern}. Host-trees are given in Table \ref{tab:host_trees}. For host-tree biomass data, we utilized multiple sources. In British Columbia, we used species-specific estimates of total live above-ground biomass from \citet{beaudoin2014mapping}. In the United States, we calculated biomass by multiplying relative abundances from Individual Tree Species Parameter (ITSP) maps \citep{ellenwood2015national} by total biomass data from \citet{blackard2008mapping}.

Our methodology involved classifying all bark beetles in the ADS data into three categories: irruptive (also referred to as excitable or aggressive), pulse-driven (semi-aggressive or opportunistically aggressive), and non-outbreaking (non-aggressive or lower-stem beetles). This classification is based on subjective judgments gleaned from the literature, as there is no single diagnostic feature of irruptive bark beetle species. We designated the following species as irruptive: mountain pine beetle, European spruce beetle, southern pine beetle, western pine beetle, and western balsam bark beetle, supported by \citet{singh2024understanding}, \citet{howe2022landscape}, \citet{howe2022numbers}, \citet{lantschner2023spatiotemporal}, \citet{biedermann2019bark}, \citet{koontz2021cross}, and \citet{weed2015population}. Pulse-driven species were defined as species that could outbreak but were not known to be irruptive (using Table 1 in \citet{lantschner2023spatiotemporal}, Table 8.1 in \citet{six2015dendroctonus}, and Table 4.1 in \citet{weed2015population}) Non-outbreaking species are provided by \citet{howe2022numbers}; all remaining beetle species in the ADS data were assumed to be non-outbreaking.

\begin{landscape}
\small
\begin{longtable}{>{\raggedright\arraybackslash\linespread{0.8}\selectfont}p{3cm}>{\raggedright\arraybackslash\linespread{0.8}\selectfont}p{3cm}>{\raggedright\arraybackslash\linespread{0.8}\selectfont}p{3cm}>{\raggedright\arraybackslash\linespread{0.8}\selectfont}p{3cm}>{\raggedright\arraybackslash\linespread{0.8}\selectfont}p{4cm}} 
\label{tab:host_trees}
\textbf{BB Common Name} & \textbf{BB Genus Species} & \textbf{Host Common Name} & \textbf{Host Genus Species} & \textbf{Notes} \\ \hline
\endfirsthead
\hline
\textbf{BB Common Name} & \textbf{BB Genus Species} & \textbf{Host Common Name} & \textbf{Host Genus Species} & \textbf{Notes} \\ \hline
\endhead
\hline
\endfoot
\hline
\endlastfoot
roundheaded pine beetle & Dendroctonus adjunctus & Mexican white pine & Pinus ayacahuite & Mexico \& CentAm \\
 &  & Mexican mountain pine & Pinus hartwegii & Mexico \& CentAm \\
 &  & Chihuahua pine & Pinus leiophylla & Mexico \& S. NM/AZ \\
 &  & Montezuma pine & Pinus montezumae & Mexico \& CentAm \\
 &  & ponderosa pine & Pinus ponderosa & BC \& W. USA \\
 &  & smooth-bark Mexican pine & Pinus pseudostrobus & Mexico \& CentAm \\
 &  & thinleaf pine & Pinus maximinoi & Mexico \& CentAm \\
\hline
western pine beetle & Dendroctonus brevicomis & ponderosa pine & Pinus ponderosa & BC \& W. USA \\
\hline
Jeffrey pine beetle & Dendroctonus jeffreyi & Jeffrey pine & Pinus jeffreyi & CA, NV, OR, NW Mexico \\
 &  & ponderosa pine & Pinus ponderosa (very rare) & BC \& W. USA \\
\hline
lodgepole pine beetle & Dendroctonus murrayanae & jack pine & Pinus banksiana & CAN: BC-NS, USA: MN-ME \\
 &  & lodgepole pine & Pinus contorta & W. NorAm \\
 &  & eastern white pine & Pinus strobus & NL to great lakes \\
\hline
mountain pine beetle & Dendroctonus ponderosae & Engelmann spruce & Picea engelmanii & W. NorAm \\
 &  & whitebark pine & Pinus albicaulis & W. Can \& W. USA \\
 &  & foxtail pine & Pinus balfouriana & CA only \\
 &  & jack pine & Pinus banksiana &  \\
 &  & lodgepole pine & Pinus contorta &  \\
 &  & Coulter/big-cone pine & Pinus coulteri & S. CA \& NW Mexico \\
 &  & pinyon pine & Pinus edulis & SW USA \\
 &  & limber pine & Pinus flexilis & W NorAm \\
 &  & Jeffrey pine & Pinus jeffreyi (rare) &  \\
 &  & sugar pine & Pinus lambertiana & OR to NW Mexico \\
 &  & single-leaf pinyon pine & Pinus monophylla & ID to NW Mexico \\
 &  & western white pine & Pinus monticola & W. Can \& USA \\
 &  & ponderosa pine & Pinus ponderosa & BC \& W. USA \\
 &  & Chihuahua white pine & Pinus strobiformis & Mexico \& SW USA \\
\hline
douglas-fir beetle & Dendroctonus pseudotsugae & western larch/tamarack & Larix occidentalis (rare) & NW Can \& USA \\
 &  & bigcone spruce/bigcone Douglas-fir & Pseudotsuga macrocarpa & CA only \\
 &  & Douglas-fir & Pseudotsuga menziesii & W. NorAm \\
 &  & western hemlock & Tsuga heterophylla & NW Can and USA incl AK \\
\hline
spruce beetle & Dendroctonus rufipennis & spruce species & All Picea species in its range (mostly engelmann and white spruce) &  \\
\hline
eastern larch beetle & Dendroctonus simplex & eastern larch/tamarack & Larix laricina & All Can, AK \& NE USA \\
\hline
red turpentine beetle & Dendroctonus valens & white fir & Abies concolor & W. USA \\
 &  & eastern larch/tamarack & Larix larciana &  \\
 &  & Norway spruce & Picea excelsa & Introduced NE USA \& Can \\
 &  & white spruce & Picea glauca & All Can, AK, \& N. USA \\
 &  & red spruce & Picea rubens & NE USA, Quebec-NS \\
 &  & lodgepole pine & Pinus contorta &  \\
 &  & Coulter/big-cone pine & Pinus coulteri &  \\
 &  & shortleaf pine & Pinus echinata & SE USA \\
 &  & pinyon pine & Pinus edulis &  \\
 &  & Mexican mountain pine & Pinus hartwegii &  \\
 &  & Jeffrey pine & Pinus jeffreyi &  \\
 &  & sugar pine & Pinus lambertiana &  \\
 &  & Lawson's pine & Pinus lawsonii & Mexico \\
 &  & Chihuahua pine & Pinus leiophylla &  \\
 &  & western white pine & Pinus monticola &  \\
 &  & Sierra lodgepole pine & Pinus murrayana &  \\
 &  & Mexican yellow pine & Pinus oocarpa & Mexico \& CentAm \\
 &  & ponderosa pine & Pinus ponderosa & BC \& W. USA \\
 &  & smooth-bark Mexican pine & Pinus pseudostrobus &  \\
 &  & Monterey pine & Pinus radiata & Coastal CA \& Mexico \\
 &  & red pine & Pinus resinosa & E Can \& NE USA \\
 &  & pitch pine & Pinus rigida & E. USA \\
 &  & California foothill pine & Pinus sabiniana & California only \\
 &  & Chihuahua white pine & Pinus strobiformis &  \\
 &  & eastern white pine & Pinus strobus &  \\
 &  & Scotch pine & Pinus sylvestris & Introduced \\
 &  & thinleaf pine & Pinus maximinoi &  \\
 &  & Virginia pine & Pinus virginiana & E. USA \\
\hline
NA & Dryocoetes autographus & fir species & Abies spp. &  \\
 &  & Engelmann spruce & Picea engelmanii &  \\
 &  & white spruce & Picea glauca &  \\
 &  & lodgepole pine & Pinus contorta &  \\
 &  & western white pine & Pinus monticola &  \\
 &  & eastern white pine & Pinus strobus &  \\
 &  & Douglas-fir & Pseudotsuga menziesii &  \\
 &  & western hemlock & Tsuga heterophylla &  \\
\hline
western balsam bark beetle & Dryocoetes confusus & pacific silver fir & Abies amabilis (rare) & W. BC, WA, OR \\
 &  & white fir & Abies concolor (rare) &  \\
 &  & subalpine fir & Abies lasiocarpa &  \\
 &  & Engelmann spruce & Picea engelmanii &  \\
\hline
pinyon ips & Ips confusus & Colerado pinyon pine & Pinus edulis &  \\
 &  & single-leaf pinyon pine & Pinus monophylla &  \\
\hline
emarginate Ips & Ips emarginatus & lodgepole pine & Pinus contorta (rare) &  \\
 &  & Jeffrey pine & Pinus jeffreyi &  \\
 &  & western white pine & Pinus monticola &  \\
 &  & ponderosa pine & Pinus ponderosa & BC \& W. USA \\
\hline
blue spruce engraver & Ips hunteri & Engelmann spruce & Picea engelmanii (rare) &  \\
 &  & blue spruce & Picea pungens &  \\
\hline
California fivespined Ips & Ips paraconfusus & knobcone pine & Pinus attenuata &  \\
 &  & lodgepole pine & Pinus contorta (rare) &  \\
 &  & Coulter/big-cone pine & Pinus coulteri &  \\
 &  & Jeffrey pine & Pinus jeffreyi &  \\
 &  & sugar pine & Pinus lambertiana &  \\
 &  & ponderosa pine & Pinus ponderosa & BC \& W. USA \\
\hline
northern spruce engraver & Ips perturbatus & white spruce & Picea glauca &  \\
\hline
pine engraver & Ips pini & jack pine & Pinus banksiana &  \\
 &  & lodgepole pine & Pinus contorta &  \\
 &  & Jeffrey pine & Pinus jeffreyi &  \\
 &  & ponderosa pine & Pinus ponderosa & BC \& W. USA \\
 &  & eastern white pine & Pinus strobus &  \\
\hline
western cedar bark beetle & Phloeosinus spp. & Rocky Mountain juniper & Juniperus scopulorum &  \\
 &  & Utah juniper & Juniperus osteosperma &  \\
 &  & one-seed juniper & Juniperus monosperma &  \\
\hline
NA & Pityophtorus pseudotsugae & white fir & Abies concolor &  \\
 &  & grand fir & Abies grandis &  \\
 &  & subalpine fir & Abies lasiocarpa &  \\
 &  & red fir & Abies magnifica &  \\
 &  & Engelmann spruce & Picea engelmanii &  \\
 &  & sugar pine & Pinus lambertiana &  \\
 &  & Douglas-fir & Pseudotsuga menziesii &  \\
 &  & western hemlock & Tsuga heterophylla &  \\
\hline
fir root bark beetle & Pseudohylesinus granulatus & pacific silver fir & Abies amabilis &  \\
 &  & grand fir & Abies grandis &  \\
 &  & subalpine fir & Abies lasiocarpa &  \\
 &  & red fir & Abies magnifica &  \\
 &  & noble fir & Abies procera &  \\
 &  & western hemlock & Tsuga heterophylla &  \\
\hline
Douglas-fir pole beetle & Pseudohylesinus nebulosus & Douglas-fir & Pseudotsuga menziesii &  \\
 &  & western hemlock & Tsuga heterophylla (rare) &  \\
\hline
silver fir beetle & Pseudohylesinus sericeus & pacific silver fir & Abies amabilis &  \\
 &  & grand fir & Abies grandis &  \\
 &  & noble fir & Abies procera &  \\
 &  & Douglas-fir & Pseudotsuga menziesii &  \\
 &  & western hemlock & Tsuga heterophylla &  \\
\hline
Douglas-fir engraver & Scolytus unispinosus & Douglas-fir & Pseudotsuga menziesii &  \\
\hline
fir engraver & Scolytus ventralis & white fir & Abies concolor &  \\
 &  & grand fir & Abies grandis &  \\
 &  & red fir & Abies magnifica &  \\
\hline
Mediterranean oak borer & Xyleborus monographus & blue oak & Quercus douglasii &  \\
 &  & valley oak & Quercus lobata &  \\
 \hline
southern pine beetle & Dendroctonus frontalis & shortleaf pine & Pinus echinata &  \\
 &  & loblolly pine & Pinus taeda &  \\
 &  & pitch pine & Pinus rigida &  \\
 &  & pond pine & Pinus serotina &  \\
 &  & Virginia pine & Pinus virginiana & 
\end{longtable}
\end{landscape}

\section{Additional supplementary figures} \label{Supplementary Figures}

\begin{figure}[H]
\centering
\includegraphics[scale = 1]{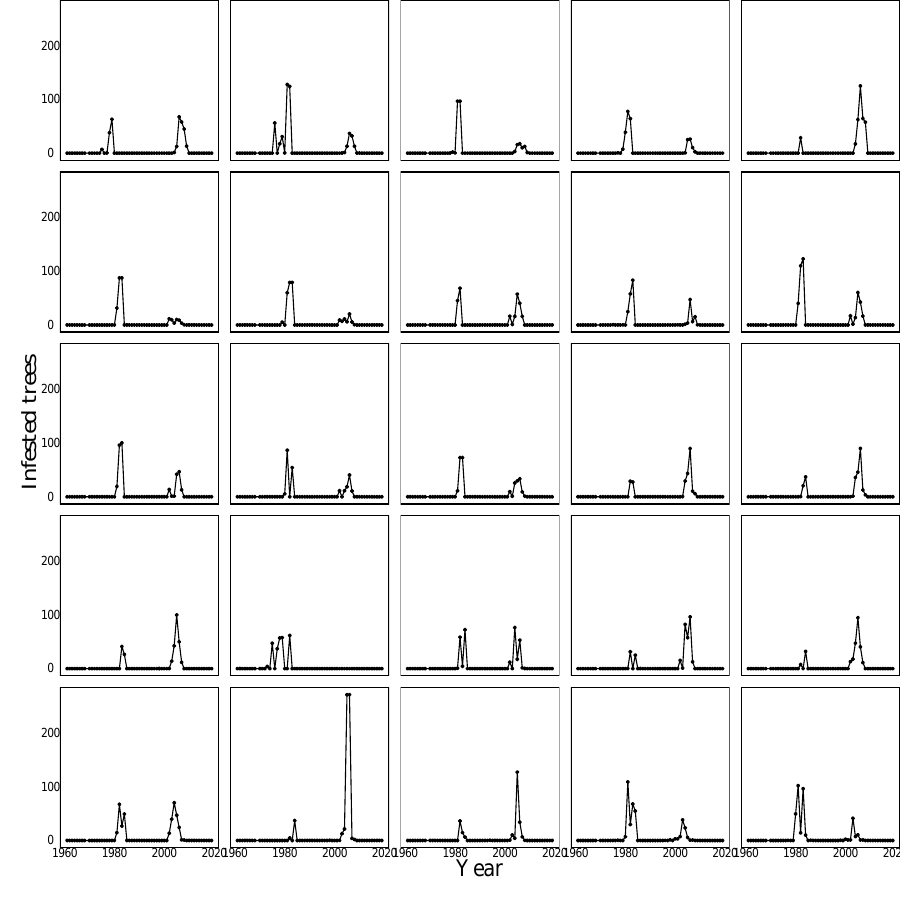}
\caption{Sample time series of MPB-infested trees from British Columbia's Aerial Overview Surveys. Each panel is a randomly selected $\text{1 km}^2$ patch from a subset of locations where cumulative infestations were in the upper quartile. To estimate the number of infested trees, we multiply the proportion of infested stems by $\frac{\text{Live aboveground biomass (units: Mg/ha)}}{10}$, which is an approximation for pine stem density that was validated by \citeauthor{koch2021signature} (\citeyear{koch2021signature}, Appendix A), using the data and analyses of \citet{nigh2008density}. The biomass data is sourced from \citet{beaudoin2014mapping}.}
\label{fig:sample_time_series}
\end{figure}

\begin{figure}[H]
\centering
\makebox[\textwidth]{\includegraphics[scale=0.9]{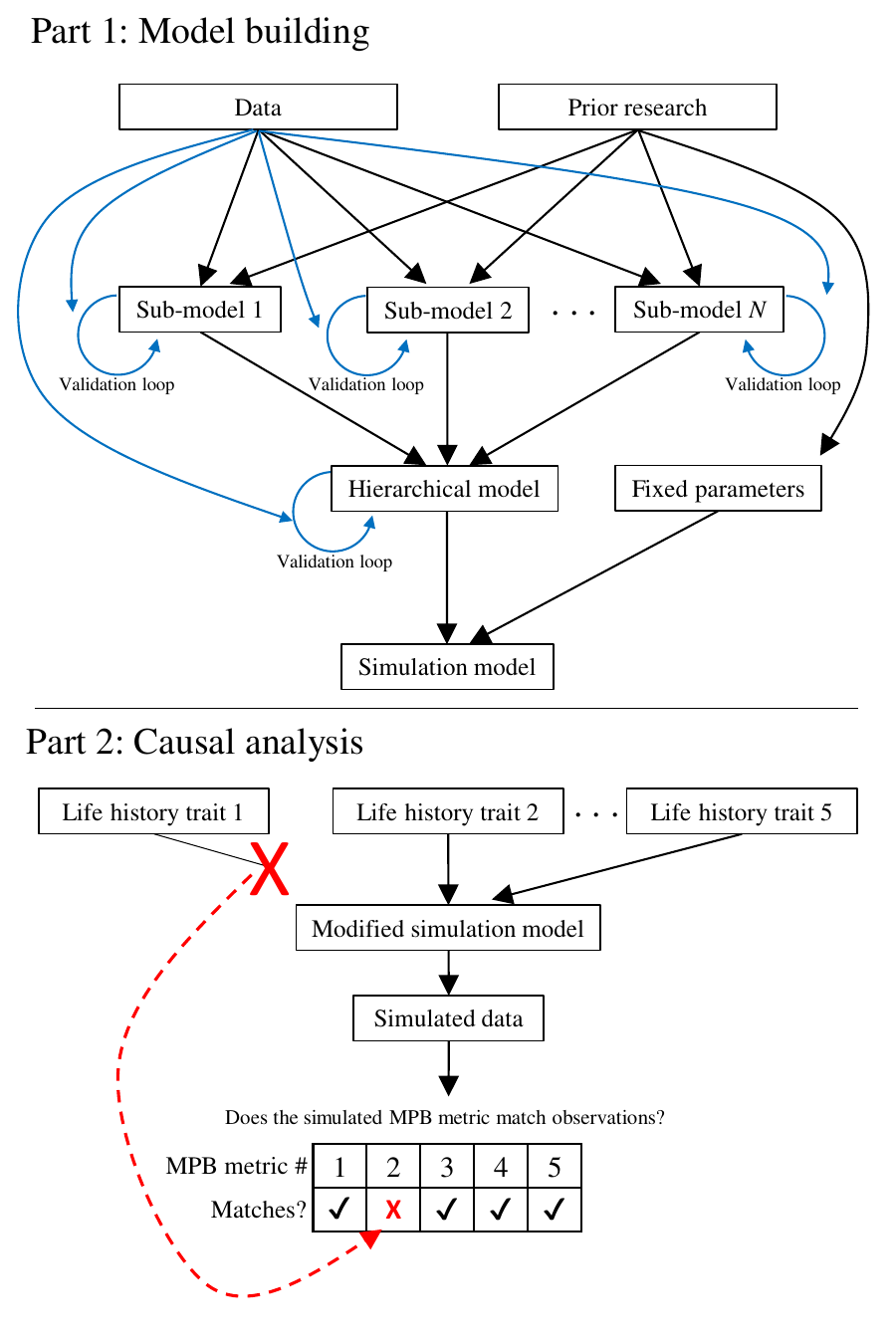}}
\caption{Graphical representation of our inferential approach. First, we develop and validate a realistic model of MPB dynamics. Second, we perform a causal analysis in which we turn biological features on or off. In the example of the causal analysis, the red crosses respectively indicate that feature \# 1 has been off, and that the simulated system does not match reality with respect to Metric \#2 (e.g., outbreak periodicity). The dashed red arrow indicates that feature \#1 is responsible for realistic values of Metric \#2. }
\label{fig:paper_concept_fig}
\end{figure}

\begin{figure}[H]
\centering
\includegraphics[scale = 1]{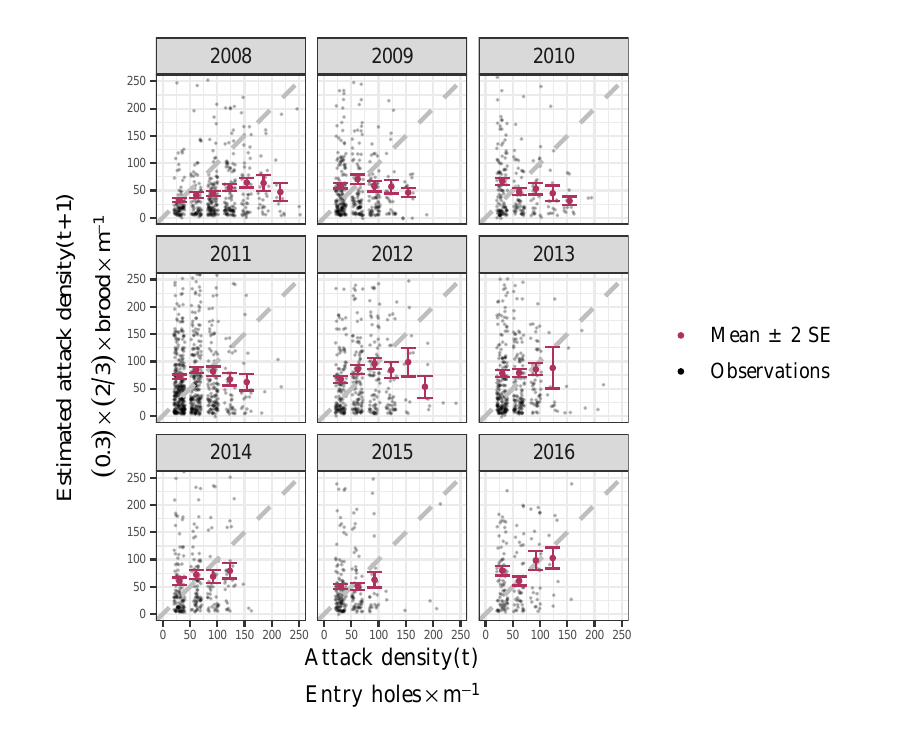}
\caption{Population maps provide evidence for compensatory density dependence at the tree level. The data comes from Alberta's disk surveys (described by \citealp{government2016mountain}, and published by \citealp{goodsman2018effect}). The estimate for attack density at time $t+1$ involves the constants $(2/3)$ (the average proportion of the brood that will be female) and $0.3$ (an estimate of dispersal-phase mortality from the \cite{klein1978attack} dataset, see Fig.\,\ref{fig:mort2}).}
\label{fig:disk_data_pop_map}
\end{figure}

\begin{figure}[H]
\centering
\includegraphics[scale = 1]{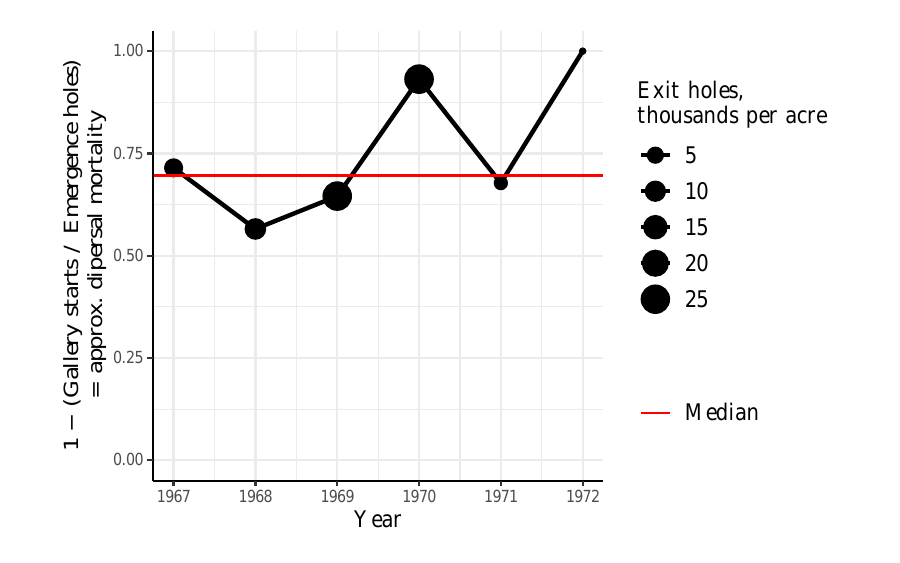}
\caption{An approximation of dispersal mortality as a function of time in the \citeauthor{klein1978attack} data. The dispersal mortality is approximated as the quotient of gallery starts and preceding emergence holes. The quotient captures both dispersal mortality and mortality from unsuccessful attacks (i.e., the Allee effect, \eqref{eq:allee}). Therefore, the approximation will be better when emergence density is larger (i.e., before 1971), because the Allee effect will be less important and the relative influence of inter-patch dispersal will be less important. The first three years can be used to approximate the baseline dispersal mortality (i.e., $1-s_0$ in \eqref{eq:disp_surv}).}
\label{fig:mort2}
\end{figure}

\begin{figure}[H]
\centering
\includegraphics[scale = 1]{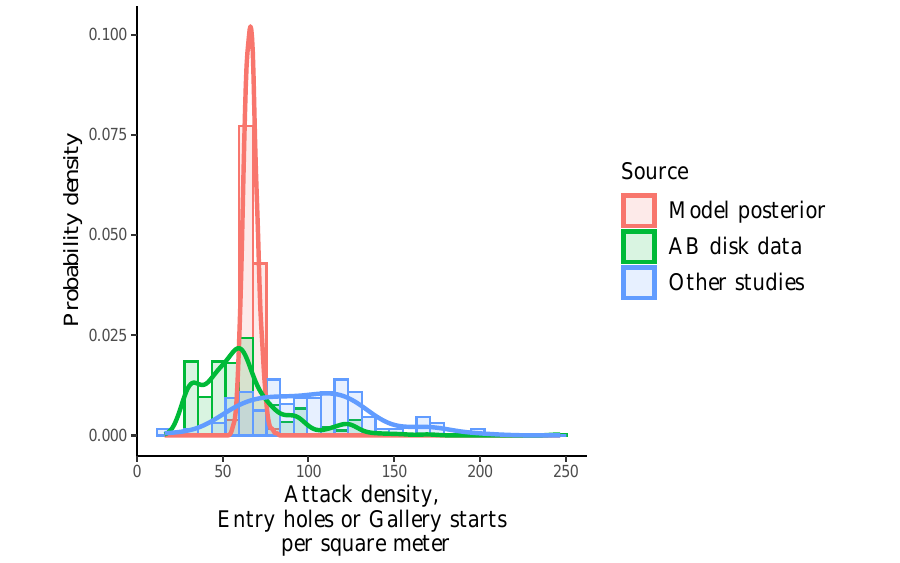}
\caption{Frequency distributions of attack density. The \textit{Model posterior} category gives the marginal posterior of $1/\gamma$ from our statistical model of the \citeauthor{klein1978attack} outbreak. The \textit{AB disk data} category contains the within-site means of attack density from 300 different sites across Alberta \citep{government2016mountain}. The \textit{Other studies} category includes estimates obtained via entry holes or gallery starts from a number of studies, mostly United States Forest Service report: \citet{tishmack2005mountain, mccambridge1967nature, negron2018biological, reid1963biology, raffa1983role, rasmussen1980emergence, shepard1965distribution, knight1959measuring, de1939biology, safranyik1988estimating, safranyik1971some, safranyik1968development, parker1979mountain, schmid1972emergence, blackman1931black, whiteside1937progress, washburn1959mountain, author1963mountain, peterman1974some, klein1978attack, safranyik1991unseasonably}.}
\label{fig:hist_attack}
\end{figure}

\begin{figure}[H]
\centering
\includegraphics[scale = 1]{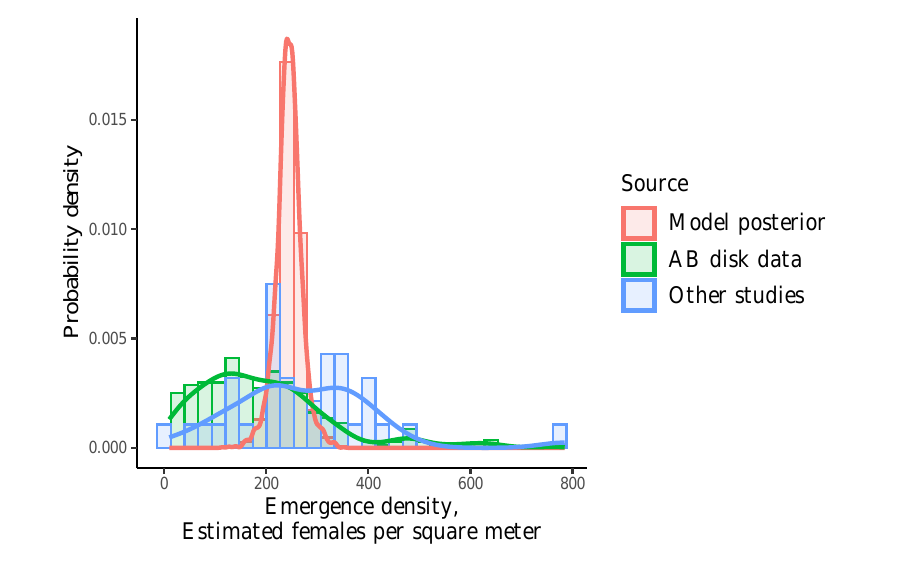}
\caption{Frequency distributions of female emergence density. Females are estimated with $(2/3) \times (3/2) \times$ emergence holes, or with $(2/3) \times$ emerging beetles or spring brood size. Note that (2/3) is the typical proportion of females in a brood \citep{reid1962biology, cole1976mathematical}, and (3/2) is the average number of beetles emerging from each hole \citep[Fig. 13]{safranyik1985relationship, peterman1974some}. The \textit{Model posterior} category gives the marginal posterior of $\zeta_0 + \zeta_1 * 10$, the emergence density on a 10-inch DBH tree; 10 inches is approximately the average DBH of MPB-killed trees in the \citeauthor{klein1978attack} dataset. The \textit{AB disk data} category contains estimates of emergence from 300 different sites across Alberta \citep{government2016mountain}. The \textit{Other studies} category includes estimates from a number of studies, mostly United States Forest Service reports: \citet{tishmack2005mountain, negron2018biological, reid1963biology, raffa1983role, rasmussen1980emergence, de1939biology, safranyik1988estimating, cole1969mountain, schmid1972emergence, whiteside1937progress, beal1938progress, peterman1974some, klein1978attack, safranyik1985relationship,safranyik1991unseasonably}.}
\label{fig:emerge_hist}
\end{figure}

\begin{figure}[H]
\centering
\includegraphics[scale = 0.6]{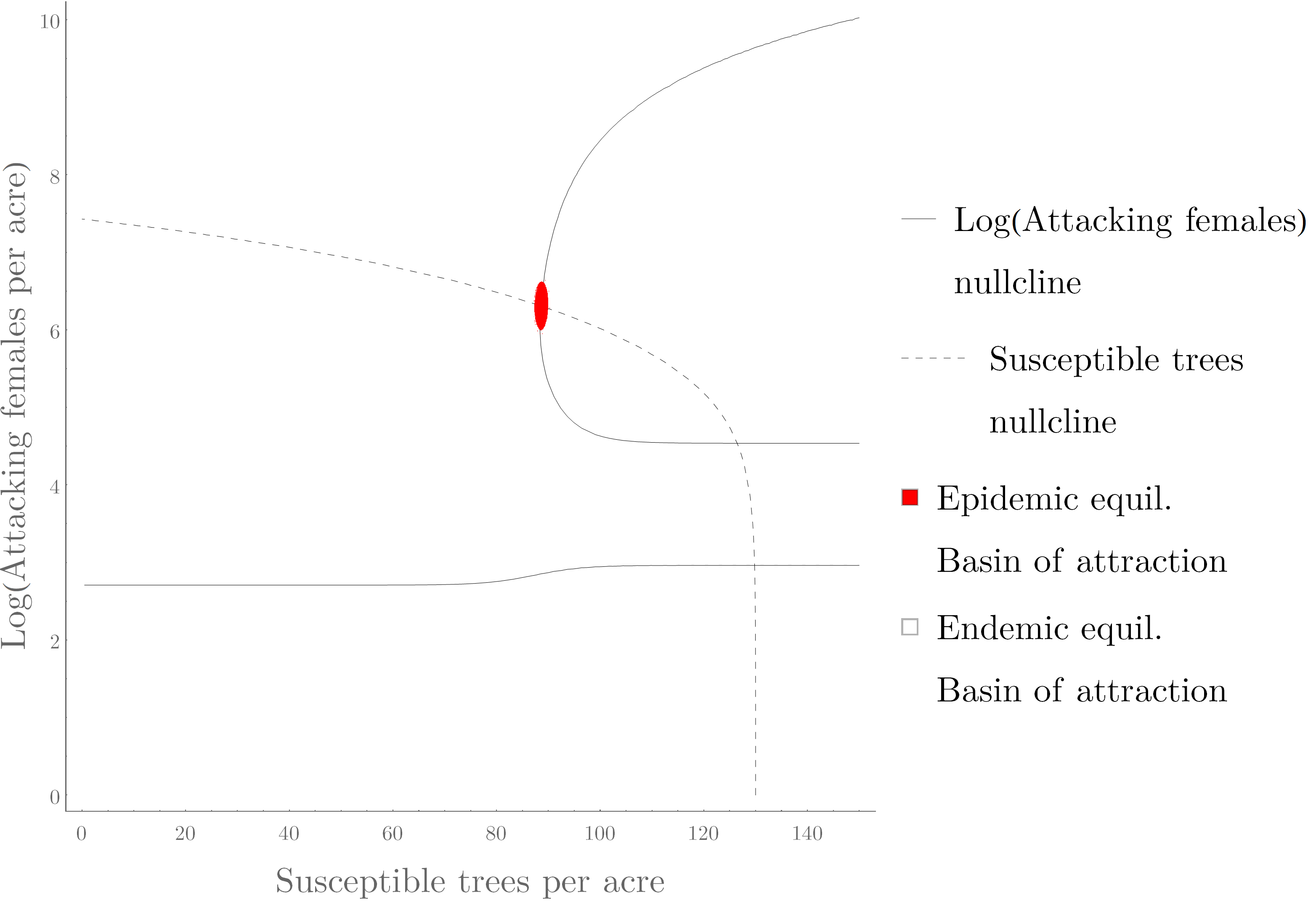}
\caption{The stable epidemic equilibrium (i.e., the intersection of the nullclines with the highest beetle density) has a very small basin of attraction. All other initial conditions lead to the endemic equilibrium. }
\label{fig:phase_boa}
\end{figure}

\end{appendices}

\bibliographystyle{apalike}
\bibliography{mpb_refs.bib}


\end{document}